\documentclass[final,3p,times]{elsarticle}

\journal{***} 

\usepackage{float} 
\usepackage{amsmath}
\usepackage{amssymb}
\usepackage{subcaption}
\usepackage{color}
\usepackage{tabularx}
\usepackage[export]{adjustbox}
\usepackage{longtable}
\renewcommand{\vec}[1]{\underline{\boldsymbol{#1}}}
\newcommand{\ten}[1]{\underline{\underline{\boldsymbol{#1}}}}
\newcommand{\mat}[1]{\mathbf{#1}}

\usepackage{url,color,bm,lineno,xspace,soul,ulem}

\usepackage{ifthen}
\newboolean{correction_mode}
\setboolean{correction_mode}{false}

\ifthenelse{\boolean{correction_mode}}
{
    \newcommand{\vyrm}[1]{{\color{myred}\sout{#1}}}
    \newcommand{\jvrm}[1]{{\color{myred1}\sout{#1}}}

    \newcommand{\vycm}[1]{{\color{mygreen}VY: #1}}
    \newcommand{\jvcm}[1]{{\color{mygreen1}\textbf{JV: #1}}}

    \newcommand{\abrm}[1]{{\color{red}\sout{#1}}}
    \newcommand{\abrcm}[1]{{\color{mygreen}ABR: #1}}
    \newcommand{\abrin}[1]{\color{blue}#1\xspace\color{black}}
    
}
{
    \newcommand{\vyrm}[1]{}
    \newcommand{\jvrm}[1]{}

    \newcommand{\vycm}[1]{}
    \newcommand{\jvcm}[1]{}

    \newcommand{\abrm}[1]{}
    \newcommand{\abrcm}[1]{}
    \newcommand{\abrin}[1]{\color{black}#1\xspace\color{black}}
    
}

\begin{document}
\begin{frontmatter}
\title{Stabilized MorteX method for mesh tying along embedded interfaces}
\author[cdm,st]{Basava Raju Akula}
\ead{basava-raju.akula@mines-paristech.fr}
\author[st]{Julien Vignollet}
\ead{julien.vignollet@safrangroup.com}
\author[cdm]{Vladislav A. Yastrebov\corref{cor1}}
\ead{vladislav.yastrebov@mines-paristech.fr}
\ead[url]{www.yastrebov.fr}
\cortext[cor1]{Corresponding author}
\address[cdm]{MINES ParisTech, PSL Research University, Centre des Mat\'eriaux, CNRS UMR 7633, BP 87, 91003 Evry, France}
\address[st]{Safran Tech, Safran Group, 78772 Magny-les-Hameaux, France}

\begin{abstract}
We present a unified framework to tie overlapping meshes in solid mechanics
applications. This framework is a combination of the X-FEM method and the mortar
method, which uses Lagrange multipliers to fulfill the tying constraints.  As
known, mixed formulations are prone to mesh locking which manifests itself by
the emergence of spurious oscillations in the vicinity of the tying interface.
To overcome this inherent difficulty, we suggest a new coarse-grained
interpolation of Lagrange multipliers. This technique consists in selective
assignment of Lagrange multipliers on nodes of  the mortar side and in non-local
interpolation of the associated traction field. The optimal choice of the
coarse-graining spacing is guided solely by the mesh-density contrast between
the mesh of the mortar side and the number of blending elements of the host
mesh. The method is tested on two patch tests (compression and bending) for
different interpolations and element types as well as for different material and
mesh contrasts. The optimal mesh convergence and removal of spurious
oscillations is also demonstrated on the Eshelby inclusion problem for high
contrasts of inclusion/matrix materials. Few additional examples confirm the
performance of the elaborated framework.
\end{abstract}

\begin{keyword}
    mesh tying \sep embedded interface  \sep MorteX method \sep stabilization \sep mortar method \sep X-FEM
\end{keyword}

\end{frontmatter}

\section{Introduction} \label{sec:intro}
\begin{figure}[htb!]
    \centering
    \includegraphics[width=.75\textwidth]{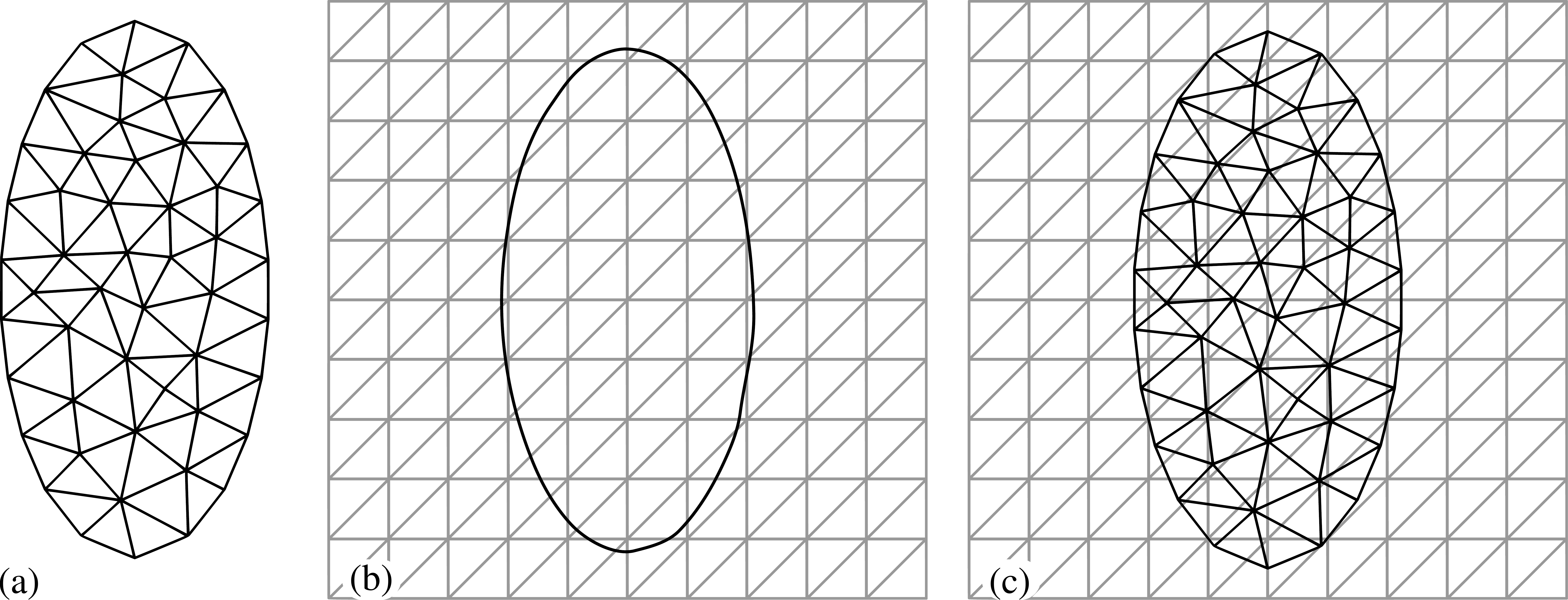}
    \caption{Illustration of meshes with (a) fitted boundary, (b) embedded boundary (for example, level set), (c) embedded mesh.}
    \label{fig:bf_ib_meshes}
\end{figure}

The finite element method (FEM) is used to solve a wide range of physical and
engineering problems. Based on a variational formulation and a discretized
representation of the geometry, this method is  extremely flexible in handling
complex geometries, non-linear and heterogeneous constitutive equations and
multi-physical/multi-field problems.  A classification of finite element models
can be proposed based on the strategy to represent the boundary of the
computational mesh. Classical FE meshes fall into the category of ``boundary
fitted'' (BF) methods, where the boundaries of the physical and computational
domains coincide [Fig.~\ref{fig:bf_ib_meshes}(a)]. Alternatively, for
``embedded/immersed boundary'' (EB) methods, the computational domain is a mesh or a
Cartesian grid hosting another physical domain [Fig.~\ref{fig:bf_ib_meshes}(b)].
Note that material properties or even the governing equations of the host medium and the embedded one can be different.
Within the EB method, the geometry contour can be embedded either fully or
partially.  
The BF methods [see Fig.~\ref{fig:bf_ib_meshes}(a)] can be used to solve
boundary value problems where the boundary conditions are prescribed on surfaces
explicitly represented by a mesh.  The EB methods  [see
Fig.~\ref{fig:bf_ib_meshes}(b)] can handle a broader class of problems.  First,
it can be used to solve the same boundary value problems as BF methods but with
the boundary represented by a rather general level-set function without explicit
discretization of the physical domain. Second, the embedded boundary can serve
as a material interface to include features such as inclusions, voids or even
cracks for linear/quadratic background meshes~\cite{fries_higher-order_2018}. In addition, within the so-called
CutFEM~\cite{burman_cutfem:_2015,claus_stable_2018}, debonding of the
interface can be taken into account as well as contact between faces. This class
of methods can be easily used to create complex
geometries~\cite{belytschko_structured_2003}, however the prescription
of boundary conditions on the embedded surface is not
straightforward~\cite{duboeuf_embedded_2017,duboeuf_embedded_2017-1}.  A
particular combination of BF and EB methods shown in
Fig.~\ref{fig:bf_ib_meshes}(c) deals simultaneously with two or several
superposed meshes, which can represent different physics or physical properties.
This framework is suited for applications involving fluid-structure interactions
(FSI)~\cite{baaijens_fictitious_2001,fournie_fictitious_2014,puso_embedded_2015}, where the
background mesh represents the fluid and the embedded mesh represents the solid.

In this work, we place ourselves in the context of continuum solid mechanics and
the finite element method to tie two overlapping meshes: one of these is a
boundary-fitted mesh and is referred to as the \emph{patch} mesh, which is fully or partly embedded in a non-boundary fitted \emph{host} mesh (the physical domain and the mesh do not coincide).
Note
that there is no restriction on the number of patches that can be embedded into
a host domain.  The primary issue addressed here is the continuity of
fields across the embedded boundaries. The standard methods to impose stiff
continuity or Dirichlet boundary conditions along embedded surfaces include the
penalty methods, the method of Lagrange multipliers, Nitsche methods and
their variants~\cite{moes_imposing_2006,sanders_methods_2009,haslinger_new_2009,ramos_new_2015}.
Here we consider a framework combining the features of the mortar domain
decomposition method and the extended finite element methods (X-FEM) to impose
continuity constraints. We refer to this framework as the MorteX method.

The X-FEM is an enrichment method based on the partition of unity
(PUM)~\cite{babuska_partition_1997}. 
In this framework, embedded surfaces, cracks or material interfaces can be modeled without explicit mesh conformity.
In X-FEM, it is achieved without compromise on the
optimal convergence by means of enrichment functions which are
added to the finite element approximation using the framework of PUM.  The X-FEM methods are extensively used in applications
 such as fracture mechanics, voids and
 inclusions modeling, \abrin{modeling discontinuities in the porous medium,
 crack propagation using cohesive elements}, simulation of shock
 wave front and oxidation front propagation, and other applications
    involving discontinuities both strong and
    weak~\cite{daux_arbitrary_2000,sukumar_modeling_2001,diez_stable_2013,gross_extended_2007,
    ji_hybrid_2002,savvas_homogenization_2014,faivre_2d_2016,ferte_3d_2016,sanchez-rivadeneira_stable_2019}. 

The mortar methods provide us with a comprehensive framework for mesh tying.
They are a subclass of domain decomposition methods
(DDM)~\cite{wohlmuth_discretization_2001,gosselet_non-overlapping_2006,keyes_domain_2007,mathew_domain_2008},
that are tailored for non-conformal spatial interface
discretizations~\cite{bernardi_new_1994}, and were originally introduced as DDM
for spectral elements~\cite{belgacem_spectral_1994,bernardi_coupling_1990}. The
    coupling and tying of different physical models, discretization schemes,
    and/or non-matching discretizations along interfaces can be ensured by
    mortar methods.  The mathematical optimality and applicability of the mortar
    methods in spectral and finite element frameworks were studied extensively
    for elliptic problems
        in~\cite{bernardi_coupling_1990,belgacem_spectral_1994,wohlmuth_discretization_2001}.
        The mortar methods have  been successfully adapted to solve contact
        problems~\cite{belgacem_mortar_1998,mcdevitt_mortar-finite_2000,puso_mortar_2004,gitterle_finite_2010,farah_mortar_2018}.

To ensure displacement continuity across the tied sub-domains, the mortar methods employ Lagrange multiplier
fields; as such it is mixed finite element formulation.
It is known that the choice of Lagrange multipliers interpolations strongly
affects the mesh convergence rate and may lead to loss of accuracy in the
interfacial tractions
reported for mixed variational formulations as a result of non-satisfaction of
Ladyzhenskaya-Babu\v{s}ka-Brezzi (LBB)  also called inf-sup
condition~\cite{babuska_finite_1973,brezzi_mixed_2012}.  Issues resulting from
the imposition of Dirichlet boundary conditions using Lagrange multipliers
methods has been a topic of interest in various domains, such as the classical
FEM~\cite{barbosa_finite_1991}, Interface-enriched Generalized Finite
Element Method (IGFEM)~\cite{ramos_new_2015}, the fictitious domain
methods~\cite{burman_fictitious_2010}, the mesh free
methods~\cite{fernandez-mendez_imposing_2004}, etc.  This problem has also been
dealt extensively within the context of the X-FEM. In
\cite{moes_imposing_2006,bechet_stable_2009,hautefeuille_robust_2012}, the
authors propose a strategy to construct an optimal 
Lagrange-multiplier space for the embedded interfaces which permits to apply
Dirichlet boundary conditions.  As opposed to the strategy of modifying the
Lagrange multiplier spaces, the authors in~\cite{sanders_methods_2009} propose a
stabilization method to mitigate the oscillatory behaviour of the standard  spaces. In the current work, we extend
the strategy of modified Lagrange multiplier spaces, which hereinafter will be referred to as
coarsening of Lagrange multiplier spaces. This strategy allows us to address
specific problems  of mesh-locking, which are inherent to mortar methods for overlapping
domains, in particular in presence of a strong contrast of material properties
in the vicinity of the interface.  

Few contributions harnessing the advantages of  the mortar method and the X-FEM (but in a different way from what is presented here) are
listed below.  In~\cite{chahine_etude_2008,chahine_non-conformal_2011} the
authors used the mortar methods to ensure weak continuity conditions across the interface between a coarser mesh domain and non-intersecting
finer mesh surrounding the crack, which in turn is represented by the X-FEM formulation. 
The tying in this case is limited to the interface with matching geometries but non-conformal discretizations, which is a classical application of the mortar method.  A
dual mortar contact formulation integrated into X-FEM fluid-structure
interaction approach is introduced in~\cite{mayer_3d_2010}. There, the combined
X-FEM fluid-structure-contact interaction method (FSCI) allows to compute contact
of arbitrarily moving and deforming structures embedded in a fluid.  
    
The proposed method of coupling mortar and X-FEM competes with the volumetric
coupling of the Arlequin method~\cite{dhia_arlequin_2005} and the Polytope FEM
for embedded interfaces~\cite{zamani_embedded_2011}. The Arlequin method
    involves superpositioning of the mechanical states in transition
    zone, and energy redistribution between these states using weight functions. The
    Polytope FEM involves the decomposition of elements cut by the embedded
    interface into new polytope elements. This is achieved by
    the creation of new degrees of freedom (DoFs) along
    the interface. In~\cite{sanders_nitsche_2012}, the authors used
    the Nitsche method to impose tying constraints for the overlapping
    domains  circumventing mesh-locking; this method however requires
    appropriate stabilization. 

\begin{figure}[htb!]
    \centering
    \includegraphics[width=\textwidth]{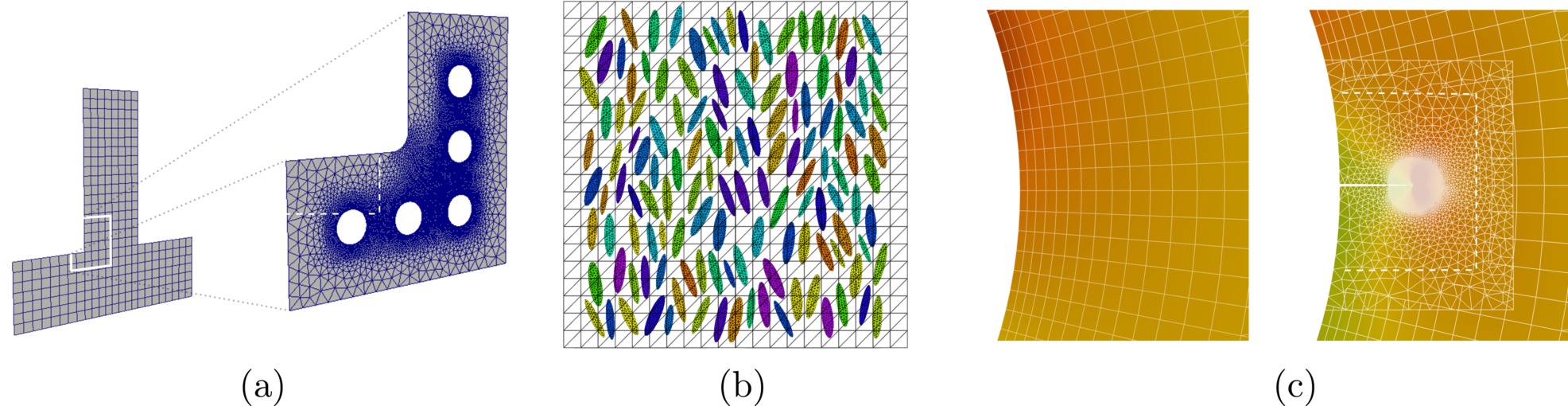}
    \caption{Few applications: (a) substructuring and inclusion of arbitrary geometrical features, (b) microstructure
    modeling, (c) localized mesh refinement, for example, around an inserted crack.}
    \label{fig:mortex_applictions}
\end{figure}

With the emphasis laid on the interface discretizations handled by the
combination of the X-FEM and the mortar methods, many applications could be
cited:  sub-structuring, inclusion of arbitrary geometrical features into
the existing mesh, meshing complex micro-structures, localized mesh
refinement near crack tips, general static and dynamic mesh refinement (Fig.~\ref{fig:mortex_applictions}).  However,
applications are not limited to mesh tying. In analogy to mortar methods,
which were extended to contact problems~\cite{belgacem_mortar_1998,fischer_frictionless_2005,fischer_mortar_2006,popp_mortar_2012}, the method elaborated here can be used to solve
contact problems between a virtual surface (represented by the X-FEM) and an
explicitly represent surface of the homologue
solid~\cite{yastrebov_numerical_2013}.  In a separate
paper~\cite{akula_mortex_2018} we present the extension of this framework to contact problems, which also allows to treat
efficiently wear problems. 

The paper is organized as follows. In Section~\ref{sec:strong_form} we present
the formulation of the interfacial mesh tying problem of overlapping domains;
its weak form using the method of Lagrange multipliers is derived in Section~\ref{sec:weak_form}.  
Section~\ref{sec:methodology} presents the core methodologies of the X-FEM in terms of inclusion/void modeling and of  the mortar
discretization for mesh tying problems.  The effect of the underlying host mesh
interpolations on the tying problem is presented as a case study in
Section~\ref{sec:effect_host_mesh_interpolation}. 
Two main remedies for mesh locking are presented, namely the triangulation of the blending elements and coarse-graining of Lagrange multipliers, which is detailed
in Section~\ref{sec:cgi}. In Section~\ref{sec:validation} we consider two patch tests (tension/compression and bending) to emphasize the inherent problems of
embedded boundaries in mixed finite element method. In Section~\ref{sec:eshelby} we study the mesh convergence
properties of the method using the Eshelby inclusion problem. 
In Section~\ref{sec:numerical_examples} we illustrate the method's performance on various applications.  In Section~\ref{sec:conclusion} we give the concluding
remarks and state the prospective works.

\section{Mesh tying problem \label{sec:strong_form}} 

\begin{figure}[htb!]
\centering
\includegraphics[width=.85\textwidth]{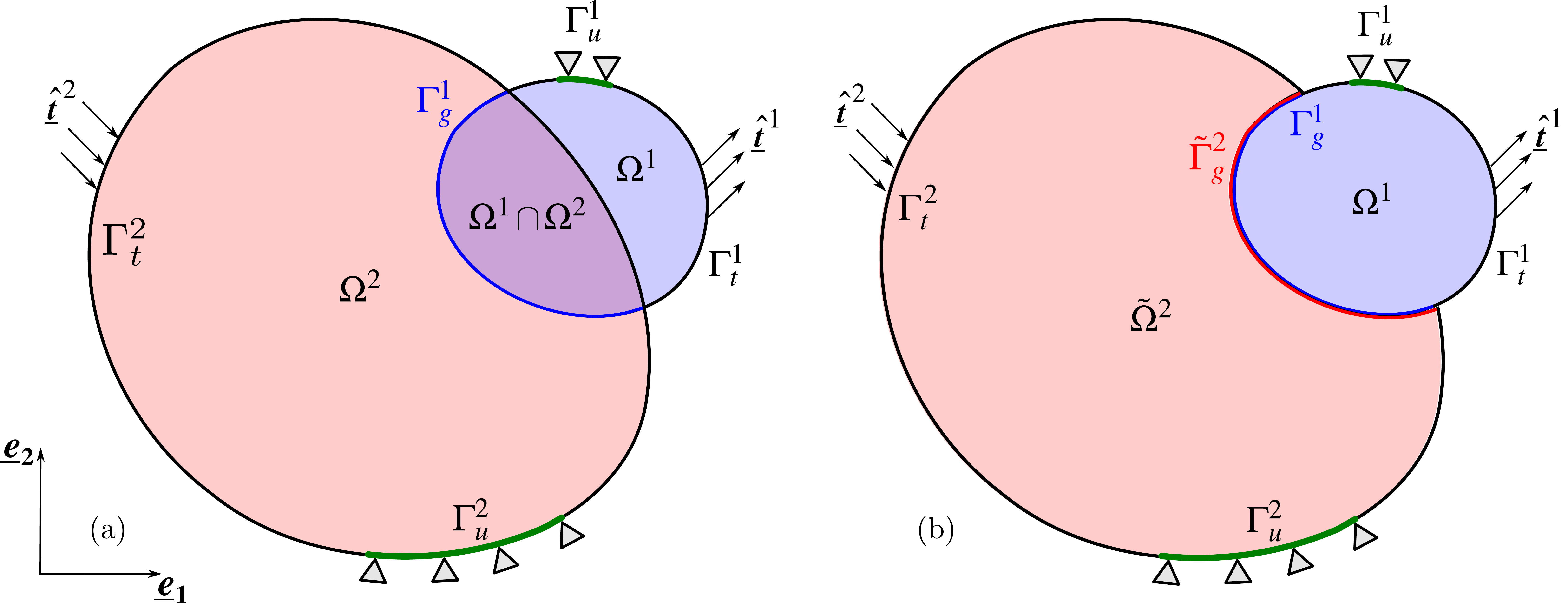}
\caption{(a) Continuum setting of the two overlapping domains with applied boundary
conditions; (b) an equivalent continuum problem without overlap.}
\label{fig:problem_setting}
\end{figure}

We consider two open domains $\Omega^1$ and $\Omega^2$ with an overlap region
$\Omega^1\cap\Omega^2$ [Fig.~\ref{fig:problem_setting}(a)].  Solid $\Omega^1$
has only outer surfaces which are split into Dirichlet, Neumann and tying
boundaries $\Gamma_u^1,\Gamma_{t}^1,\Gamma_g^1$, respectively, such
that $\Gamma_u^1\cup\Gamma_{t}^1\cup\Gamma_g^1=\partial\Omega^1$.  We refer to
the domain $\Omega^2$ as the ``host'' domain as it hosts the partially embedded
domain $\Omega^1$.  In addition to outer boundaries (the Dirichlet boundary
$\Gamma_{u}^2$ and Neumann boundary $\Gamma_{t}^2$), the host domain contains an
embedded boundary $\tilde\Gamma_g^2 = \Gamma_g^1\cap\Omega^2$ (the subscript
$"g"$ refers to the ``gluing'').  We assume that the solids are glued together
along the interface formed by the boundaries $\tilde\Gamma_g^2$ and
$\Gamma_g^1$; the physics in the overlap zone is determined by solid $\Omega^1$.
Therefore, in the reference configuration $\tilde\Gamma_g^2 = \Gamma_g^1$ and
must remain so in any configuration.

The mesh tying problem is concerned with the enforcement of
displacement continuity along the interface $\Gamma_g^1$ and
$\Gamma_g^1$.
The standard boundary value problem
(BVP) involves ensuring the balance of linear momentum along with the imposed
boundary conditions for the two bodies ($i=1,2$):
\begin{align}
    \nabla\cdot\ten{\sigma}^1+\vec{f}_v^1 &= 0\quad
    \mbox{in}\,\Omega^1,\quad
    \nabla\cdot{\ten{\sigma}}^2+{\vec{f}}_v^2 = 0\quad
    \mbox{in}\,\tilde\Omega^2
    \label{eq:strong_form_linear_momentum}\\
    {\ten{\sigma}}^i\cdot\vec{n}^i &=
    \vec{\hat{t}}^i\quad\mbox{at}\,\,\Gamma_t^i,
    \label{eq:strong_form_NeumannBC}\\
    {\vec{u}}^i &=
    \vec{\hat{u}}^i\quad\mbox{at}\,\,\Gamma_u^i,\quad\label{eq:strong_form_DirichletBC}\\
\end{align}
where $\ten\sigma$ is the Cauchy stress tensor, $\vec{f}_v^i$ represent the density of body forces,
$\vec{n}^i$ is the unit outward normal to $\Omega^i$ and 
$\vec{\hat t}^i$, $\vec{\hat{u}}^i$ are the prescribed tractions and displacements.
$\tilde\Omega^2\,=\Omega^2\setminus\bar\Omega^1$ is the effective
non-overlapping region of the host domain volume
[Fig.~\ref{fig:problem_setting}(b)], where the bar-notation denotes the open domain united with its
closure, i.e. $\bar\Omega^1=\Omega^1\cup\partial\Omega^1$. 
The solution is searched in the appropriate functional space $\mathcal{U}^0 = \{u^i\in H^2(\Omega^i)\,|\,u^i=\hat{u}^i\,\mbox{on}\,\Gamma_u^i\}$, where $H^2$ is second order Sobolev space.
Displacement continuity between the two domains is enforced as:
\begin{equation}
    \vec{u}^1\,=\,{\vec{u}}^2\quad\mbox{along the interface formed by the boundaries } \tilde\Gamma_g^2 \mbox{ and } \Gamma_g^1.
    \label{eq:eq_constraints}
\end{equation}
 An equivalent continuum problem is depicted in Fig.~\ref{fig:problem_setting}(b).

\section{Weak form}\label{sec:weak_form}

Eqs.~\eqref{eq:strong_form_linear_momentum}-\eqref{eq:strong_form_DirichletBC}
represent the strong form of the standard solid mechanics BVP, and
Eq.~\eqref{eq:eq_constraints} represents the tying constraints.  The weak form
of the BVP is accordingly divided into the standard structural part and the
contribution from the tying.  In the weak form, the requirements
on the solution
functional space $\mathcal{U}^i$ is slightly relaxed compared to the strong
form, and an additional field of test functions is selected from another
functional space $\mathcal{V}^i$ given respectively by :
\begin{align}
\mathcal{U}^i &= \{\vec u^i\in H^1(\Omega^i)\,|\,\vec u^i=\hat{\vec u}^i\,\mbox{
    at }\,\Gamma_u^i\}\\
\mathcal{V}^i &= \{\delta \vec u^i\in H^1(\Omega^i)\,|\,\delta \vec
u^i=0\,\mbox{ at }\,\Gamma_u^i\},
\label{eq:weak_form_structural}
\end{align}
where $H^1(\Omega^i)$ denotes the standard first order Sobolev space. 
In infinitesimal strain formulation, the virtual work $\delta W_s$ corresponding to the structural part is given by: 

\begin{align}
    \delta W_s =
    \underbrace{\int\limits_{\Omega^1}\ten{\sigma}^1:\delta\ten\varepsilon^1\,dV+\int\limits_{\tilde\Omega^2}{\ten{\sigma}}^2:\delta\ten\varepsilon^2\,dV}_{\delta
    W_{\text{int}}}-\underbrace{\int\limits_{\Omega^1\cup\tilde\Omega^2}\vec{f}_v^i\cdot\delta\vec
    u^i\,dV - \int\limits_{\Gamma^i_t}\hat{\vec t}^i\cdot\delta{\vec
    u}^i\,dS}_{\delta W_{\text{ext}}},
    \label{eq:virtual_work_structural}
\end{align}
where $\ten\varepsilon$ is the infinitesimal strain tensor, and $\delta W_{\text{int}},\delta W_{\text{ext}}$ are the change in internal energy and the virtual work of forces, respectively.
Note that the integration of the virtual internal energy of the host domain is carried out only over the effective volume $\tilde\Omega^2$.
The equality constraints~\eqref{eq:eq_constraints} are enforced
using the method of Lagrange multipliers, whose functions as well
as their variations are selected from the space $H^{-1/2}$. The multipliers
are defined over the surface
$\Gamma_g^1$, which was selected from practical consideration of storing
Lagrange multipliers on the explicitly represented surface	:
\begin{equation}
    \vec\lambda,\;\delta\vec\lambda \in  H^{-1/2}(\Gamma_g^1),
\end{equation}
where $H^{-1/2}$ is the fractional Sobolev space defined over the tying boundary.
The Lagrangian of the constrained problem leads to a mixed variational saddle point problem which includes both structural $ \delta W_s$ and tying $\delta W_{g}$ contributions. The variation of the Lagrangian is given by:
\begin{equation}
    \delta\mathcal L(\vec u,\vec \lambda) = \delta W_s +\delta W_{g} = \delta W_s +
     \int\limits_{\Gamma_{g}^1} \left[\vec\lambda\cdot(\delta\vec
     u^1-\delta{\vec u}^2) + 
    \delta\vec\lambda\cdot(\vec
    u^1-{\vec u}^2)\,\right] d\Gamma = 0,
    \label{eq:vw_cont}
\end{equation}
where the integration is carried out over the boundary $\Gamma_g^1$ .
The first term in square brackets in Eq.~\ref{eq:vw_cont} represents the virtual work
contribution from the tying interface and second term represents the weak form
of the equality constraints.

\section{Methodology\label{sec:methodology}}

The evaluation of the internal virtual
work restricted to the effective volume of the host solid
$\tilde\Omega^2$ is accomplished with the X-FEM
method. The mortar method is extended to enforce the displacement equality
constraint over the interface between the overlapping domains, i.e. between the
boundary of the embedded domain (patch) $\Gamma^1_g$ and the corresponding virtual surface $\tilde\Gamma_g^2$ 
of the host solid. 

\begin{figure}[htb!]
    \centering
    \includegraphics[width=\textwidth]{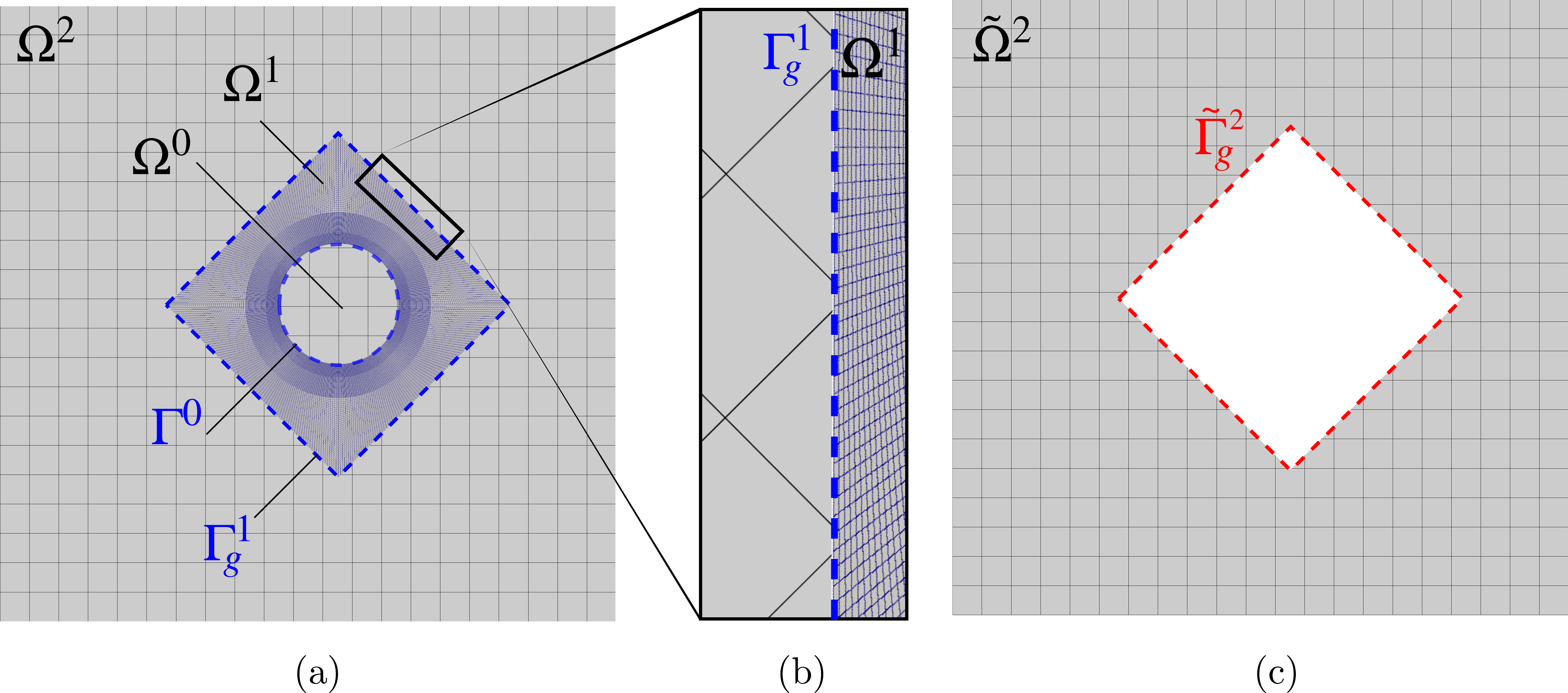}
    \caption{(a) Two overlapping meshes: the host $\Omega^2$ and the patch
    $\Omega^1$ with a circular hole $\Omega^0$ are tied along interface
    $\Gamma^1_g$; (b) zoom on the interface between the host and patch meshes; (c) effective volume of the host mesh $\tilde{\Omega}^2=
    \Omega^2\setminus\{\bar\Omega^0\cup\bar\Omega^1\}$.}
    \label{fig:plate_with_hole}
\end{figure}

The main features of the proposed method are illustrated on an example shown in
Fig.~\ref{fig:plate_with_hole}.  It represents the discretized finite-element setting for the
overlapping domains.  A discretized\footnote{Note that for brevity and
simplicity hereinafter we preserve the same notations for discretized entities
as were introduced in the continuous problem statement.} square patch with a circular
hole $\Omega^0_{}\cup\Omega^1_{}$ with surfaces 
$\Gamma_{}^0=\partial\Omega^0$ and $\Gamma^1_{g}
=\partial\left\{\Omega^1_{}\cup\bar\Omega^0\right\}$ is embedded into a host mesh
$\Omega^2$. As before the bar notation is used to denote the open domain united with its closure, here $\bar\Omega^0 = \Omega^0\cup\partial\Omega^0$.
Note that the necessity to consider explicitly the hole as an extra domain comes from the particularities of the problem and was introduced for the sake of avoidance of misinterpretation: the physics inside the contour $\Gamma^1_g$ is fully determined by the patch with a circular hole.
The intersection of the patch's boundary $\Gamma^1_g$ and the
host domain represents the virtual surface $\tilde\Gamma^2_g=\Gamma^1_g\cap\Omega^2$.
The X-FEM is used to account for the virtual work only in the effective domain
volume $\tilde\Omega^2 =
\Omega^2\setminus\{\bar\Omega^0\cup\bar\Omega^1\}$. The mortar method is
is brought into play to tie together the two domains
$\Omega^1$ and $\tilde\Omega^2$ along the interface made of $\Gamma^1_g$ and $\tilde\Gamma^2_g$.
Note that in the presented example, the tying boundary is fully embedded.
%

\subsection{Extended finite element method\label{sec:selective integration}}

\begin{figure}[htb!]
    \centering
    \includegraphics[width=1\textwidth]{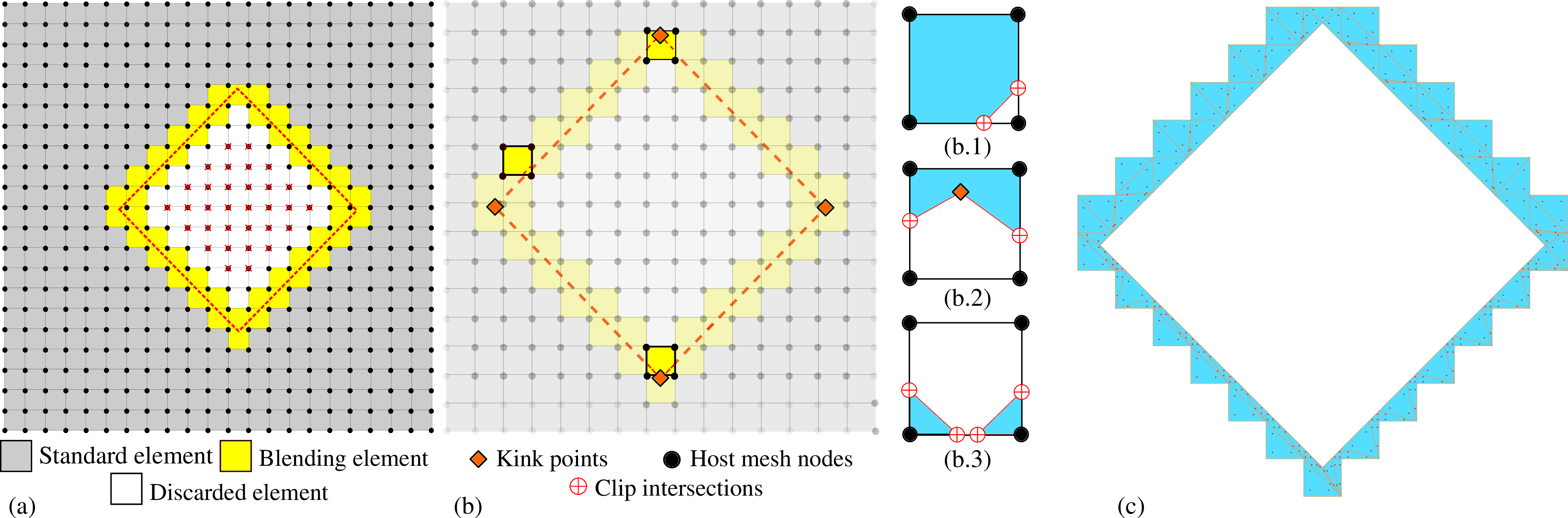}
    \caption{(a) Element classification in X-FEM framework; (b) clipping of blending elements by $\tilde\Gamma^2_g$, the volume
    colored in blue in (b.1-3) is the effective volume of integration
    ($\tilde{\Omega}^e$): (b.1) a convex polygon, (b.2) a non-convex polygon,
    (b.3) disjoint polygons; (c) selective integration is carried out over re-triangulated blending elements with reinitialized Gauss integration points (shown in red).}
    \label{fig:various_possible_clipping}
\end{figure}

The virtual surface $\tilde\Gamma^2_g$ of the host domain is treated as an internal
discontinuity.
 This is modeled within the X-FEM framework, thereby
nullifying the presence of the overlap region
$\left(\bar\Omega^1\cup\bar\Omega^0\right)\cap\Omega^2$ in the domain
$\Omega^2$ [see Fig.~\ref{fig:plate_with_hole}(c)]. 
The X-FEM relies on enhancement of the FEM shape functions used to
interpolate the displacement fields. Here the enrichment functions
describing the field behavior are incorporated locally into the finite element
approximation. This feature allows the resulting displacement to capture discontinuities.  
The subdivision of the host mesh is
defined by indicator function $\phi({\vec
X}): \mathbb R^{\mathrm{dim}}\to \{0,1\}$ (where ${\vec X}$ is
the spatial position vector in the reference configuration 
in domain $\Omega^2$)~\cite{sethian_level_1999}. The
indicator function is non-zero only in the non-overlapping part of domain
$\Omega^2$: $$
  \phi(\vec X) = \begin{cases}
                    1,&\mbox{ if }\vec X\in \tilde{\Omega}^2;\\
                    0,&\mbox{ elsewhere.}
                 \end{cases}
$$
The discontinuity surface $\tilde\Gamma^2_g$ can be seen as a level-set defined as follows:
$$
  \tilde\Gamma^2_g = \left\{\vec X \in \Omega^2: \; \nabla\phi(\vec X) \ne 0\right\}
$$
As a result, the indicator function $\phi(\vec X)$
partitions the elements of the host domain $\Omega^2$ into three distinct
categories [Fig.~\ref{fig:various_possible_clipping}(a)], 
 namely standard elements, blending elements and discarded elements.

In practice, the enrichment of shape functions in case of void/inclusion problem
can be  simply replaced by a selective integration
scheme~\cite{sukumar_modeling_2001}.  For the standard elements, there is no
change in volume of integration and the discarded elements are simply excluded
from the volume integration procedure. In order to obtain the effective volume
of integration for each blending element, we
perform the clipping of the blending elements by the discretized surface
$\tilde\Gamma^2_g$ [Fig.~\ref{fig:various_possible_clipping}(b)] .  The clipping of a single element could
result in one or several various polygons\footnote{Hereinafter, we assume that
all elements use first order interpolation, therefore all edges of elements are
straight. It enables us to assume that an intersection or difference of elements
can be always represented as one or several polygons.} both convex, and
non-convex , which represent the effective volumes of integration. 

To selectively integrate the internal virtual work in the effective volume only,
the resulting polygons are virtually remeshed into standard convex elements (for
example, triangles). Note that this remeshing is merely performed to use a Gauss
quadrature for integration [Fig.~\ref{fig:various_possible_clipping}(c)], and does not imply the creation of
additional degrees of freedom, and as such does not change the topological
connectivity of nodes.
The displacement field is evaluated using the
standard shape functions and the original DoFs; only the integration is changed.  To
carry out this remeshing, we applied the ear clipping triangulation algorithm~\cite{mei_ear-clipping_2013} to the
polygons.  The DoFs associated with the
elements outside the integration domain $\tilde\Omega^2$ [marked
with red crosses in Fig.~\ref{fig:various_possible_clipping}(a)] are removed from the
global system of equations. 

\subsection{MorteX discretization}

Within the mortar discretization framework, the tied domains are classified into
mortar and non-mortar sides. The superscript $"1"$ refers to the mortar side of
the interface and $"2"$ to the non-mortar side; the former stores the Lagrange
multipliers (dual DoFs) in addition to displacement degrees of freedom (primal
DoFs).  
If the host is selected as a mortar side, the context of the problem becomes
similar to the one considered
in~\cite{moes_imposing_2006,bechet_stable_2009,hautefeuille_robust_2012}, where
it was shown that 
strong restrictions apply
on the choice of Lagrange multiplier spaces in order to fulfill the inf-sup
condition. The algorithm for
construction of such spaces is not straightforward.  Therefore, to avoid these
difficulties, we select the patch side as the mortar surface, which provides us
with a more flexible setting. This choice was already reflected in the fact that
the host boundary was chosen as the integration side for tying
conditions~\eqref{eq:vw_cont}.  However, under specific problem settings, the
choice of employing standard interpolation functions for the Lagrange multipliers on the
embedded interface still leads to spurious oscillations of interfacial
tractions.  Remedies for this problem will be discussed in
Section~\ref{sec:validation}.

Displacements on the mortar side $\Gamma^1_g$ are given  by classical one-dimensional shape functions
with the interpolation order equal to that of the underlying mesh:
\begin{equation}
  \vec u^1(\xi^1,t) = N^1_m(\xi^1)\vec u^m(t),\quad m \in [1,\mathrm M]
  \label{eq:primal_shape}
\end{equation}
where $\mathrm M$ is the number of nodes per mortar element's edge and 
$\xi^1\in[-1;1]$ is the parametric coordinate of the mortar side. The displacements along the
virtual surface $\tilde\Gamma^2_g$ running through the host mesh elements
are
characterized by the two dimensional host mesh interpolations, and can be
expressed as follows:
\begin{equation}
\vec u^2(\zeta,t) =  N^2_i\left(\mu^2(\zeta),\eta^2(\zeta)\right)\vec u^i(t),\quad i \in [1,\mathrm N]
  \label{eq:primal_shape2}
\end{equation}
where $\zeta$ is the one-dimensional parametric coordinate of integration segment of non-mortar side, $\mu^2,\eta^2 \in [-1;1]$ are the
classical two-dimensional parametric coordinates of the host element, and $\mathrm N$ is the number of nodes of this element.
The use of two-dimensional interpolation on the non-mortar side marks the difference between the classical mortar and the presented MorteX frameworks.
The Lagrange multipliers (defined on the
mortar side) are interpolated using shape functions $\Phi$:
\begin{equation}
   \vec\lambda(\xi^1,t)= \Phi_{l}(\xi^{1})\,\vec\lambda^{l}(t) ,\quad l \in [1,\mathrm L]
    \label{eq:dual_shape}
\end{equation}
where $\mathrm L$ can be less than or equal  to $\mathrm M$. It enables to
select shape functions for dual variables independently of the primal shape
functions.

\begin{figure}[htb!]
\centering
\includegraphics[width=\textwidth]{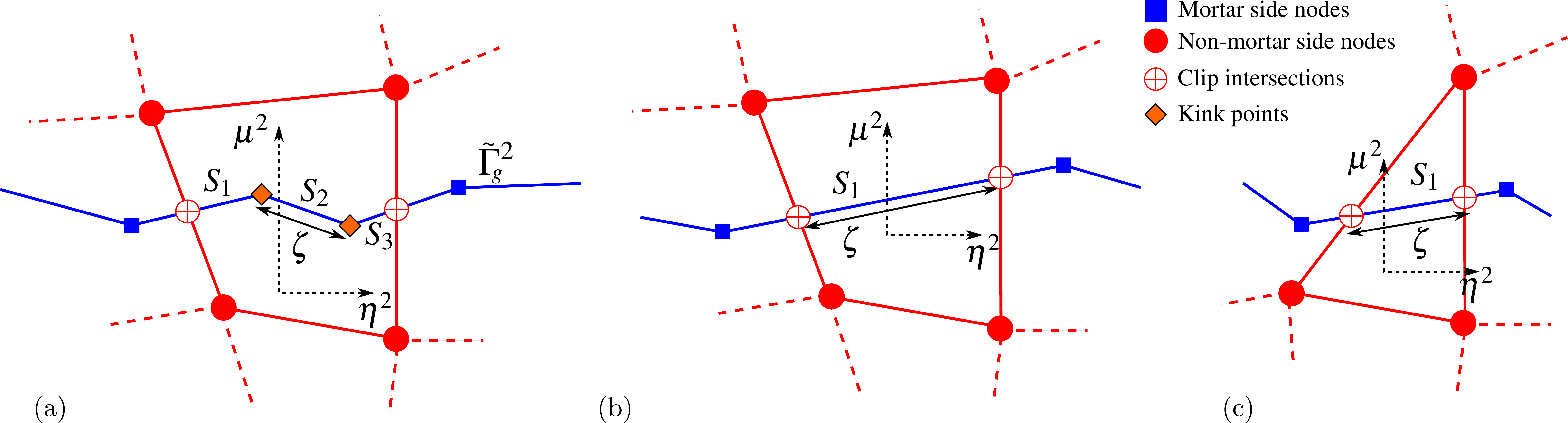}
\caption{
Illustration of a single host element intersected by the virtual
surface $\tilde\Gamma^2_g$: (a) quadrilateral host element intersected by several
mortar side segments; (b) quadrilateral host element intersected by a single
mortar segment; (c) triangular host element intersected by a single mortar
segment.}
\label{fig:mortar_discretization_interpol}
\end{figure}

Few remarks could be made here.  We need first to introduce the notion of
segment: it is a straight line whose vertices can either be a clip
intersection\footnote{The clip intersection points are located on the edges of
blending elements intersected by the virtual surface $\tilde\Gamma^2_g$.} or a
kink point, the later being a node of the mortar side lying inside the host
element.  Note that segments are defined both on mortar and non-mortar
sides as they coincide.  Since a host element can be intersected by several
patch segments $S_{i}$ [Fig.~\ref{fig:mortar_discretization_interpol} (a)],
the functions $\mu^2(\zeta)$, $\eta^2(\zeta)$ can be  piece-wise
smooth, which implies that the underlying displacement (and
coordinate) can also be piece-wise smooth. 
Second remark: if the host elements are
quadrilateral, then each segment interpolation is given by
$p_1\times(p_2+1)$, where $p_1$ is the mortar interpolation order, and $p_2$ is
the non-mortar interpolation order; the later is augmented by one since the
virtual interface passes inside the element, where the interpolation order in
quadrilateral 2D elements is one order higher than along the edges.
For a triangular host mesh, the interpolation order is
simply the product of interpolation orders of the host and patch
meshes $p_1\times p_2$. To give an example, let us assume that the patch mesh is
linear and the host mesh is linear quadrilateral; then  let us
imagine that a host element is cut into two parts by a (straight) patch segment.
Then the displacement in the host element along this straight segment will be
second order polynomial function of the parameter $\zeta$
[see Fig.~\ref{fig:mortar_discretization_interpol}(b)]. However, the
displacement along such a cut would remain linear for triangular linear host elements
[see Fig.~\ref{fig:mortar_discretization_interpol}(c)].

\subsubsection{Mortar interface element}

A mortar element is formed with a single mortar segment (on the patch side) and
a single non-mortar element (on the host side).   Each tying element consists of
$(\mathrm M+\mathrm N)$  nodes, $\mathrm M$ from the mortar segment, and
$\mathrm N$ from non-mortar element, and stores $(\mathrm L \times
\mathrm{dim})$ Lagrange multipliers, where $(\mathrm{dim})$ is the spatial
dimension. The choice of $\mathrm L$ is guided by the inf-sup condition
requirement of the discrete Lagrange multiplier spaces (usually $\mathrm L \leq
\mathrm M$). 

Substituting the interpolations~\eqref{eq:primal_shape}, \eqref{eq:primal_shape2}, \eqref{eq:dual_shape}
into the weak form~\eqref{eq:vw_cont} and extracting only the terms related to the mesh tying of a single mortar element, we obtain
\begin{equation}
    \label{eq:vw_discrete}
\end{equation}
\begin{equation}
        \delta W_{g}^{\mathrm{el}} =\left( \vec\lambda^l \cdot\delta \vec u^m +
        \delta\vec\lambda^l\cdot\vec u^m\right) D_{lm} - \left( \vec\lambda^l
        \cdot\delta \vec u^i + \delta \vec\lambda^l \cdot \vec u^i \right)
        M_{li},\quad l\in[1,\mathrm L],\; i\in[1,\mathrm N],\; m\in[1,\mathrm M] 
    \label{eq:vw_discrete}
\end{equation}
where $D_{lm}$ and $M_{ln}$ are the mortar integrals evaluated over the
mortar-side segments $S^{\mathrm{el}}\subset \left( \Gamma^1_g \cap
\tilde\Omega^e_{i}\right)$, where $\tilde\Omega^e_{i}$ is the current host
element forming the mortar element.\\
\begin{tabularx}{\textwidth}{XX}
\begin{equation}
    D_{lm}
    = \int\limits_{S^{\mathrm{el}}}\Phi_l^1(\xi^1)N_m^1(\xi^1)\,d\Gamma,\quad
    \label{eq:mortar_integrals_D}
\end{equation}&
\begin{equation}
    M_{li}=\int\limits_{S^{\mathrm{el}}}\Phi_l^1(\xi^1)
    N_i^2\left(\mu^2(\zeta^2),\eta^2(\zeta^2)\right)\,d\Gamma.
    \label{eq:mortar_integrals_M}
\end{equation}
\end{tabularx}
The nodal blocks of the mortar  matrices denoted as $\mat{D}$ ($\mathrm L\times \mathrm M$)
and $\mat{ M}$ ($\mathrm L\times \mathrm N$) can be expressed as:
\begin{equation}\label{eq:operators_D}
    \mat{D}(l,m) = D_{lm}\ten I
\end{equation}
\begin{equation}\label{eq:operators_M}
    \mat{M}(l,i) = M_{li}\ten I
\end{equation}
where, 
$\ten I$ is the identity tensor of the spatial dimension of the problem.
Using matrix notations, Eq.~\eqref{eq:vw_discrete} reads
\begin{equation}\label{eq:mortel_residual}
     \delta W_{g}^{\mathrm{el}}  
     = \begin{bmatrix}
           \mat D^\intercal \cdot\mat L \\ -\mat M^\intercal \cdot\mat L \\ \mat D\cdot\mat U^1 - \mat M\cdot\mat U^2
          \end{bmatrix}^\intercal\cdot
\begin{bmatrix}
           \delta\mat U^1 \\ \delta\mat U^2 \\ \delta\mat L
          \end{bmatrix}
\end{equation}
where arrays $\mat U^1,\mat U^2,\mat L$ store current values of associated nodal primal (on mortar and non-mortar sides) and dual (mortar) DoFs:
$$
\mat U^1 = \left[\vec u^1, \dots, \vec u^N\right]^\intercal,\quad \mat U^2 =
\left[\vec u^1, \dots, \vec u^M\right]^\intercal,\quad \mat L = \left[\vec
\lambda^1, \dots, \vec \lambda^L\right]^\intercal
$$
whereas their variations are denoted $\delta \mat U^1,\delta \mat U^2,\delta \mat L$.
The tangent operator for the mortar interface element is obtained by taking the derivatives of~\eqref{eq:mortel_residual} with respect to its DoFs:
\begin{equation}
\mat K= 
    \begin{bmatrix} 
        \mat 0 & \mat 0 & \mat D^{\intercal}\\ 
        \mat0 & \mat 0 & -\mat M^{\intercal}\\ 
        \mat D & -\mat M & 0
    \end{bmatrix}.
    \label{eq:kmort}
\end{equation}
This tangent operator 
has zero blocks for primal DoFs, non-zero blocks to link primal with dual
DoFs, and zero blocks on the main diagonal, which is 
the typical structure of a saddle-point system. 

\subsubsection{Evaluation of mortar integrals}\label{subsub:integ_eval}

\begin{figure}[htb!]
\centering
\includegraphics[width=0.75\textwidth]{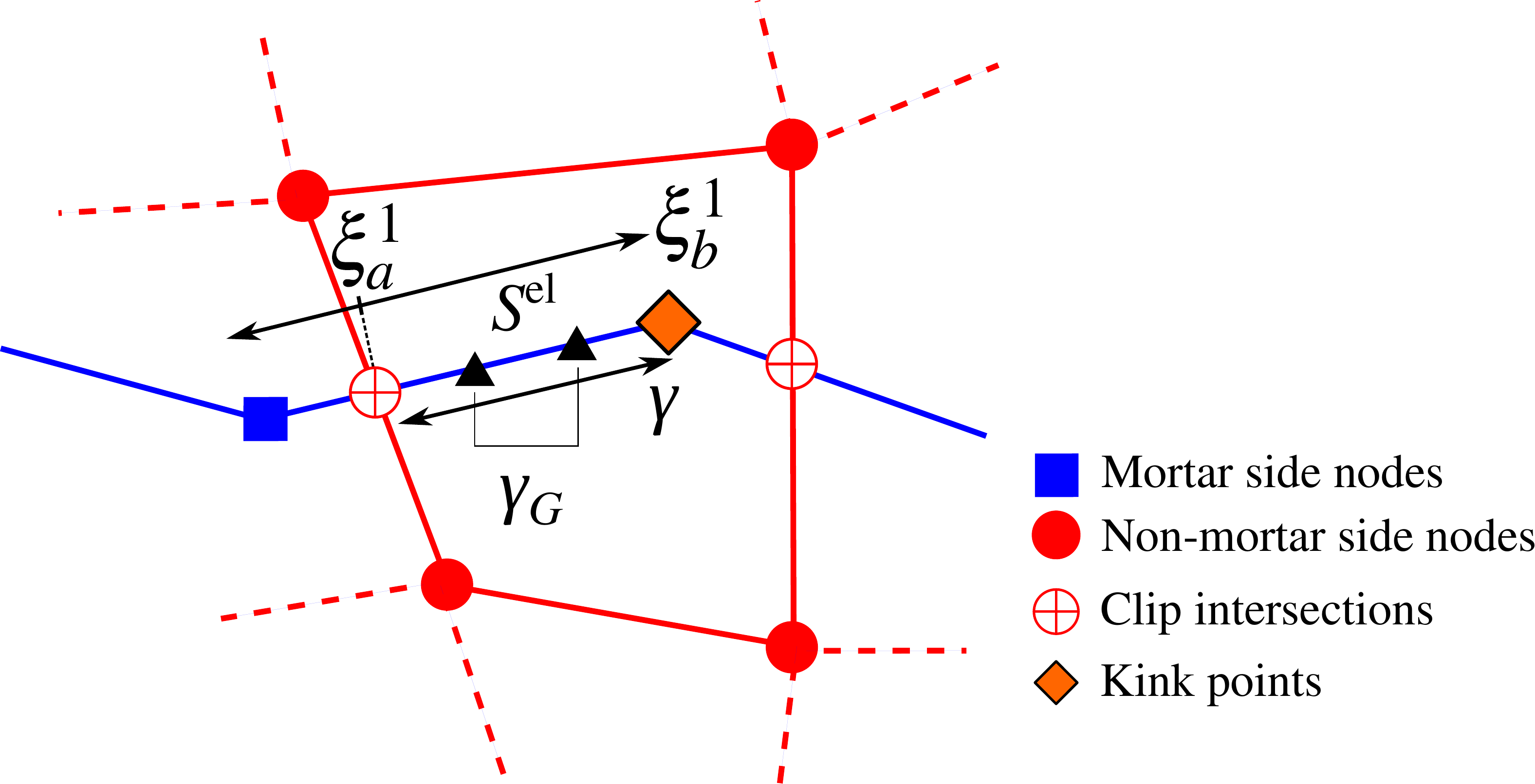}
\caption{Example of mortar integration domain defined by segment $S^{\mathrm{el}}$ connecting a clip intersection $\xi^1_a$ and the kink point $\xi^2_b$, Gauss points are shown as triangles.}
\label{fig:classic_mortar_projections}
\end{figure}

The evaluation of mortar integrals is performed over segments $S^{\mathrm{el}}$
(see
Figs.~\ref{fig:mortar_discretization_interpol},\ref{fig:classic_mortar_projections}).
The evaluation of the  integrals $D_{lm}$~\eqref{eq:mortar_integrals_D} is
straightforward, as it involves the product of shape functions from 
the mortar side only. In contrast, the integral
$M_{li}$~\eqref{eq:mortar_integrals_M} combines shape functions from both
mortar and non-mortar sides. The evaluation of this integral requires a mapping
between the mortar and non-mortar sides.

The integration domain is parametrized by $\gamma\,\in[-1,1]$ (Fig.~\ref{fig:classic_mortar_projections}), which needs to be linked with the parametrization of the mortar side, which is given by:
\begin{equation} \label{eq:classical_map_xi_eta}
    \xi^1(\gamma) =
    \frac{1}{2}(1-\gamma)\xi_{a}^{1}\,+\,\frac{1}{2}(1+\gamma)\xi_{b}^{1},
\end{equation}
where $\xi_a^1$ and $\xi_b^1$ define the limits of the integration on the mortar side as shown in Fig.~\ref{fig:classic_mortar_projections}.
To evaluate the integrals using the Gauss quadrature, we need to find the
location of Gauss points $\gamma_G$ in terms of mortar ($\xi^1_G$) and
non-mortar ($\mu^2_G,\nu^2_G$) parametrization. While the former is
straightforward using~\eqref{eq:classical_map_xi_eta}, the latter can be done by solving the following equation:
\begin{equation}
  N^1_m(\xi^1_G) \vec X^1_m = N^2_i(\mu_G,\eta_G)\vec X^{2}_i,
\end{equation}
where the physical location of the Gauss point is given by $\vec X_G = N^1_m(\xi^1_G) \vec X^1_m$.
With these notations, the mortar integrals can now be evaluated
using the Gauss quadrature rule as:
\begin{equation}
    {D_{lm} =
    \int\limits_{S^{\mathrm{el}}}\Phi_l(\xi^1)N_m^1(\xi^1)\,d\Gamma=}
    \sum_{G=1}^{N_G} w_{G}\Phi_{l}(\xi^{1}_{G})N_{m}^{1}(\xi^{1}_{G})J_{\mbox{\tiny seg}}(\xi_{G}^{1}) 
    \label{eq:D_op_eval}
\end{equation}
\begin{equation}
    {M}_{li}=\int\limits_{S^{\mathrm{el}}}\Phi_l(\xi^1)N_i^2(\mu^2,\eta^2)\,d\Gamma
    =\sum_{G=1}^{N_G} w_{G}\Phi_{l}(\xi^{1}_{G})N_{i}^{2}\big(\mu^{2}_G,\eta^{2}_G\big)J_{\mbox{\tiny
    seg}}(\xi_{G}^{1}) 
    \label{eq:M_op_eval}
\end{equation}
where as previously $l\in [1,L]$, $m\in[1,M]$, $n\in[1,N]$ and $N_G$ is the
number of Gauss integration points, $w_G$ are the Gauss weights, $J_{\mbox{\tiny
seg}}$ is the Jacobian of the mapping from the parent space
$\xi^1$ to the
real space  including the adjustment of the integral limits:  
\begin{equation}
    J_{\mbox{\tiny seg}}(\xi^{1}) = \bigg|\frac{\partial
    N_{m}}{\partial\xi^{1}} \frac{\partial \xi^1}{\partial \gamma}\vec X_{m}\bigg|.
    \label{eq:2D_jacob}
\end{equation}

\section{Intra-element interpolation of displacements in the host mesh\label{sec:effect_host_mesh_interpolation}}
\begin{figure}[htb!]
   \centering
   \includegraphics[width=\textwidth]{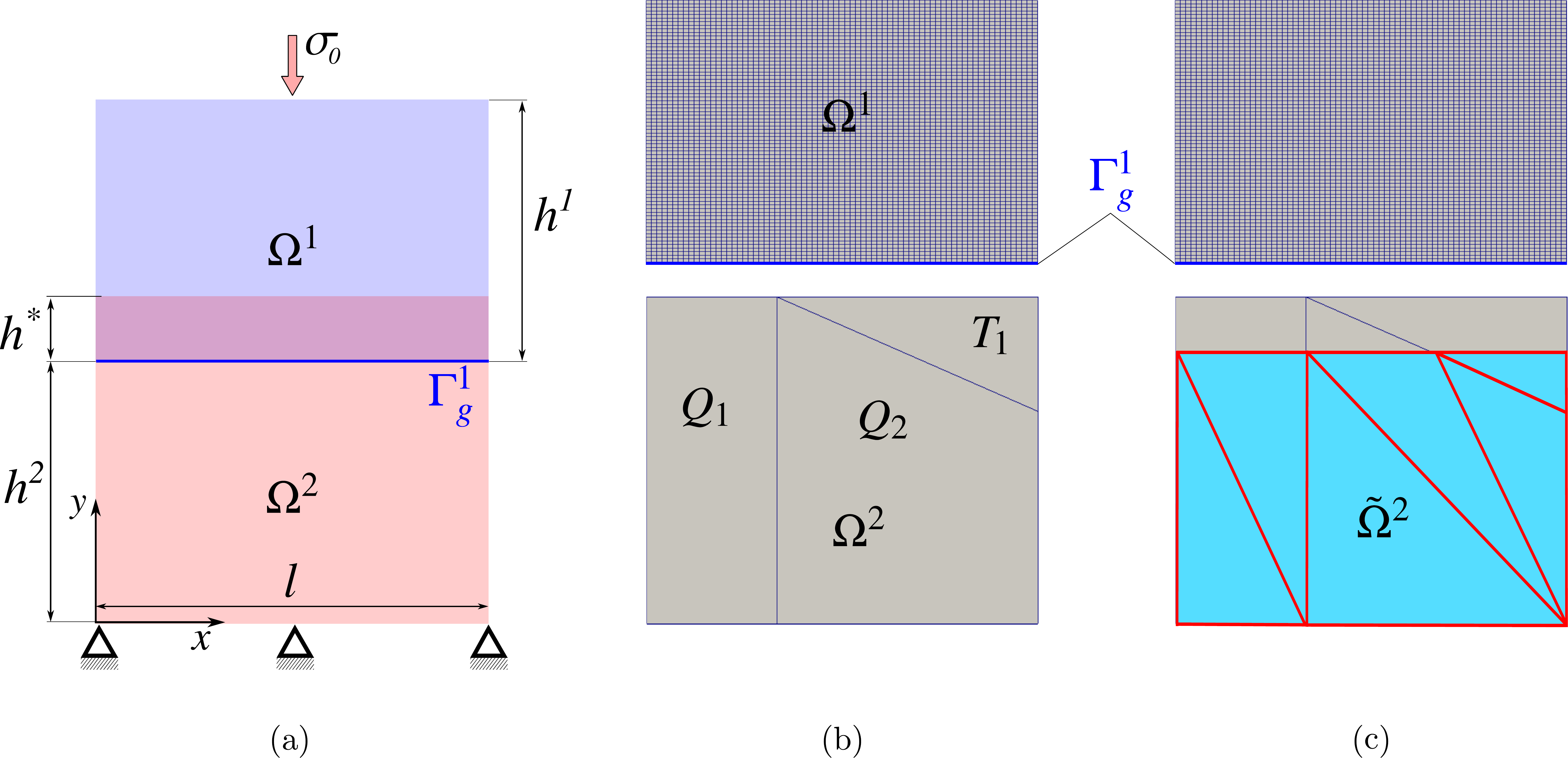}
   \caption{Illustration of the effect of underlying mesh interpolations: (a)
   problem setting: tying of two rectangular overlapping domains, (b) discretized
   patch (upper solid) and host (lower solid) domains, (c) triangulated
   effective volume of the host domain $\tilde{\Omega}^2$; note that in (b,c)
   the two meshes are shown separated only for convenience.}
   \label{fig:quad_vs_tria}
\end{figure}

For the overlapping domains the coupling is made between the patch mesh boundary
and virtual surface running through the host mesh elements.  This is reflected
in the mortar matrix $\mat M$~\eqref{eq:M_op_eval} that contains the integral of
a product of volumetric (in the host mesh) and surface (patch mesh) shape
functions.  To demonstrate the effect of the interpolation choice, we use the
set-up shown in Fig.~\ref{fig:quad_vs_tria} (a). The following dimensions are
used: $h^1=1$ mm, $h^2=1.25$ mm, $h^*=0.25$ mm and $l=1.5$ mm. A uniform
pressure of $\sigma_0=1$ MPa is applied on the top surface of $\Omega^1$.  The
meshes $\Omega^1$ and $\Omega^2$ are tied along the interface $\Gamma_{g}^1$.

\begin{figure}[htb!]
\centering
\includegraphics[width=.5\textwidth]{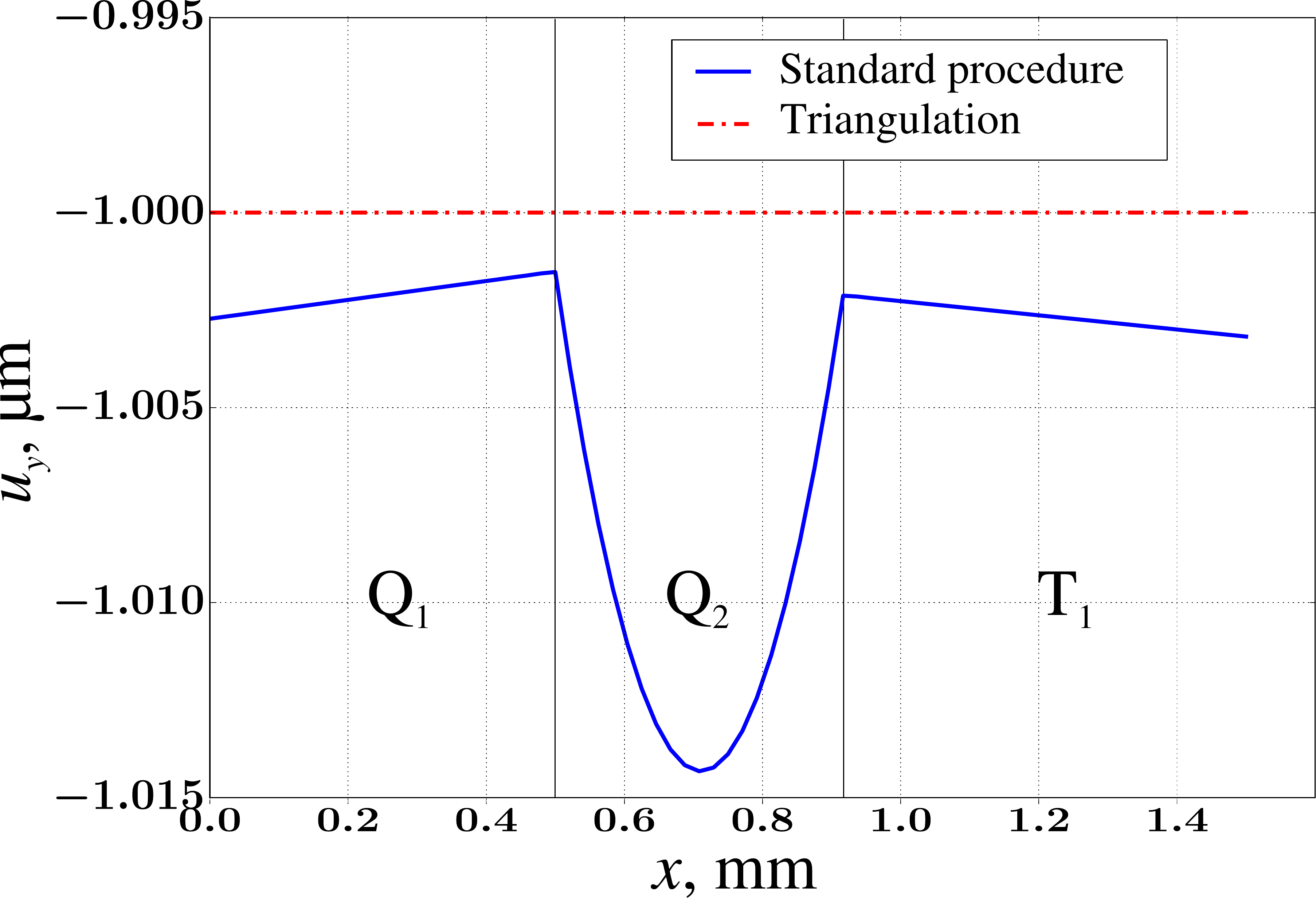}
\caption{Displacement $u_y$ profile along nodes of $\Gamma_{g,h}^1$ for the set-up shown in Fig.~\ref{fig:quad_vs_tria}: direct tying using the MorteX method results in non-linear displacement field (solid line);  triangulation of blending elements
results in a perfect match (dash-dotted line) with the reference solution  $u_y=-1$ $\mu$m.
}

\label{graph:quad_vs_tria}
\end{figure}

The domain $\Omega^2$ is discretized with a triangular (T1) and two
quadrilateral (Q1, Q2) elements, all elements use first order interpolation, and the patch domain is discretized into a rectangular
elements [Fig.~\ref{fig:quad_vs_tria}(b)].  The two domains are made of the same
linearly elastic material {($E=1$ GPa, $\nu=0$)}; the reference solution for the
selected boundary conditions is a uniform stress field
($\sigma_{xx}=\sigma_{xy}=0$ and $\sigma_{yy}=\sigma_0$)  whether it be under
plane strain or plane stress formulation. Of course, the displacement along the
tying line is uniform. 
However, as shown in Fig.~\ref{graph:quad_vs_tria} the selected discretization
does not allow to obtain the reference solution.
The solid line in Fig.~\ref{graph:quad_vs_tria} shows the resulting vertical
displacement along the tying line $\Gamma_{g}^1$; it consists of a combination
of linear and non-linear portions. 
We hypothesize that the inability to reproduce the reference solution is somehow
related to the interpolation order of displacement in host elements.  As seen
from the figure, the order of the solution is matching the maximum available
interpolation order of displacements $p_{\max}$.
This maximum interpolation order is one ($p_{\max}=1$) along any straight line
inside a linear triangle (T1).
However, this order raises to two ($p_{\max}=2$) along straight lines inside
quadrilateral elements, as long as both parent coordinates $\mu$ and $\eta$
change along these lines (Q2).
This is not the case in Q1 ($p_{\max}=1$), where one of parent coordinates
remains constant along the virtual interface, i.e. $\mu^2(\zeta) =
\mathrm{const}$ or $\eta^2(\zeta) = \mathrm{const}$.

This observation motivates us to test triangulation of blending element to limit the maximum interpolation order in host element to one ($p_{\max}=1$).
This operation does not change the number of DoFs and is easy to handle in practice.
This procedure enables us to obtain the reference solution in the considered case as demonstrated in Fig.~\ref{graph:quad_vs_tria}. The general applicability of this method will be tested in the following sections.

\section{Coarse Grained Interpolation (CGI) of Lagrange multipliers}\label{sec:cgi}
\begin{figure}[htb!]
\centering
\includegraphics[width=.75\textwidth]{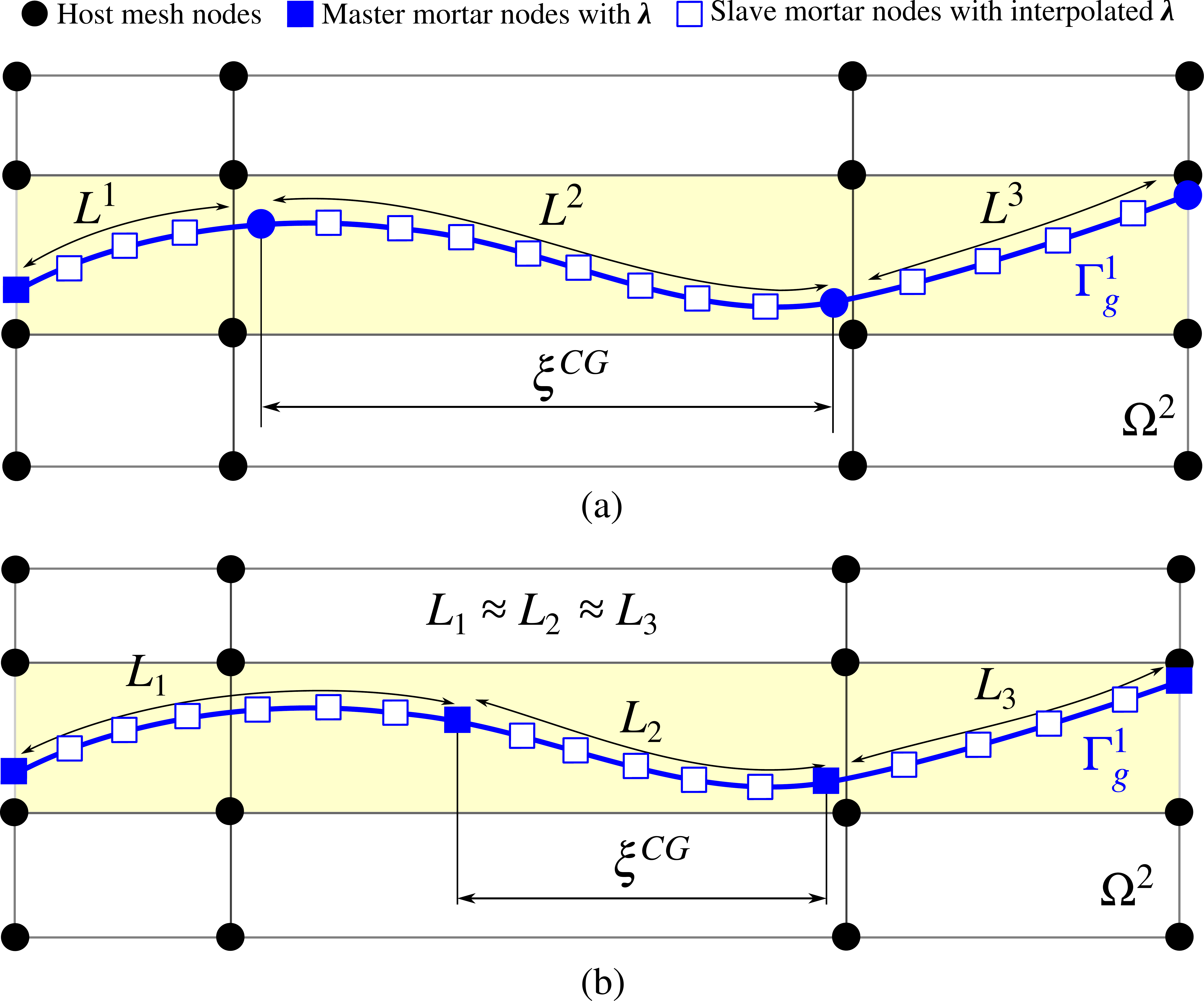}
\caption{An example illustrating the embedded surface $\Gamma_{g}^1$ cutting
through the blending elements (shaded in yellow) of the coarser host mesh. The
coarse graining of Lagrange multipliers can be implemented with respect to the
local (a) or global (b) contrast in mesh densities.
}
\label{fig:coarse_grain_concept_avg}
\end{figure}

The stability of the proposed mixed formulation is guided by the requirement to
satisfy the inf-sup condition~\cite{babuska_finite_1973}, which is not a trivial
task.  For example,  the use of Lagrange multipliers with standard interpolation
leads to non-physical oscillations along the interface when used to enforce
Dirichlet boundary conditions~\cite{barbosa_finite_1991}.  Having a much stiffer
patch than the host material represents a case approximately similar to the
imposition of Dirichlet boundary conditions, therefore in most presented
examples stiffer patch will be used.  The technique presented
in~\cite{moes_imposing_2006,bechet_stable_2009} for the X-FEM framework, involves
coarsening of the Lagrange multipliers to avoid spurious oscillations.  This is
achieved by algorithmically selecting nodes referred to as "winner nodes" along
the Dirichlet boundary.  The algorithm favors nodes close to the 
boundary, or
nodes from which many edges intersecting the boundary emanate.  The Lagrange
multiplier space is built using the winner nodes.  The size of the multiplier
space is the size of the winner nodes set. Inspired from this technique, we
suggest to coarse grain the interpolation of Lagrange multipliers (dual DoFs) to
avoid spurious oscillations.  In Fig.~\ref{fig:coarse_grain_concept_avg}, for
example, the number of mortar nodes (each of which carries Lagrange multipliers in the
standard approach) per host element considerably outnumbers the
associated DoFs to render the associated constraints independent of physical
deformations the host elements allow~\cite{sanders_nitsche_2012}. Therefore the
system shown in this figure is overconstrained and requires an appropriate
stabilization.  Coarse graining of Lagrange multiplier interpolation functions
enables to reduce the number of constraints and thus improves the problem
stability.  In this approach, not every mortar node is equipped with a Lagrange multiplier. 
Therefore, the
interpolation functions become non-local, i.e. they
span more than one patch segment. 
For this purpose, we choose a 1D parametric space $\xi^{CG} \in [-1,1]$,
spanning multiple mortar-side segments. Such parametrization can be chosen such
that length $L^i$ of the corresponding super-segment in the physical space is
comparable to the size of host elements.  As shown in
Fig.~\ref{fig:coarse_grain_concept_avg}(a), the mortar-surface is segmented into
three super-segments of lengths $L^1, L^2$ and $L^3$. The end nodes of these
segments are termed the ``master'' nodes (they carry the dual DoFs
$\vec\lambda$), other mortar-nodes are termed ``slave'' nodes.  We introduce the
local coarse-graining parameter $\kappa$ that determines the number of segments
contained in a super-segment, and thus $(\kappa-1)$ determines the number of
slave nodes per super-segment.  In Fig.~\ref{fig:coarse_grain_concept_avg}(a),
the coarse-graining parameter takes the values $\kappa=4,9,5$, for the
super-segments of lengths $L^1, L^2$ and $L^3$, respectively. 

In theory, the coarse graining is achieved through defining dual DoFs only on
master nodes.  In practice, this can also be done by 
keeping the Lagrange multipliers at
all mortar nodes and using a multi-point constraint (MPC) to enforce a linear
interpolation between the master nodes.  Hence, for a given slave node, the dual
DoFs are given by:
\begin{equation}
\vec\lambda_{\mbox{\tiny slave}}(\xi^{CG}) =
\Phi^{CG}_{l}(\xi^{CG})\vec\lambda_{\mbox{\tiny master}}^{\,l},\quad
l=1,2,
\label{eq:cgi_lam_interpol}
\end{equation}
The parametrization of the super-segment can be chosen such that $\forall i$:
$(\xi^{CG}_{i+1} - \xi^{CG}_{i})/L_{i+1,i} = \mathrm{const}$, where the
numerator represents the spacing between two mortar (patch) nodes in the
coarse-grained parametric space and $L_{i+1,i}$ corresponds to the physical
length of the corresponding segment.  An average ratio of number of mortar
segments per number of blending elements, which is termed ``mesh contrast'' $m_c$,
can be used to guide the selection of the coarse-graining parameter $\kappa$:
for example, $m_c=6$ (18 mortar segments per 3 host elements) in the example
    shown in Fig.~\ref{fig:coarse_grain_concept_avg}.  The coarse-graining
    parameter $\kappa$, for an open mortar surface, can take values in the range
    $\kappa\in[1,N_m]$, where $N_m$ is the number mortar segments; for a closed
    mortar surface the upper limit is one less.

The optimal choice of coarse-graining parameter $\kappa$ is studied on particular problem settings in Sections~\ref{sec:validation} and \ref{sec:eshelby} for open and closed mortar surfaces. 
The limit case $\kappa=1$  corresponds to the standard Lagrange interpolation (SLI). 
For the case of approximately regular discretization on both sides, a global coarse-graining parameter can be chosen, and its value is set to be approximately equal to the mesh contrast parameter $\kappa\approx m_c$ as shown in Fig.~\ref{fig:coarse_grain_concept_avg}(b). In case of non-regular mesh discretizations on mortar or/and on host sides, the coarse-graining parameter should be selected element-by-element according to the local mesh contrast as shown in Fig.~\ref{fig:coarse_grain_concept_avg}(a).
However, in all the  examples considered below, we use regular discretizations, and thus the global coarse-graining parameter will be used.
    
\section{Patch tests\label{sec:validation}}

\begin{figure}[htb!]
    \centering
    \includegraphics[width=\textwidth]{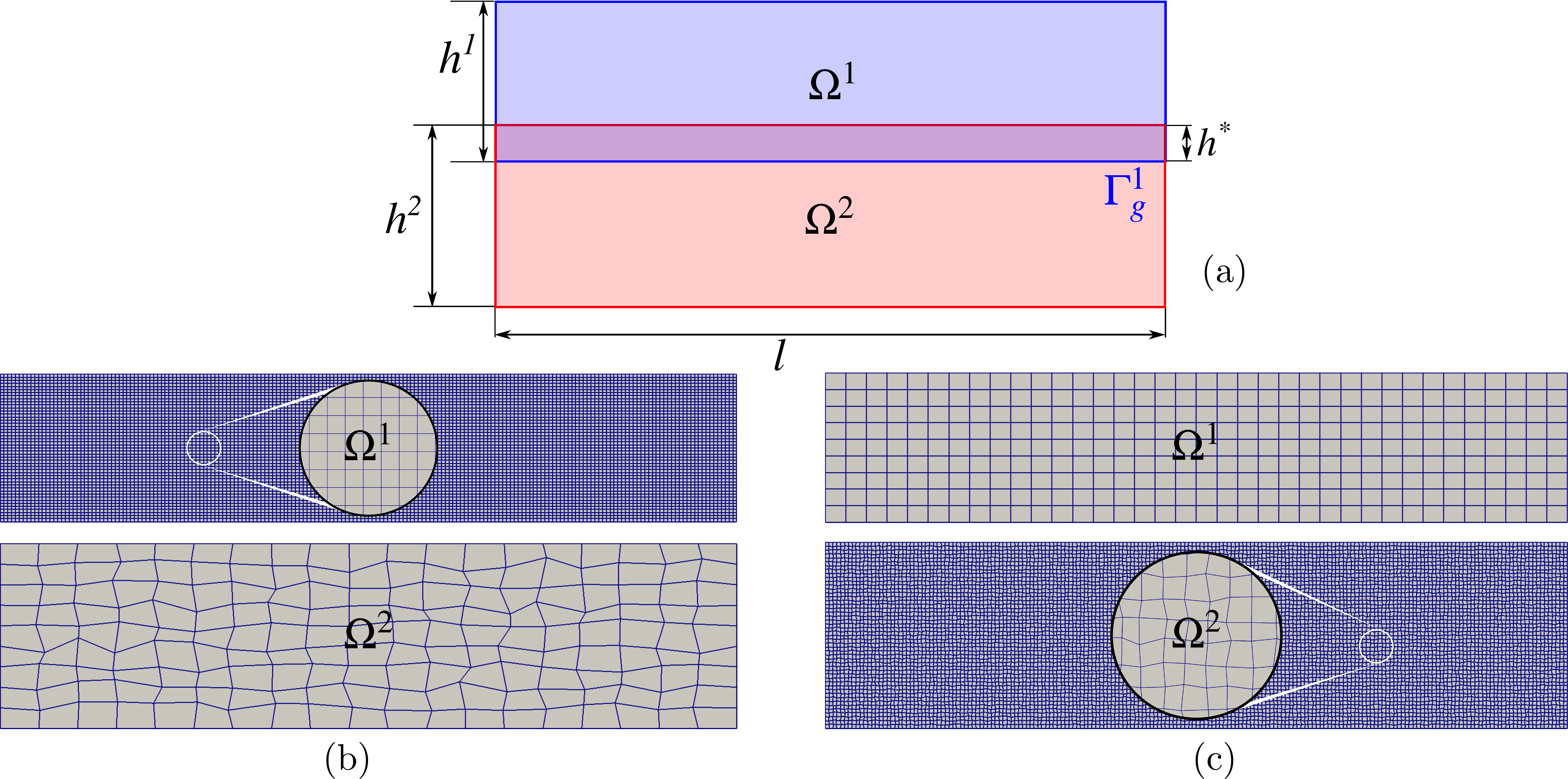}
    \caption{Validation tests set-up: (a) problem setting $h^1=1.0$ mm, $h^2=1.25$ mm, $h^*=0.25$ mm, $l=5.0$ mm, the elastic contrast between the patch and the host is given by $E^1/E^2=1000$; finite-element discretizations of the patch and host solids are shown in (b) for Case 1 ( the patch mesh is finer than that of the host, $m_c\approx 11$, $N_m = 191$), and in (c) for Case 2 (the host mesh is finer than that of the patch, $m_c\approx 0.1$, $N_m = 35$).}
    \label{fig:validation_setup}
\end{figure}

In this section, the algorithms introduced in the previous sections, implemented
in the finite element suite Z-set~\cite{besson_large_1997}, are tested on simple
problems of tied overlapped domains of different discretizations and different
material contrasts subject to bending or tensile/compressive boundary
conditions: the patch mesh is $\Omega^1$ and the host mesh is $\Omega^2$.
Linear elastic material properties are used for both the patch ($E^1,\nu^1$) and
the host ($E^2,\nu^2$). 
The geometric set-up of the patch and host domains are
illustrated in [Fig.~\ref{fig:validation_setup} (a)].  The following two extreme
cases will be considered:
{\bf Case 1.} a finer
and stiffer patch mesh is superposed onto the host mesh, and {\bf Case 2.} a coarser and stiffer patch mesh is
superposed onto the host mesh.

These two particular cases are chosen for the validation since they are
prone to severe manifestations of the mesh locking~\cite{sanders_nitsche_2012} as in the case of enforcing
Dirichlet boundary conditions using  Lagrange
multipliers along embedded surfaces~\cite{moes_imposing_2006,bechet_stable_2009,hautefeuille_robust_2012}.
Additionally, the host domain is meshed with ''distorted`` quadrilaterals 
which is classical in patch test studies to exacerbate potential anomalies. 
Moreover, as was shown by a simple example
Section~\ref{sec:effect_host_mesh_interpolation} the tying along distorted quadrilateral elements is prone to considerable errors if the mortar-type tying is used directly.
The material contrast is introduced by choosing $E^1/E^2 =1000$, and $E^1 = 1$ GPa, both domains have the same Poisson's ratio $\nu^1=\nu^2=0.3$.
The discretizations for the  two cases are shown in
Fig.~\ref{fig:validation_setup} (b, c). 
The mesh contrast $m_c\approx 11$ and $m_c\approx 0.1$ and the number of mortar segments $N_m = 191$ and $N_m = 35$ are used for Case 1 and 2, respectively.
Note that all the stress fields $\sigma_{xx},\sigma_{yy}$ along the tying interface, which are presented on the plots below, are found by the data extrapolated from Gauss points and averaged
 at mortar nodes.
 
\begin{figure}[htb!]
    \centering
    \includegraphics[width=\textwidth]{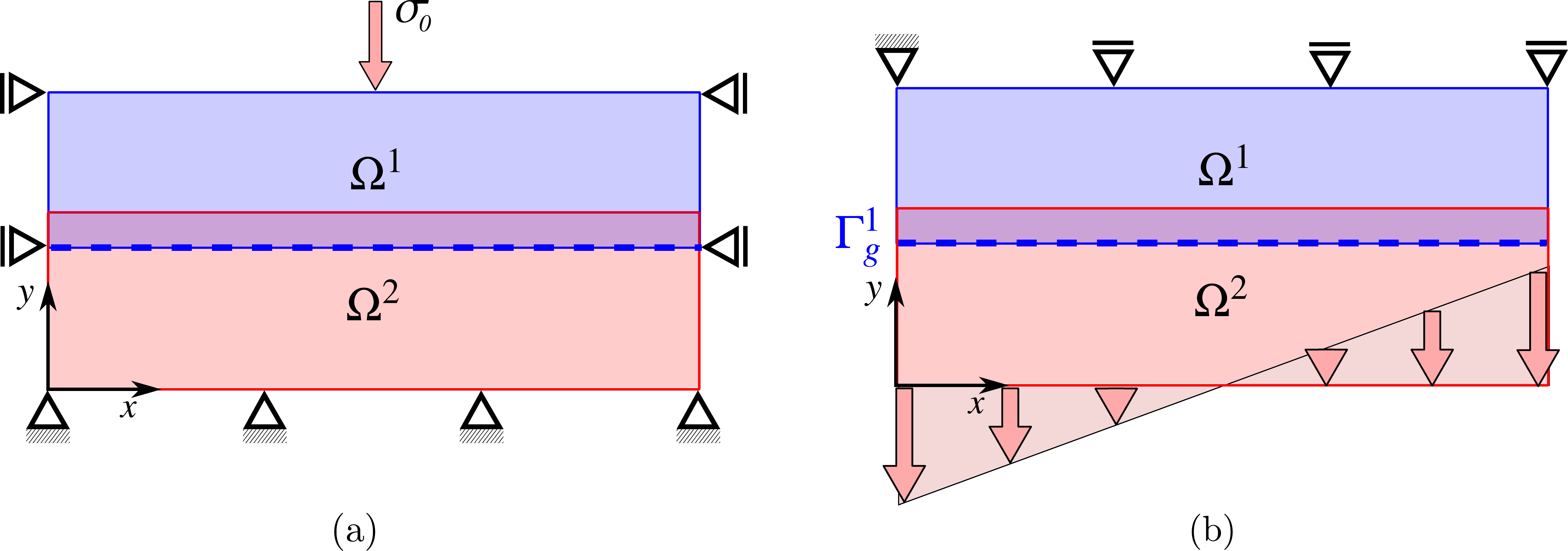}
    \caption{Validation tests boundary conditions: (a) compression patch test,
    (b) bending patch test.}
    \label{fig:validation_BCs}
\end{figure}
    
\subsection{Tension/compression patch test}

\begin{figure}[htb!]
    \centering
    \includegraphics[width=\textwidth]{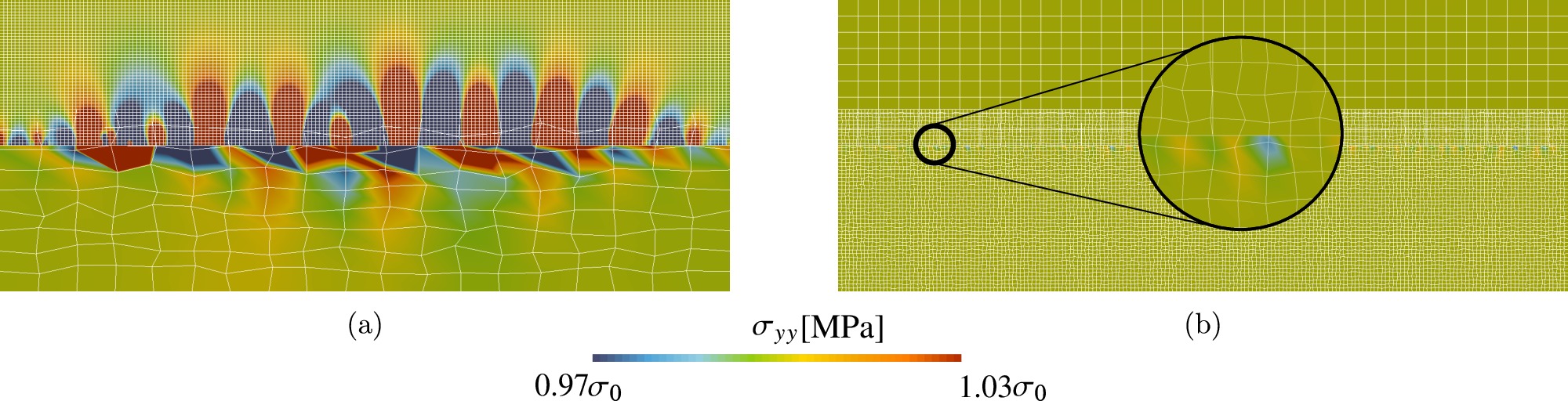}
    \caption{Compression patch test: contour plots of stress component $\sigma_{yy}$ in (a) Case 1, and (b) Case 2.} 
    \label{fig:case1_vs_case2_patch_test}
\end{figure}

\begin{figure}[htb!]
    \centering
    \includegraphics[width=\textwidth]{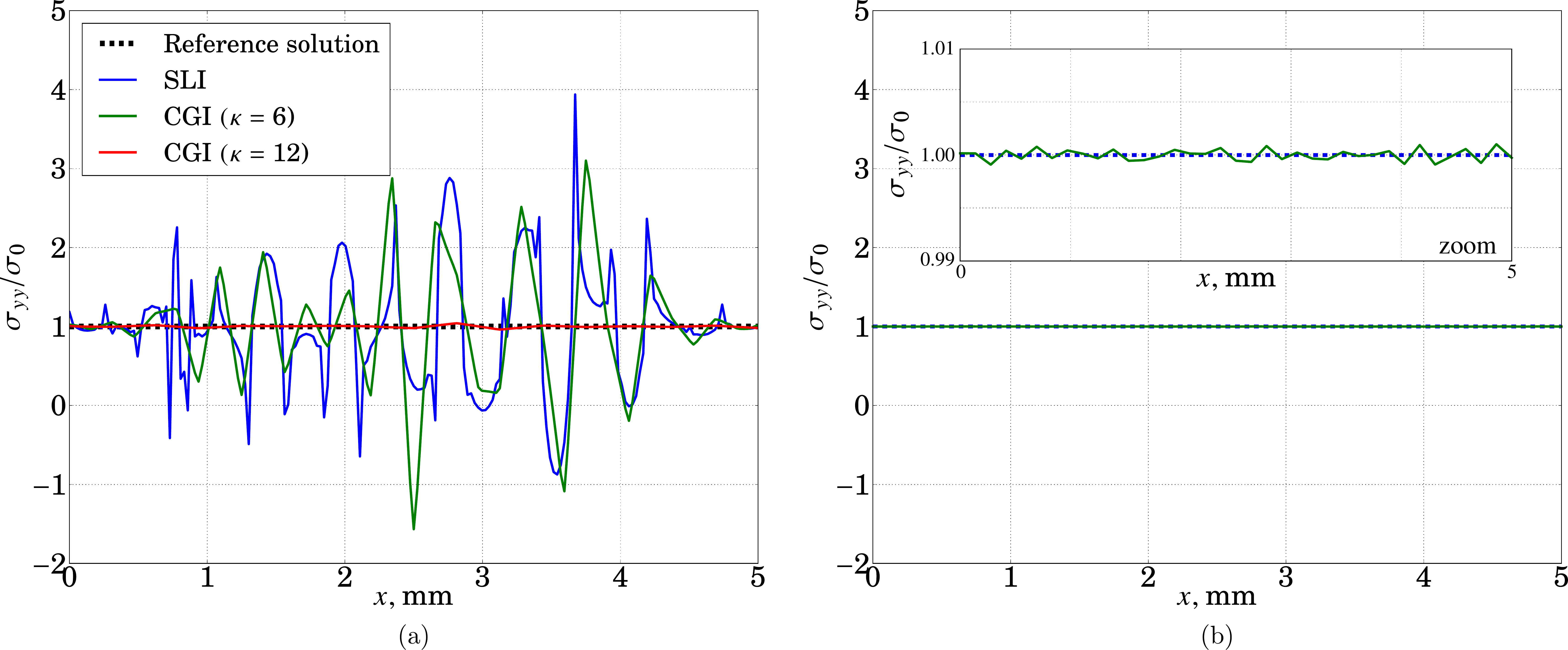}
    \caption{Compression patch test: distribution of $\sigma_{yy}$ along
    the tying interface in (a) Case 1 for standard (SLI) and coarse-grained (CGI) Lagrange interpolation, and (b) in Case 2 for SLI.}
    \label{graph:patch_test_c1_c2_plots}
\end{figure}

A uniform pressure $\sigma_0$ is applied on the top
surface, the bottom surface is fixed in all directions $\vec u=0$
[Fig.~\ref{fig:validation_BCs}(a)]. 
This is a classical patch test in contact mechanics,
which is used here to test the tying of different materials. 
This material contrast requires additional lateral conditions 
(lateral sides are fixed in normal direction $u_x=0$) to avoid singularities at extremities of the interface. 
The reference solution for $\sigma_{yy}$ is a uniform field $\sigma_{yy} = \sigma_0$. 
As expected, in case of stiffer and finer patch mesh (Case 1), spurious oscillations are observed. 
They have large amplitude that reaches $300$ \% of the reference
solution, moreover, they are not confined to the interface but propagate into the bulk [Figs.~\ref{fig:case1_vs_case2_patch_test}(a), \ref{graph:patch_test_c1_c2_plots}(a)]. 
In case of stiffer and coarser patch mesh (Case 2), the spurious oscillations are of
considerably lower amplitude  (under $1$ \%), they
are rather localized in the host mesh in 
close vicinity of the interface and do not extend in the patch mesh
[Figs.~\ref{fig:case1_vs_case2_patch_test}(b), \ref{graph:patch_test_c1_c2_plots}(b)]. 

\begin{figure}[htb!]
    \centering
    \includegraphics[width=\textwidth]{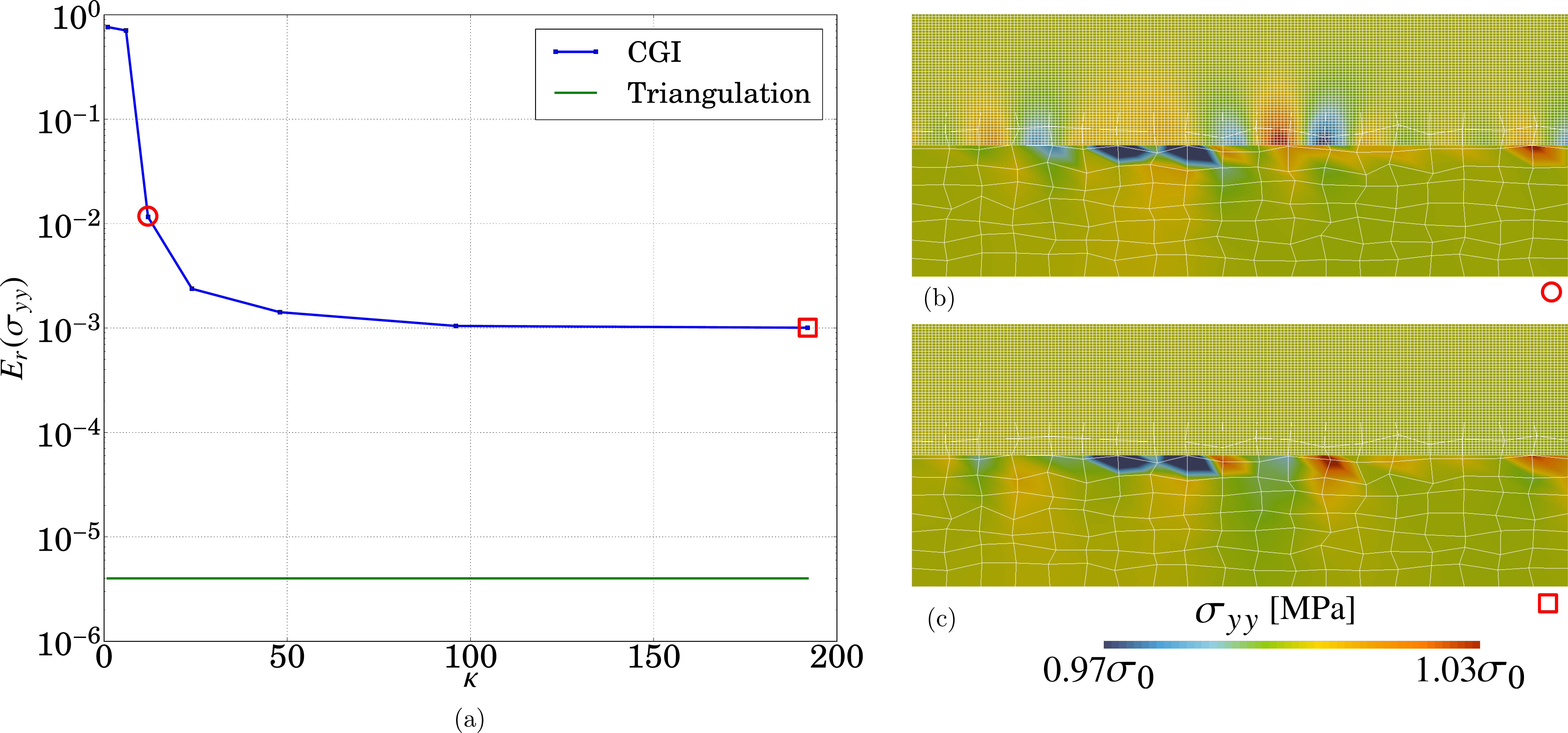}
    \caption{
    Compression patch test: (a) decay and saturation of the relative error $E_r(\sigma_{yy})$ for $\kappa=\{1,6,12,24,48,96,192\}$ for CGI in comparison with the error obtained with triangulation of blending elements; (b,c) contour plots of stress component $\sigma_{yy}$  for (b) $\kappa = 12$ and (c) $\kappa = N_m = 192$.
    }
    \label{graph:patch_test_CGI}
\end{figure}

\begin{figure}[htb!]
    \centering
    \includegraphics[width=\textwidth]{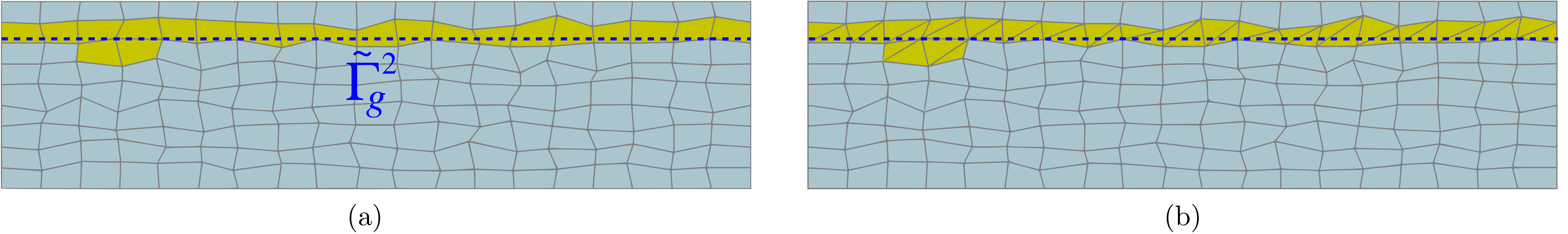}
    \caption{(a) Blending elements, (b) Triangulated blending
    elements.}
    \label{fig:pre_trian_idea}
\end{figure}

\begin{figure}[htb!]
    \centering
    \includegraphics[width=\textwidth]{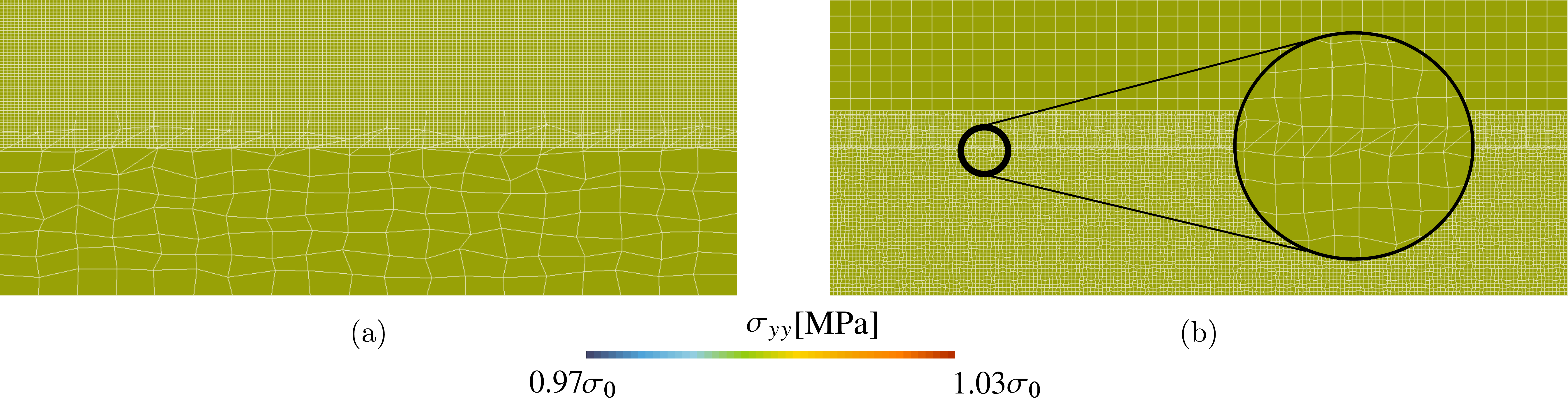}
    \caption{Patch test stress $\sigma_{yy}$ with Triangulation of blending
    elements: (a) Case 1, (b) Case 2.}
    \label{fig:patch_test_split_tria}
\end{figure}

In order to quantify the improvement achieved with the suggested coarse grained interpolation (CGI) and with triangulation technique, we
    introduce the $L^2$-norm of the error in the $\sigma_{yy}$ stress component:
    \begin{equation}\label{eq:error_norm_bending}
        E_r(\sigma_{yy}) = 
        \frac{\left\|\sigma_{yy}^{\mbox{\tiny
        ref}}-\sigma_{yy}\right\|_{L^2(\Gamma_g^1)}}{\left\|\sigma_{yy}^{\mbox{\tiny
        ref}}\right\|_{L^2(\Gamma_g^1)}},
    \end{equation}
    where the norm means $\left\|f(x)-g(x)\right\|_{L^2(\Gamma_g^1)} = \sqrt{\sum_i \left[f(x_i)-g(x_i)\right]^2}$, where $x_i \in [0,L]$ are the $x$-coordinate of mortar nodes in the reference configuration, and $L$ is the length of the surface $\Gamma_g^1$.
In Fig.~\ref{graph:patch_test_CGI} we demonstrate the performance of the CGI technique. 
As seen from the figure, the error in stress greatly reduces compared to the standard interpolation (SLI), when coarse-graining parameter $\kappa$ increases. However,
the error saturates at $\approx 10^{-3}$ and the convergence to the reference solution is missing. On the contrary, the triangulation technique (Fig.~\ref{fig:pre_trian_idea}) 
enables to achieve a superior precision as shown in Figs.~\ref{graph:patch_test_CGI}(a) and \ref{fig:patch_test_split_tria}.

\subsection{Bending patch test}\label{subsec:bending_patch_test}
\begin{figure}[htb!]
    \centering
    \includegraphics[width=\textwidth]{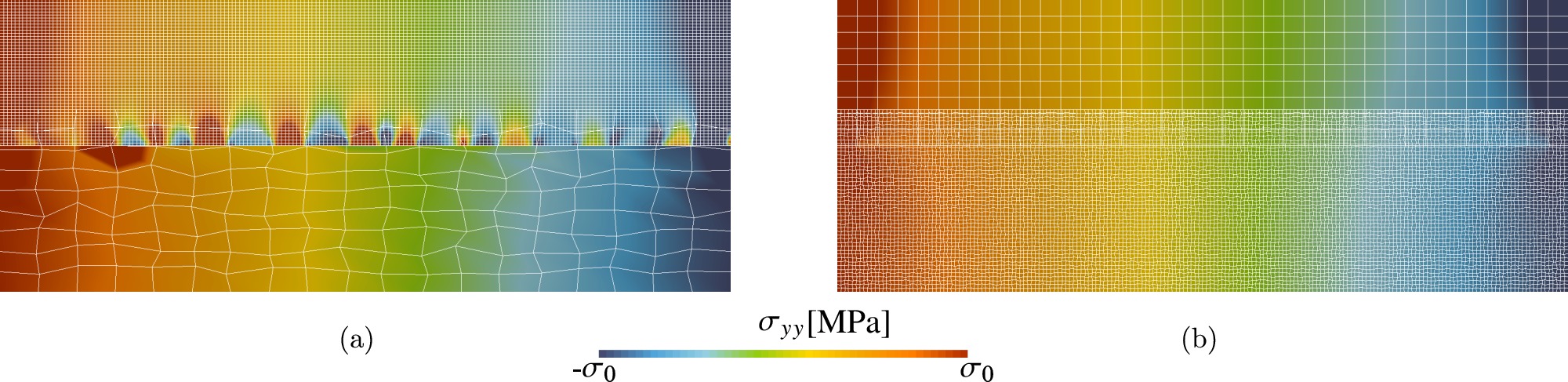}
    \caption{Bending stresses $\sigma_{yy}$: (a) Case 1, (b) Case 2.}
    \label{fig:case1_vs_case2_bending}
\end{figure}

\begin{figure}[htb!]
    \centering
    \includegraphics[width=.5\textwidth]{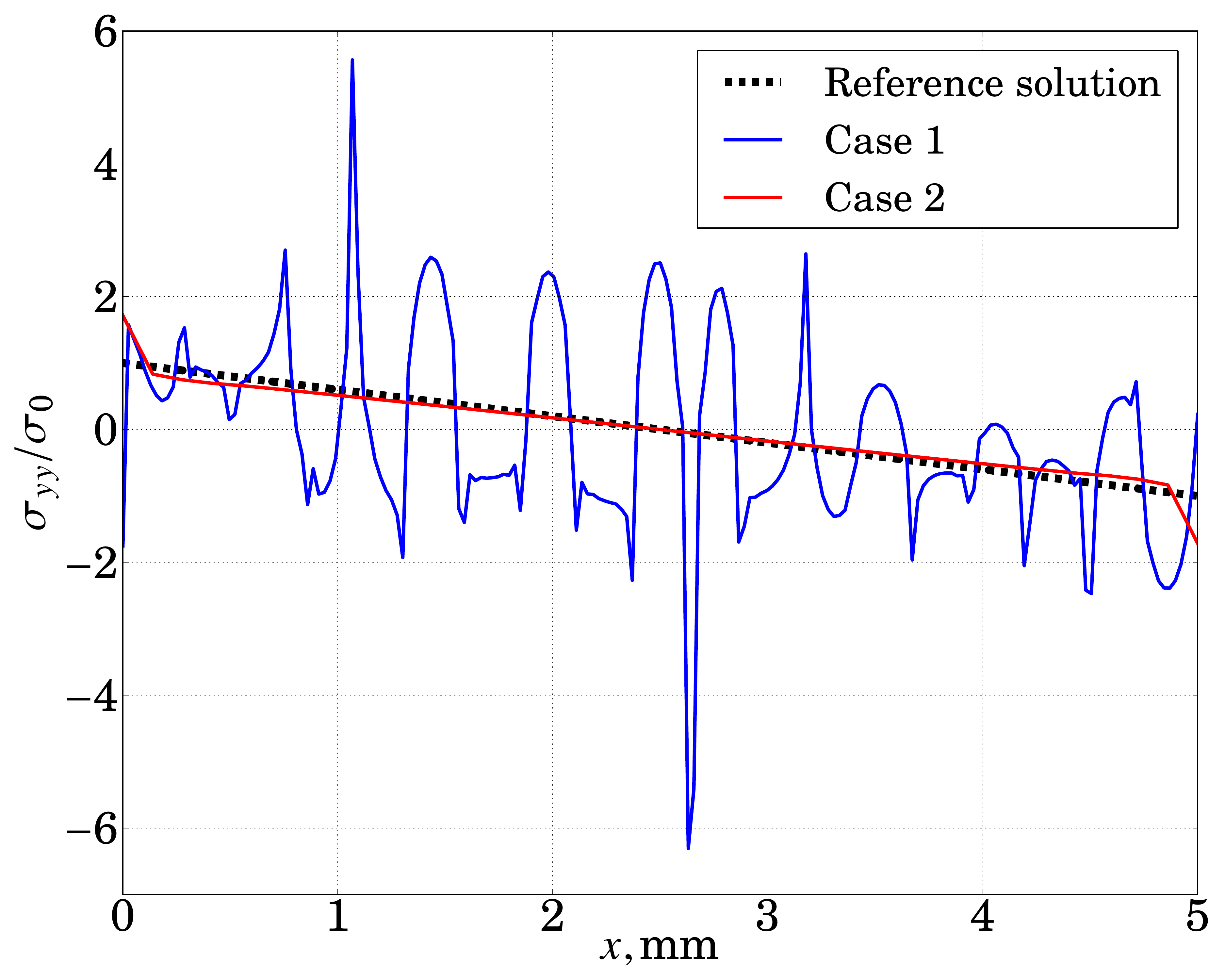}
    \caption{Bending stress $\sigma_{yy}$ along the tying interface, comparison with the reference
    solution.}
    \label{graph:case1_case2_bending_compare}
\end{figure}

A linear distribution of pressure 
\begin{equation} 
    \sigma_{yy} = 2\sigma_0(x/l-1/2)
    \label{eq:bending_stress}
\end{equation} 
with $\sigma_0=1$ MPa is
applied on the bottom surface, while keeping the top surface fixed in vertical
direction, only the corner point is fixed in horizontal direction, the lateral sides remain free $\sigma_{xy} = \sigma_{xx} = 0$
[Fig.~\ref{fig:validation_BCs} (b)].
The same linear distribution of the vertical stress
component~\eqref{eq:bending_stress}  through the two solids should take place. 
This case study (Case 1, 2) was inspired from the work~\cite{sanders_nitsche_2012},  where the authors also used the combination of the mortar method and the selective integration. 
It was shown that Case 1, in particular, results in high-amplitude spurious
    oscillations in the interface contrary to Case 2 that has a smoother
    stress profile along the interface. 
    Under the standard Lagrangian
    interpolation set-up we could reproduce similar results, see
    Fig.~\ref{fig:case1_vs_case2_bending}, \ref{graph:case1_case2_bending_compare}. 
    These oscillations could be removed with
    Nitsche method provided some adjustment of the stabilization penalty
    parameters on each side of the interface~\cite{sanders_nitsche_2012}. We
    demonstrate below that using the coarse-grained interpolation for Lagrange
    multipliers, as suggested in Section~\ref{sec:cgi}, also permits avoiding these
    oscillations  in the MorteX framework. As discussed in
    Section~\ref{sec:cgi}, the choice of the optimal coarse-graining parameter is
    governed by local or global mesh contrast, which can be easily determined either for
    every segment or for the whole interface. It renders the choice of the coarse-graining parameter $\kappa$ fully automatic. 
    Contrary to the stabilized Nitsche method, no knowledge about  local material contrast is needed.
    
\begin{figure}[htb!]
    \centering
    \includegraphics[width=\textwidth]{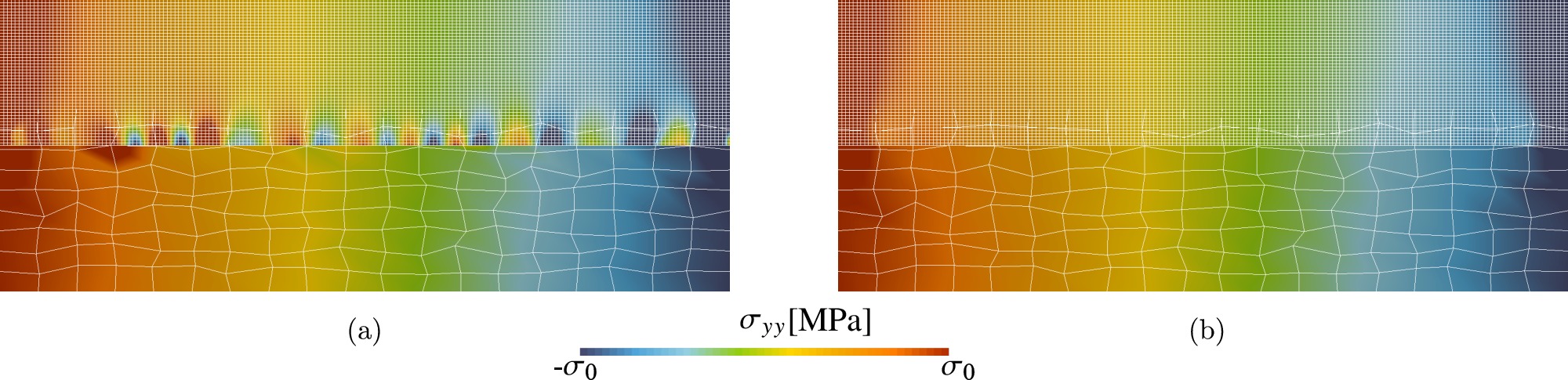}
    \caption{Coarse grained Lagrange multiplier space for Case 1: (a) $\kappa\,=\,6$,
    (b) $\kappa\,=\,12$.}
    \label{fig:spacing_bending_compare_stressY}
\end{figure}

\begin{figure}[htb!]
    \centering
    \includegraphics[width=\textwidth]{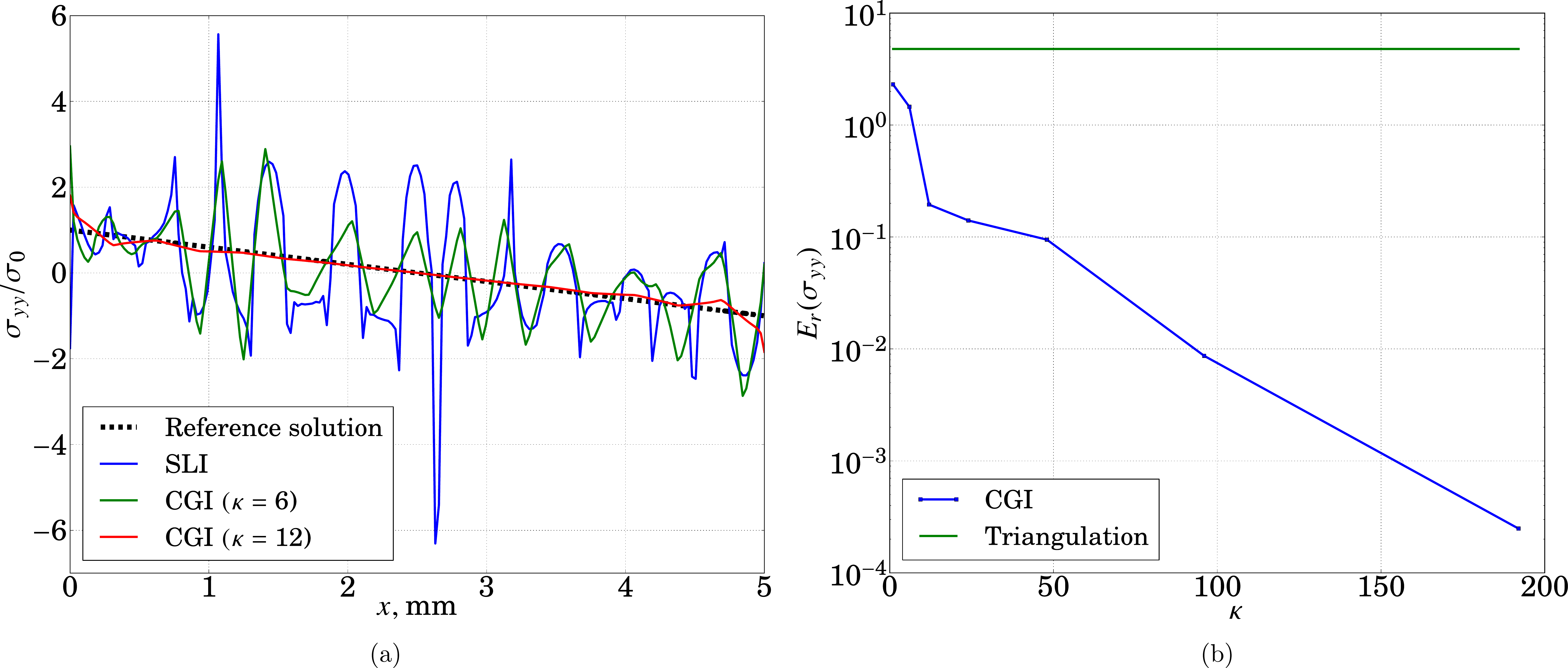}
    \caption{(a) Case 1: comparison of bending stresses ($\sigma_{yy}$) for SLI and CGI
    with the reference solution; (b) decay of the relative error $E_r(\sigma_{yy})$ for $\kappa=\{1,6,12,24,48,96,192\}$ for CGI in comparison with the error obtained with triangulation of blending elements.}
    \label{graph:beding_k_effect}
\end{figure}

\begin{figure}[htb!]
    \centering
    \includegraphics[width=\textwidth]{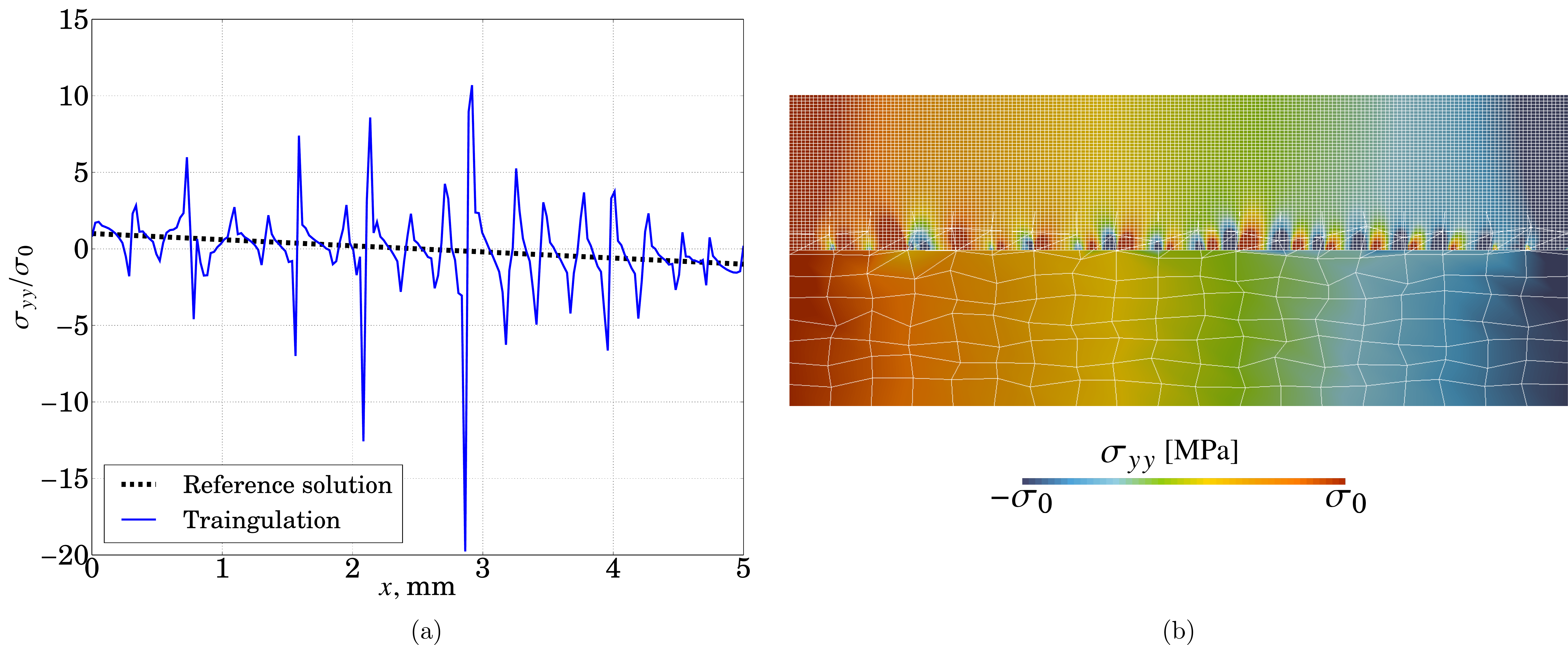}
    \caption{Bending stresses $\sigma_{yy}$ in Case 1 for triangulated blending elements: (a) stress distribution along the tying interface; (b) contour plot of the stress component.}
    \label{fig:bending_oneD_trian_stress}
\end{figure}

    In Fig.~\ref{fig:spacing_bending_compare_stressY}, the vertical stress
    component is shown for the Case 1 when the coarse-graining parameter for Lagrange
    multipliers is set to (a) $\kappa=6$ and (b) $\kappa = 12$.  It is
    clearly seen in the figures that $\kappa = 6$ does not sufficiently relaxes the over-constraining of the
    Lagrange multipliers space (we recall that $m_r=11$) to obtain a
    smooth reference solution, even though the amplitude of oscillations is
    slightly reduced compared to standard Lagrange multipliers, which can be seen in Fig.~\ref{graph:beding_k_effect}(a) where the standard solution (obtained with the standard interpolation of Lagrange multipliers SLI) is compared with the coarse-grained interpolation (CGI). With the coarse-graining parameter $\kappa = 12$ we obtain a much improved result comparable to the reference solution.
    The relative $L^2$ error $E_r(\sigma_{yy})$
    Eq.~\eqref{eq:error_norm_bending} is shown in
    Fig.~\ref{graph:beding_k_effect}(b) for different values of $\kappa$.  The
    error becomes acceptable only for $\kappa \ge m_c$, however, the accuracy of
    the solution constantly improves with a further increase of $\kappa$.  The
    fast drop of the error for $\kappa \in [1,m_c]$ is associated with the
    graduate removal of spurious oscillations in the stress distribution,
    whereas for $\kappa > m_c$  improves further the error by better
    approximation of stresses at extremities of the interface. Since the
    reference stress distribution is linear, only two Lagrange multipliers
    are sufficient to capture it, leading to an error reduction up to $\kappa = N_m$.  In
    general, as will be shown later, too coarse a
    representation of Lagrange multipliers leads to deterioration of the
    solution (see Section~\ref{sec:eshelby}).  The  triangulation of blending
    element is also tested in the bending patch test, see
    Figs.~\ref{fig:bending_oneD_trian_stress}. In contrast to the compression
    patch test, triangulation here does not help with the removal of spurious
    oscillations.

\subsection{Summary of patch tests}

\begin{figure}[htb!]
\centering
\includegraphics[width=\textwidth]{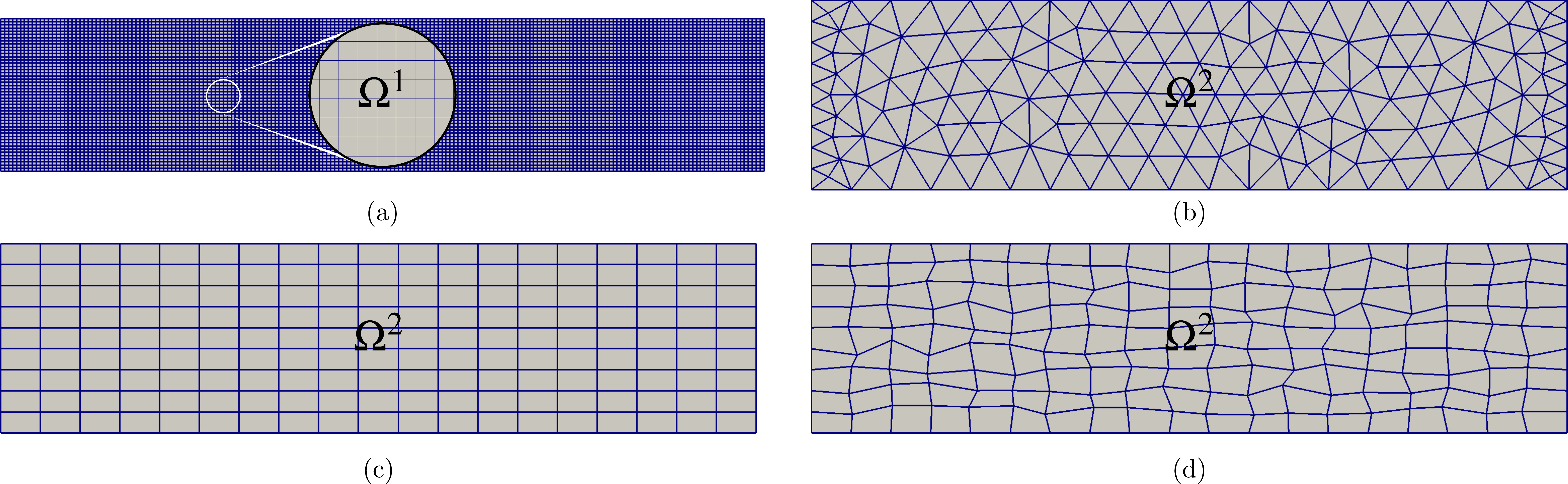}
\caption{Discretized setting of coarse host mesh and finer patch mesh
($m_c\approx10$): (a) patch mesh ($\Omega^1$), (b) host mesh with
linear triangular elements ($\Omega^2$), (c) host mesh with bilinear
quadrilateral elements ($\Omega^2$), (d) host mesh with bilinear
distorted quadrilateral elements ($\Omega^2$).}
\label{fig:table1_host_meshes}
\end{figure}

\begin{figure}[htb!]
\centering
\includegraphics[width=\textwidth]{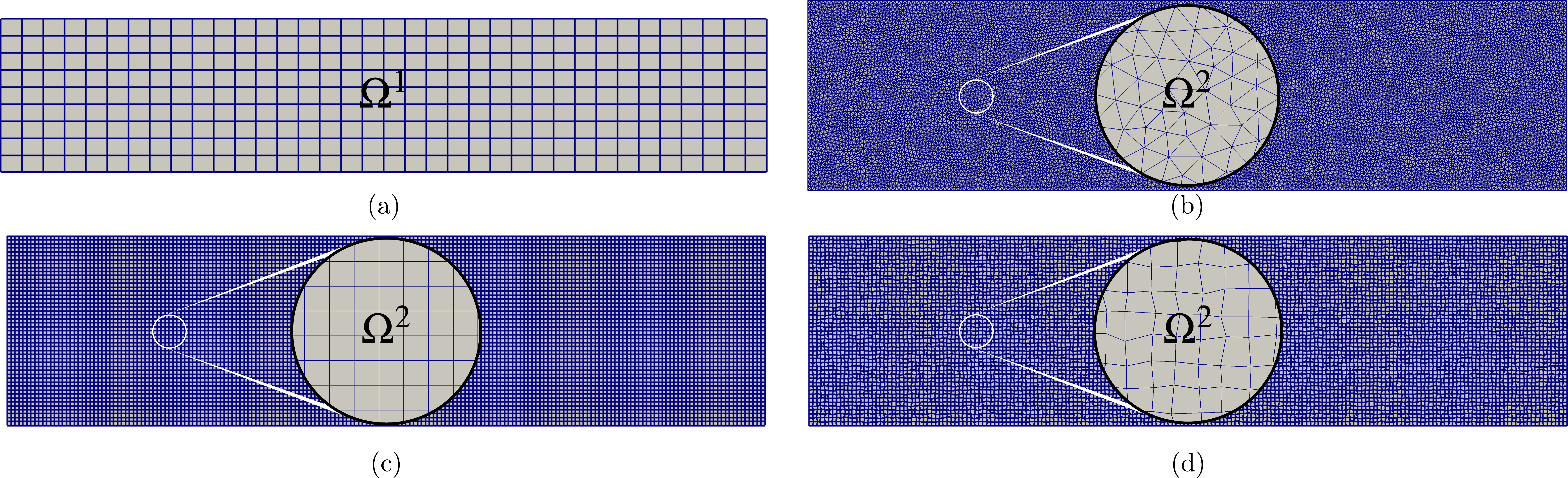}
\caption{Discretized setting of fine host mesh and coarse patch mesh
($m_c\approx0.1$): (a) patch mesh ($\Omega^1$), (b) host mesh with
linear triangular elements ($\Omega^2$), (c) host mesh with bilinear
quadrilateral elements ($\Omega^2$), (d) host mesh with bilinear
distorted quadrilateral elements ($\Omega^2$).}
\label{fig:table2_host_meshes}
\end{figure}

Here, we present the ensemble of patch-test results for various combinations of
host/patch meshes and material contrasts.  As before, we consider two cases:
Case 1 corresponds to a fine patch mesh [mesh-density contrast $m_c=10$,
Fig.~\ref{fig:table1_host_meshes}(a)] which is tied with a coarser host mesh
made of triangular, aligned or distorted quadrilateral elements
Fig.~\ref{fig:table1_host_meshes}(b,c,d), respectively.  Results of Case 1 are
presented in Table~\ref{tab:coarse_host}.  In Case 2, the patch mesh
[mesh-density contrast $m_c=0.1$, Fig.~\ref{fig:table2_host_meshes}(a)] is
coarser than the host mesh, which again can be made of triangles, aligned or
distorted quadrilateral elements, see Fig.~\ref{fig:table2_host_meshes}(b,c,d),
respectively. Results of Case 2 are presented in Table~\ref{tab:fine_host}.
Softer ($E^1/E^2=10^{-3}$) and stiffer ($E^1/E^2=1000$) patch materials are
compared to the host material were considered.
We also tested different interpolation order (p0 and p1) for standard Lagrange
interpolation (SLI), and p1-interpolation for coarse-grained interpolation (CGI)
in which the coarse-graining parameter takes its maximum value $\kappa=N_m$.
The table clearly demonstrates that the tying performance is strongly dependent
on the type of patch test. Cases which show a small error in bending test can
demonstrate a slightly higher error in compression test as in case $m_c=10$,
$E^1/E^2=1000$ for distorted quads with CGI scheme. However, with a high
fidelity it could be stated that if the tying method passes the bending patch
test then it passes the compression patch test. The inverse is, in general,
false. Interestingly, the triangulation of quadrilateral elements can
considerably increase the error in case of SLI scheme, it does not happens with
the CGI scheme. Clearly, from these tables it can be concluded that the CGI
scheme outperforms the standard SLI scheme (both p0 and p1) in all studied
combinations of mesh, element types and patch-test type (48 tests in total).

%

\begin{table}
\begin{tabular}{cccccc}
           &  				& Triangulation       & Dual  &  \\
 $E^1/E^2$ & Host-mesh type   & of blending  &  interpolation  &
 $L^2$ error $E_r (\sigma_{yy})$ & $L^2$ error $E_r (\sigma_{yy})$  \\
               &                            &
                      elements   &  & (bending patch test)& (compression patch test) \\[5pt]\hline
 1000 & Triangles & No & SLI (p0) & 1.668e+01 & 3.57e-06\\
 $\sim$ & Aligned quads & No & $\sim$ & 3.357e-01&0.00\\
 $\sim$ & Distorted quads & No & $\sim$ & 2.95e+00&1.064e+00\\
 $\sim$ & $\sim$ & Yes & $\sim$ & 7.66e+00&4.76e-05\\[5pt]\hline\\[-7pt]
 1000 & Triangles & No & SLI (p1) & 1.266e+01&3.55e-06 \\
 $\sim$ & Aligned quads & No & $\sim$ & 2.789e-01&0.00\\
 $\sim$ & Distorted quads & No & $\sim$ & 2.30e+00&7.627e-01\\
 $\sim$ & $\sim$ & Yes & $\sim$ & 4.786e+00&4e-06\\[5pt]\hline\\[-7pt]
 1000 & Triangles & No & CGI (p1) & 4.53e-05 &0.00\\
 $\sim$ & Aligned quads & No & $\sim$ & 4.51e-05&0.00\\
 $\sim$ & Distorted quads & No & $\sim$ & 2.4e-04&1.e-3\\
 $\sim$ & $\sim$ & Yes & $\sim$ & 2.3e-04&3.e-4\\[5pt]\hline\\[-7pt]
 1.e-3 & Triangles & No & SLI (p0)  & 2.192e-01&0.00 \\
 $\sim$ & Aligned quads & No & $\sim$ & 2.192e-01&0.00\\
 $\sim$ & Distorted quads & No & $\sim$ & 2.186e-01&2.9e-05\\
 $\sim$ & $\sim$ & Yes & $\sim$ & 2.192e-01&2.4e-05\\[5pt]\hline\\[-7pt]
 1.e-3 & Triangles & No & SLI (p1)  & 2.193e-01 &0.00\\
 $\sim$ & Aligned quads & No & $\sim$ & 2.194e-01&0.00\\
 $\sim$ & Distorted quads & No & $\sim$ & 2.191e-01&1.15e-05\\
 $\sim$ & $\sim$ & Yes & $\sim$ & 2.194e-01&0.00\\[5pt]\hline\\[-7pt]
 1.e-3 & Triangles & No & CGI (p1) & 4.51e-05 &0.00\\
 $\sim$ & Aligned quads & No & $\sim$ & 4.51e-05&0.00\\
 $\sim$ & Distorted quads & No & $\sim$ & 2.4e-04&2.8e-4\\
 $\sim$ & $\sim$ & Yes & $\sim$ & 2.3e-04&2.8e-4\\
\end{tabular}
 \caption{Patch test performance for overlapping domains with a finer patch ($m_c=10$).}
 \label{tab:coarse_host}
 \end{table}

\begin{table}
\begin{tabular}{cccccc}
           &  				& Triangulation       & Dual  &  \\
 $E^1/E^2$ & Host-mesh type   & of blending  &  interpolation  &
 $L^2$ error $E_r (\sigma_{yy})$ & $L^2$ error $E_r (\sigma_{yy})$  \\
               &                            &
                      elements   &  & (bending patch test)& (compression patch test) \\[5pt]\hline
  1000 & Triangles & No & SLI (p0) & 2.122e-01 & 2.34e-05\\
  $\sim$ & Aligned quads & No & $\sim$ & 2.113e-01&0.00\\
  $\sim$ & Distorted quads & No & $\sim$ & 2.114e-01&3.58e-04\\
  $\sim$ & $\sim$ & Yes & $\sim$ & 7.66e+00&4.76e-05\\[5pt]\hline\\[-7pt]
 1000 & Triangles & No & SLI (p1) & 2.493e-01&1.63e-05 \\
 $\sim$ & Aligned quads & No & $\sim$ & 2.481e-01&0.00\\
 $\sim$ & Distorted quads & No & $\sim$ & 2.483e-01&5.00e-04\\
 $\sim$ & $\sim$ & Yes & $\sim$ & 4.786e+00&4e-06\\[5pt]\hline\\[-7pt]
 1000 & Triangles & No & CGI (p1) & 6.00e-04 &7.42e-07\\
 $\sim$ & Aligned quads & No & $\sim$ & 6.00e-04&0.00\\
 $\sim$ & Distorted quads & No & $\sim$ & 6.00e-04&1.75e-06\\
 $\sim$ & $\sim$ & Yes & $\sim$ & 6.00e-04&0.00\\[5pt]\hline\\[-7pt]
 1.e-3 & Triangles & No & SLI (p0) & 1.702e-01&0.00 \\
 $\sim$ & Aligned quads & No & $\sim$ & 1.702e-01&0.00\\
 $\sim$ & Distorted quads & No & $\sim$ & 1.702e-01&1.72e-06\\
 $\sim$ & $\sim$ & Yes & $\sim$ & 2.192e-01&2.4e-05\\[5pt]\hline\\[-7pt]
 1.e-3 & Triangles & No & SLI (p1) & 1.734e-01 &0.00\\
 $\sim$ & Aligned quads & No & $\sim$ & 1.734e-01&0.00\\
 $\sim$ & Distorted quads & No & $\sim$ & 1.735e-01&0.00\\
 $\sim$ & $\sim$ & Yes & $\sim$ & 2.194e-01&0.00\\[5pt]\hline\\[-7pt]
 1.e-3 & Triangles & No & CGI (p1) & 5.5e-04 &0.00\\
 $\sim$ & Aligned quads & No & $\sim$ & 5.5e-04&0.00\\
 $\sim$ & Distorted quads & No & $\sim$ & 5.5e-04&0.00\\
 $\sim$ & $\sim$ & Yes & $\sim$ & 5.5e-04&0.00\\
 \end{tabular}
 \caption{Patch test performance for overlapping domains with a coarser patch mesh ($m_c=0.1$).}
 \label{tab:fine_host}
\end{table}

\section{Circular inclusion in infinite plane: convergence study\label{sec:eshelby}}

\begin{figure}[htb!]
\centering
\includegraphics[width=\textwidth]{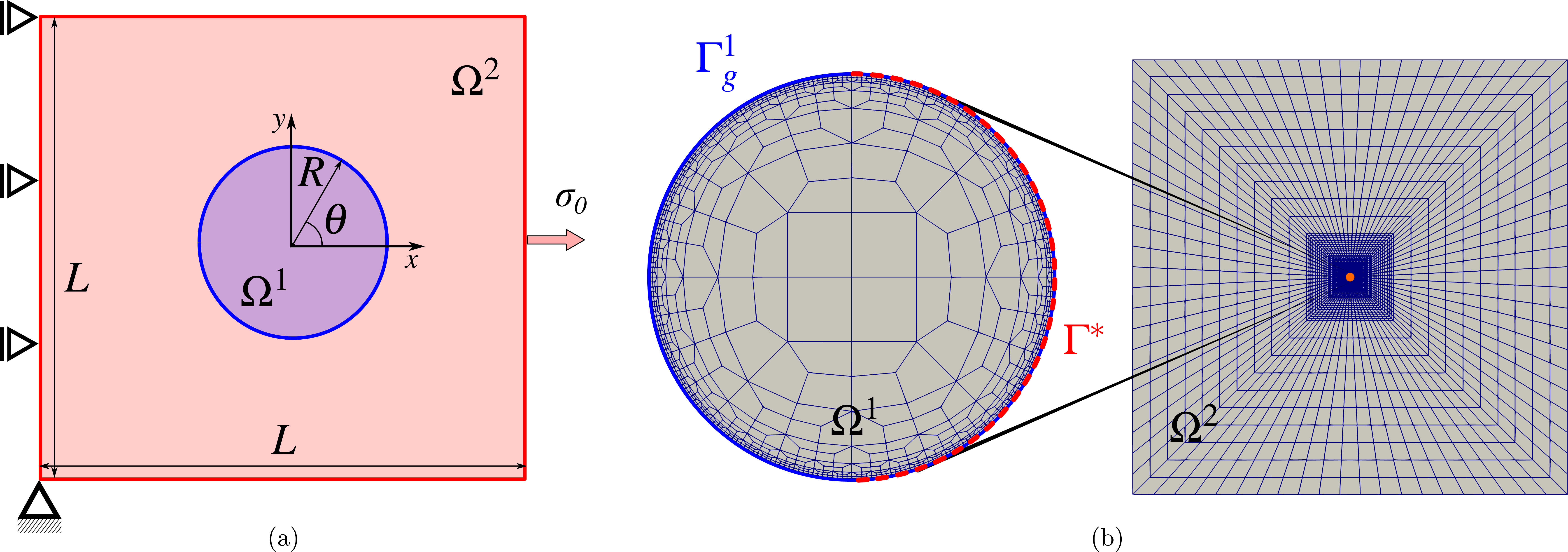}
\caption{Square matrix with circular inclusion: (a) problem setting (not to scale) with
inclusion domain $\Omega^1$ (radius $R=0.1$ mm) superposed over the matrix domain $\Omega^2$
($L=10$ mm), (b) inclusion and matrix domain discretizations, where
$\Gamma_g^1$ is the tying boundary and $\Gamma^* \subset\Gamma_g^1$
($\theta\in[-\pi/2, \pi/2]$).}
\label{fig:eshelby_geom_setup}
\end{figure}

Having demonstrated a general good performance of the coarse-graining
interpolation, here we carry out a mesh-convergence study.  We focus on the
worse case scenario (see Section~\ref{sec:validation}) when the patch mesh is
finer and stiffer than the host mesh.
%
We consider a circular inclusion embedded in an infinite softer matrix in plane
strain formulation, and subject to a uniform traction applied at
infinity~\cite{sharma_circular_1979,kachanov_handbook_2013,herve_elastic_1995}.
This particular problem represents a sub-case of a general Eshelby problem of an
ellipsoidal inclusion in a
matrix~\cite{eshelby_elastic_1959,muskhelishvili_basic_nodate}.
Fig.~\ref{fig:eshelby_geom_setup}(a) shows the used computational set-up: a
circular inclusion $\Omega^1$ (patch) with radius $R=0.1$ mm, centered at origin, 
is superposed on a matrix $\Omega^2$ (host) represented by a square of side
$L=10$ mm ($L\gg R$). Linear elastic material properties are applied to both the
inclusion ($E^1,\nu^1$) and the matrix ($E^2,\nu^2$). The inclusion is made more
rigid than the matrix by choosing $E^1/E^2=1000$, $E^1 = 1$ GPa, the same
Poisson's ratio is used for both $\nu^1=\nu^2=0.3$.  A uniform pressure
$\sigma_0=0.1$ MPa is applied on the right side as shown in
Fig.~\ref{fig:eshelby_geom_setup}(a), displacements on the left side are fixed
in horizontal direction $u_x = 0$ and the lower left corner is fixed. The
inclusion patch is tied to the host matrix along the boundary of the inclusion
$\Gamma_g^1$.

The analytical solution for the stress state inside and outside the inclusion is given below in polar coordinates ($r,\theta$)~\cite{kachanov_handbook_2013}. 
Stress components inside the inclusion ($r < R$) are given by:
\begin{align}
\sigma_{rr}^-&=\frac{\sigma_0}{2}\bigg(\beta^1+\delta^1\cos2\theta\bigg)\label{eq:esh_in_sig_rr}\\
\sigma_{\theta\theta}^-&=\frac{\sigma_0}{2}\bigg(\beta^1-\delta^1\cos2\theta\bigg)\\
\sigma_{r\theta}^-&= -\frac{\sigma_0}{2} \delta^1\sin2\theta
\label{eq:analytic_eshelby_inside}
\end{align}
where, 
\begin{equation}
\beta^1=\frac{\mu^1(k^2+1)}{2\mu^1+\mu^2(k^1-1)},\quad\delta^1 =
\frac{\mu^1(k^2+1)}{\mu^2+\mu^1k^2}.
\label{eq:esh_inside_constants}
\end{equation}
Outside the inclusion ($r>R$), the stress components are given by:
\begin{align}
\sigma_{rr}^+&=\frac{\sigma_0}{2}\bigg[1-\gamma^2\frac{R^2}{r^2}+\bigg(1-2\beta^2\frac{R^2}{r^2}-3\delta^2\frac{R^4}{r^4}\bigg)\cos2\theta\bigg]\label{eq:esh_out_sig_rr}\\
\sigma_{\theta\theta}^+&=\frac{\sigma_0}{2}\bigg[1+\gamma^2\frac{R^2}{r^2}+\bigg(1-3\delta^2\frac{R^4}{r^4}\bigg)\cos2\theta\bigg]\label{eq:esh_out_sig_tt}\\
\sigma_{r\theta}^+&=-\frac{\sigma_0}{2}\bigg(1+\beta^2\frac{R^2}{r^2}+3\delta^2\frac{R^4}{r^4}\bigg)\sin2\theta
\label{eq:esh_out_sig_rt}\\
\end{align}
where
\begin{equation}
\beta^2=-\frac{2(\mu^1-\mu^2)}{\mu^2+\mu^1k^2},\quad\delta^2=
\frac{\mu^1-\mu^2}{\mu^2+\mu^1k^2},\quad\gamma^2=\frac{\mu^2(k^1-1)-\mu^1(k^2-1)}{2\mu^1+\mu^2(k^1-1)}.
\label{eq:esh_outside_constants}
\end{equation}
For the considered plane strain formulation the material constants
$\mu^{1,2}$ and $k^{1,2}$ are given by
\begin{equation}
    \mu^{1,2} = \frac{E^{1,2}}{2(1+\nu^{1,2})},\quad k^{1,2} = 3-4\nu^{1,2}.
\end{equation}

\subsection{Mesh convergence\label{subsec:mesh_convergence}}

\begin{figure}[htb!]
\centering
\includegraphics[width=\textwidth]{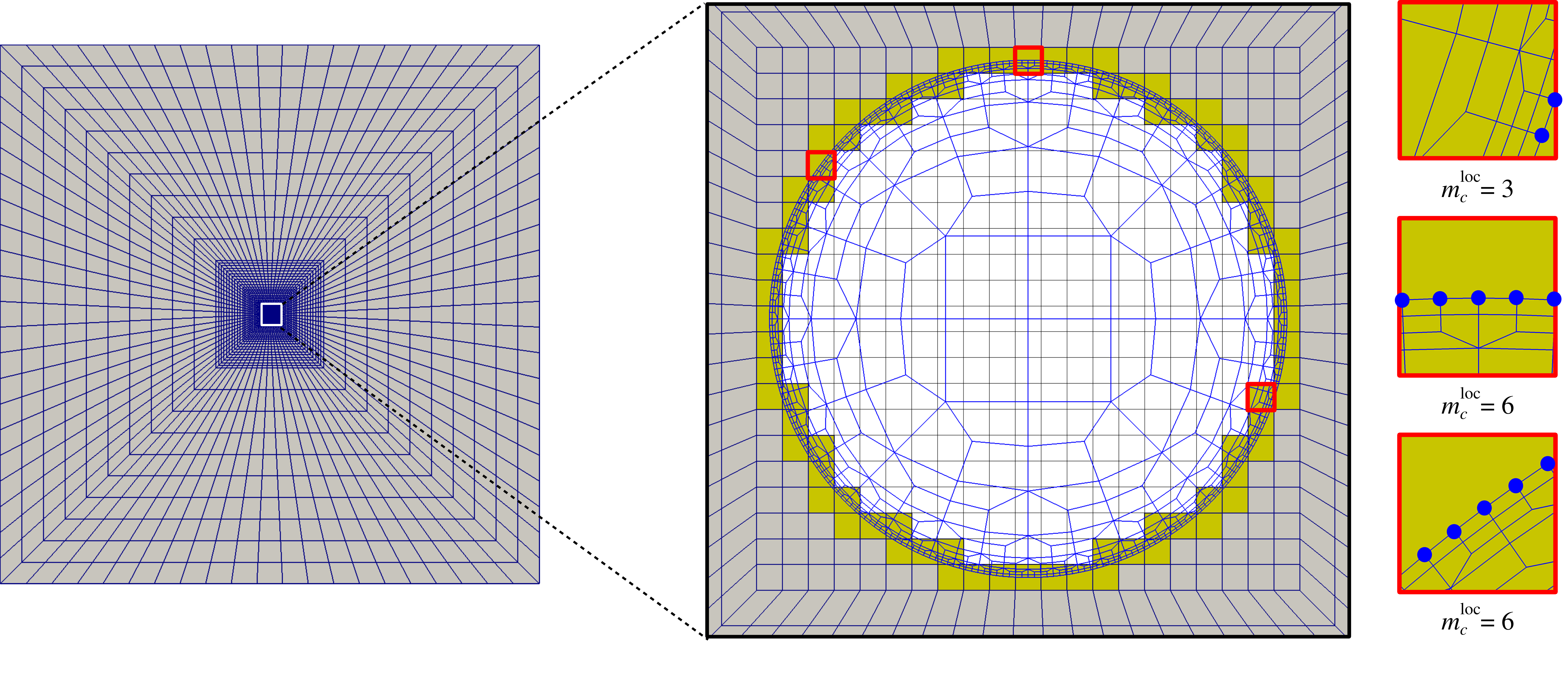}
\caption{Illustration of the local mesh density contrast. The ratio of mortar segment per number of blending elements is  $m_c \approx3$, but locally the number of mortar segments intersecting different blending elements can vary considerably $m_c^{\mathrm{loc}} = 3,6,6$ in zooms shown on  the right.}
\label{fig:eshelby_dr_show}
\end{figure}

We are particularly interested in the stress state along the tying boundary
($r=R$) where possible spurious oscillations take place.  The
distribution of the radial stress component $\sigma_{rr}(\theta)$ at $r=R$ can be obtained
either from\footnote{Note however that $\sigma_{\theta\theta}$ is not
continuous across the interface} Eqs.~\eqref{eq:esh_in_sig_rr} or 
\eqref{eq:esh_out_sig_rr}.  This analytical solution is compared with the
numerical one obtained using MorteX method along the interface
$\Gamma_{}^* \subset\Gamma_g^1$, for which $\theta\in[-\pi/2, \pi/2]$ [see
Fig.~\ref{fig:eshelby_geom_setup}(b)].  For this purpose we use the $L^2$
error norm as in~\eqref{eq:error_norm_bending} defined along $\Gamma_{}^*$:
\begin{equation}\label{eq:error_norm_eshelby}
    E_r(\sigma_{rr}) =
    \frac{||\sigma_{rr}-\lambda_{rr}||_{L^2(\Gamma_{}^*)}}{||\sigma_{rr}||_{L^2(\Gamma_{}^*)}}
\end{equation}
where $\sigma_{rr}$ is the analytical solution, and $\lambda_{rr}$ is the radial
component of Lagrange multiplier vector obtained by projecting it on the radial
basis vector of polar coordinates $\lambda_{rr} = \vec \lambda\cdot\vec e_r$.
The mesh refinement is carried out maintaining a constant mesh contrast between
the patch and the host meshes, i.e. the ratio of the mortar segments to the
number of blending elements is fixed to be $m_c\approx3$
(Fig.~\ref{fig:eshelby_dr_show}); the number of mortar segments was varied $N_m \in \{256,512,1024,2048\}$ (in
Fig.~\ref{fig:eshelby_geom_setup}(b) and \ref{fig:eshelby_dr_show} the coarsest mesh with $N_m = 256$ is shown). Four
cases are considered for this convergence study:
    (i) standard p1 interpolation (SLI) is used for Lagrange multipliers;
    (ii) blending elements are triangulated;    
    (iii) coarse grained interpolation (CGI) is used for  Lagrange multipliers with various coarse graining parameter $\kappa$;
    (iv) both triangulation and coarse graining are used.

\begin{figure}[htb!]
\centering
\includegraphics[width=\textwidth]{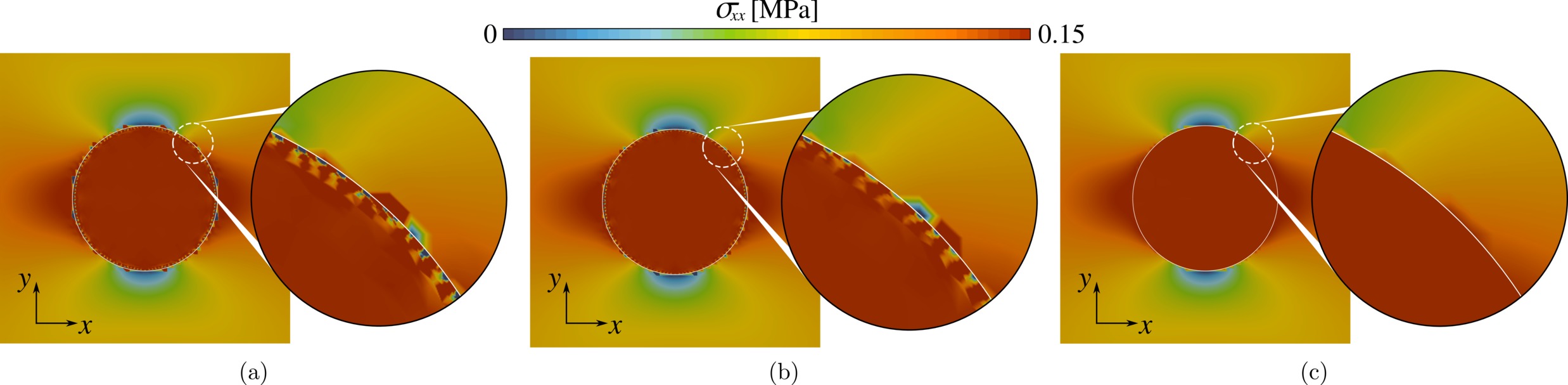}
\caption{Stress component $\sigma_{xx}$ computed using MorteX method: (a) standard interpolation (SLI), (b) coarse grained interpolation (CGI) for $\kappa=4$, (c) CGI for $\kappa = 16$.}
\label{fig:eshelby_field}
\end{figure}

\begin{figure}[htb!]
\centering
\includegraphics[width=\textwidth]{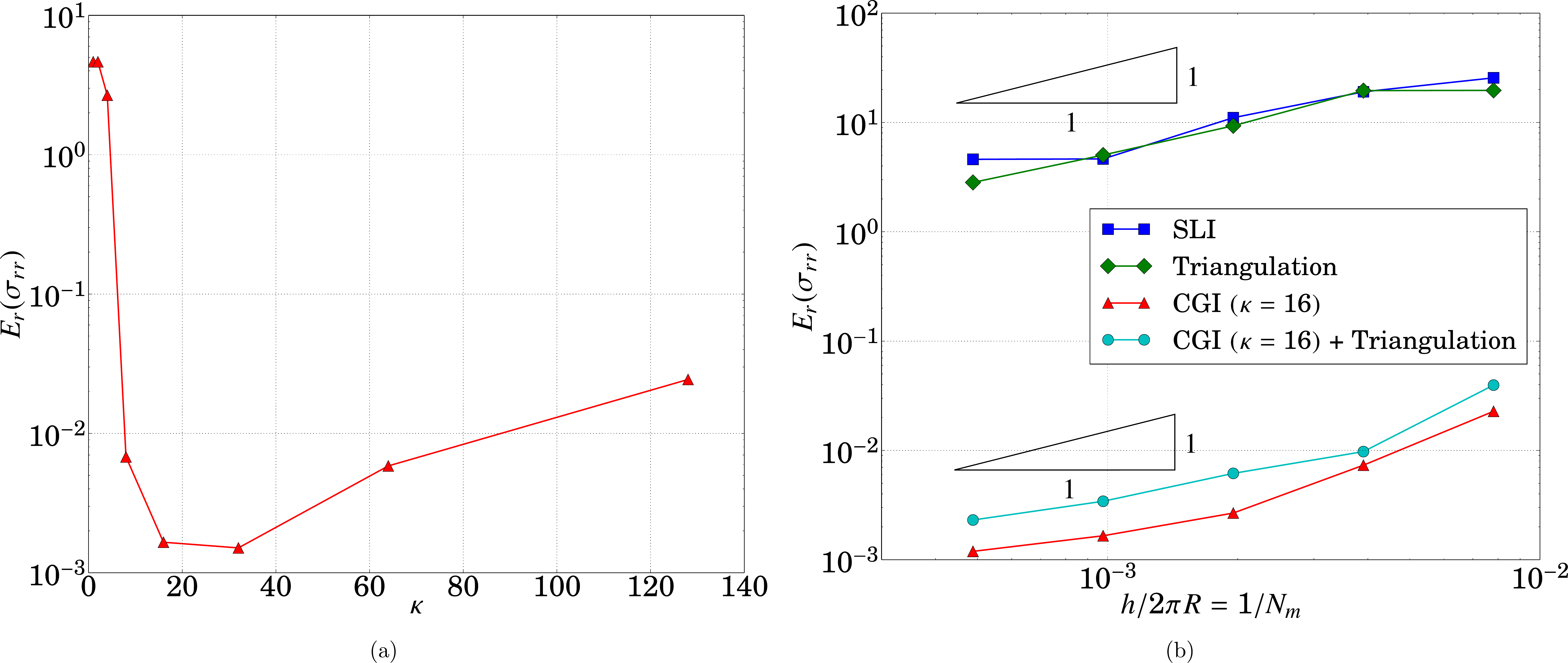}
\caption{Convergence study results for the circular inclusion problem for mesh
contrast $m_c\approx6$: (a) $E_r(\sigma_{rr})$ error change with coarse-graining
parameter $\kappa$, (b) comparison of convergence of SLI and CGI with and
without triangulation of blending elements, the mesh size along the interface is
normalized by the circumference $h/2\pi R=1/N_m$ which is equivalent to the
inverse of the number of mortar segments.}
\label{fig:eshelby_conv_k_study}
\end{figure}

\begin{figure}[htb!]
\centering
\includegraphics[width=\textwidth]{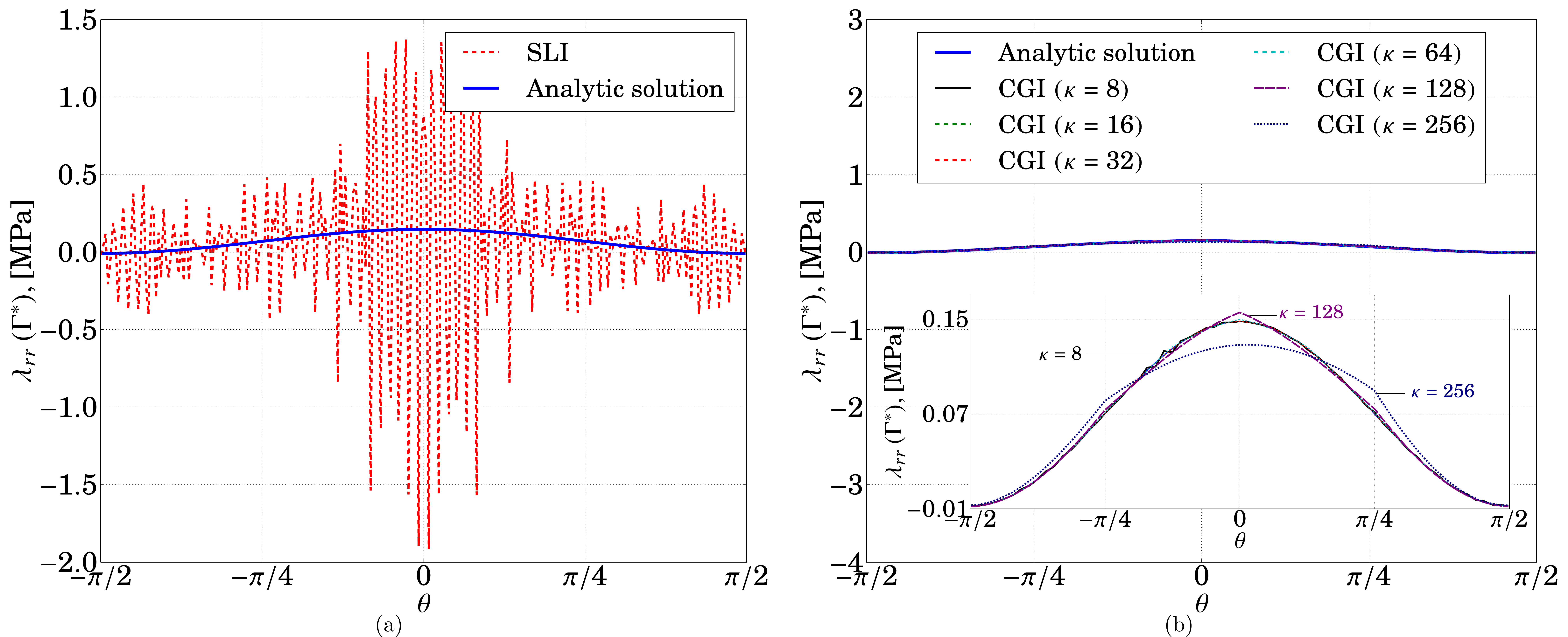}
\caption{Comparison of $\lambda_{rr}$ with analytical solution for various
values of $\kappa$ along $\Gamma^*$ (for the mesh with $N_m=1024$): (a) the standard Lagrange
multiplier spaces ($\kappa=1$), (b) the coarse grained Lagrange multiplier
solution for $\kappa\in[8,32,64,128,256]$. }
\label{fig:size_8_cgi_effect}
\end{figure}

We recall that to represent the interfacial
tractions in compression and bending patch tests one or two Lagrange
multipliers, respectively, were enough. 
In contrast to these patch tests, here the stress distribution is no
longer affine along the tying interface, 
therefore it is expected to obtain a more
practical result for the selection of the coarse graining parameter $\kappa$,
which ensures optimal convergence. 
Results obtained for mesh contrast
$m_c\approx6$ and for $N_m=1024$ and different coarse-graining are shown in
Fig.~\ref{fig:eshelby_field}. The oscillations are clearly seen near the
inclusion/matrix interface, especially inside the inclusion. However, for high
enough $\kappa$ these oscillations are completely removed and a uniform stress
field is recovered inside the inclusion.  
A quantitative convergence study
is presented in Fig.~\ref{fig:eshelby_conv_k_study}(a). It clearly demonstrates  that for
standard interpolation (SLI, i.e. $\kappa=1$) or coarse-grained interpolation
(CGI) used with small values of $\kappa = \{2,4\}$, the presence of spurious
oscillations induces very high errors in interfacial tractions [see
Fig.~\ref{fig:size_8_cgi_effect}(a)]. 
When
$\kappa={16,32}$, the error reaches its minimum. This is due to the fact that
for the given discretization, this level of coarse graining offers an
    appropriate balance between on the one hand the relaxation of the
    over-constraining of the Lagrange multipliers, and on the other hand the
    ability to accurately describe the complex traction field at the interface. For higher values
of $\kappa = \{64,128\}$ the error increases again because of too coarse
representation of interfacial tractions. It is thus expected that, in general
case, there exists a range for $\kappa$ which ensures oscillation free and
accurate enough solution. It is also expected that optimal $\kappa$ is
determined by the global mesh density contrast $m_c$. However, as demonstrated
in Fig.~\ref{fig:eshelby_dr_show}, the local  mesh density contrast can be more
pronounced than the average one, therefore it is expected that the optimal value
of coarse graining parameter $\kappa$ lies in the range $\kappa > m_c$; for the
considered case the error is minimized for $\kappa/m_c = \{2.667,5.333\}$,
probably, the optimal value lies in between.  The effect of optimal $\kappa$ is
clearly demonstrated in Fig.~\ref{fig:size_8_cgi_effect}(b) where interfacial
tractions $\lambda_{rr}$ for different $\kappa$ are plotted.

For the fixed mesh contrast $m_c=6$ and optimal $\kappa = 16$ and sub-optimal
$\kappa = 8$, the mesh convergence was carried out with meshes of different
densities $N_m \in \{128,256,512,1024,2048\}$.  In
Fig.~\ref{fig:eshelby_conv_k_study}(b) we plot the error decay with decreasing
mesh size, for which we select the length of mortar edge normalized by the total
length of the interface $h/2\pi R = 1/N_m$.  For the selected error-measure
along the inclusion/matrix interface, the standard interpolation for Lagrange
multiplier (SLI) results in optimal convergence ($E_r \sim h$).  However, even
though the convergence is optimal, the error remains very high due to the
spurious oscillations, implying that an excessively fine mesh would be required
to achieve an acceptable error. For example, to reach $E_r = 0.1$ \% in case of
SLI, 51\,200 elements on the mortar side would be needed. Moreover, for
quadrilateral host mesh in absence of triangulation of blending elements, the
convergence is lost for very fine meshes. In contrast, the coarse graining
technique (CGI) used with optimal $\kappa = 16$ results in the error below $0.1$
\%, even with the coarsest mesh used in our study $N_m = 128$. At the same time,
the optimal convergence is preserved.  As expected, the triangulation of the
blending elements slightly deteriorates the quality of the solution, but
preserves the optimality of the convergence.

\begin{figure}[htb!]
    \centering
    \includegraphics[width=\textwidth]{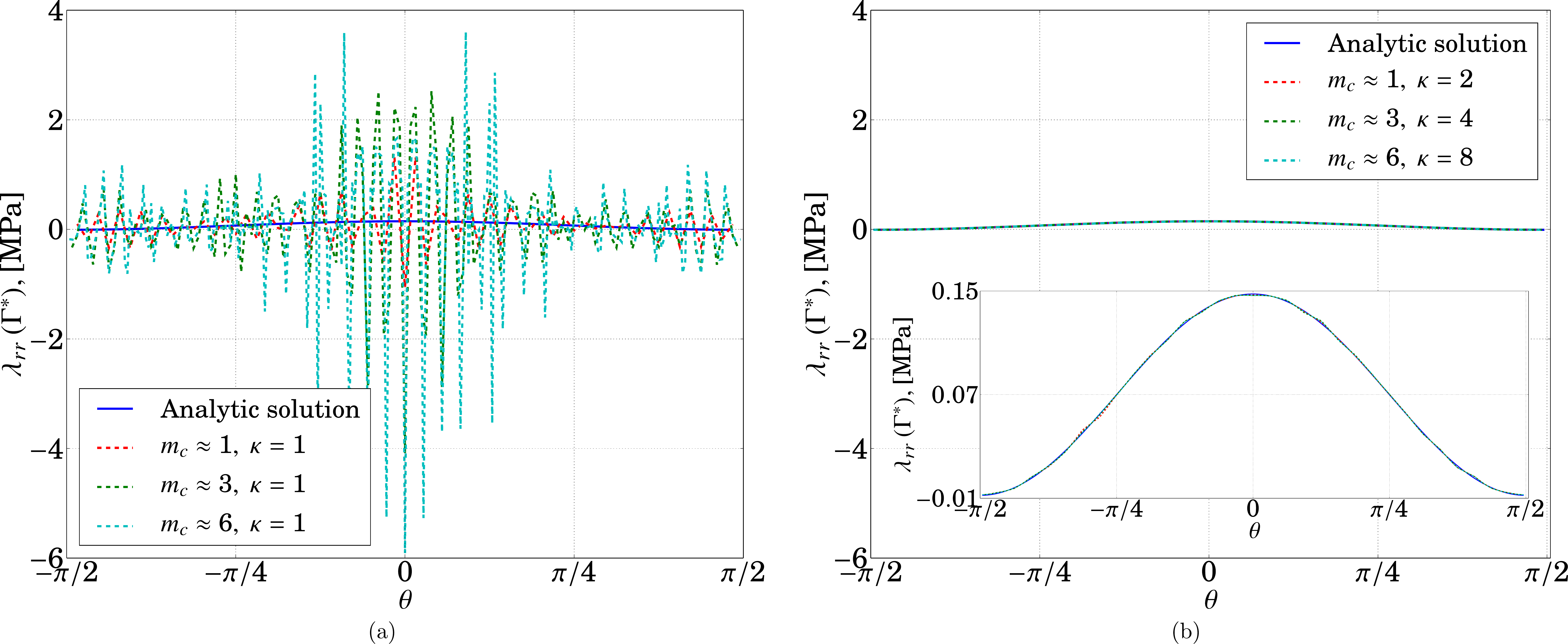}
    \caption{The effect of $m_c$ on $\lambda_{rr}$ along $\Gamma_{}^*$:
    (a) with standard Lagrange multiplier spaces ($\kappa=1$), (b) with coarse
    grained Lagrange multiplier spaces ($\kappa{\approx}m_c$).	}
    \label{fig:d_r_effect_eshelby}
\end{figure}

The effect of the parameter $m_c$ on the amplitude
of spurious oscillations is demonstrated in Fig.~\ref{fig:d_r_effect_eshelby}.  For a fixed host-mesh discretization we increase the
number of mortar edges $N_m=[64,128,256]$, which correspond to $m_c \approx \{1,3,6\}$.
Fig.~\ref{fig:d_r_effect_eshelby}(a) demonstrates the increase in the amplitude of oscillations with increasing mesh contrast $m_c$ for SLI interpolation. 
The removal of spurious oscillations with the CGI scheme is shown in Fig.~\ref{fig:d_r_effect_eshelby}(b) for reasonable choice of coarse graining parameter $\kappa=\{2,4,8\}$ for  $m_c \approx \{1,3,6\}$, respectively.

\section{Numerical examples\label{sec:numerical_examples}}
In this section we illustrate the method of mesh tying along embedded interfaces
in light of potential applications. In all the presented examples, we use a
linear elastic material model under plane strain assumption.  All the problems
are solved in the in-house finite element suite
Z-set~\cite{besson_large_1997}. All triangular and quadrilateral elements used in simulations
possess three and four Gauss points for integration, respectively.
The
triangulated blending elements also use three Gauss points.  The
MorteX interface uses three integration points to evaluate the MorteX
integrals.

\subsection{Plate with a hole} 

\begin{figure}[htb!]
   \centering
   \includegraphics[width=\textwidth]{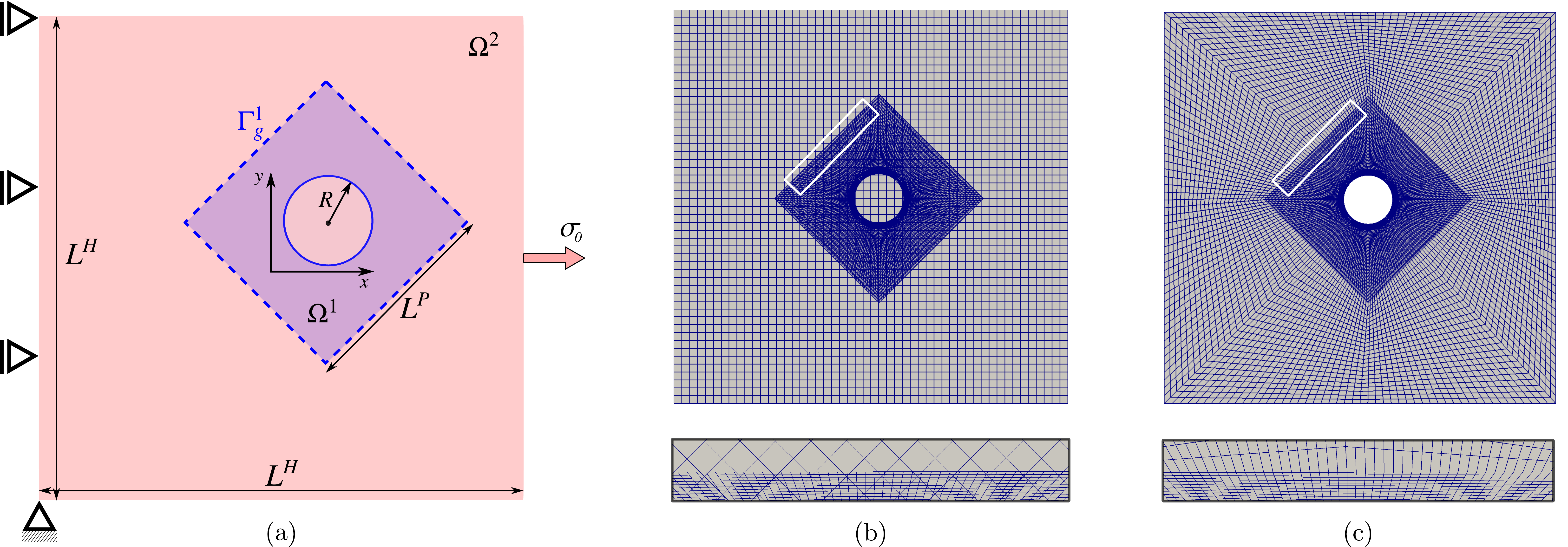}
   \caption{Plate with a hole: (a) overlapping domain setting, (b) discretized
   overlapping domains, (c) monolithic discretization used to obtain the reference solution.}
   \label{fig:plate_with_hole_setup}
\end{figure}

As a first example, we solve the problem of a square plate with an embedded square patch
containing a circular hole, which was used to illustrate the method in Section~\ref{sec:methodology} (Fig.~\ref{fig:plate_with_hole}).  
This example demonstrates the ease with which arbitrary geometrical features can be included into the host mesh. 
Classically, in the X-FEM method a void can be easily
included in the host mesh, however, in the vicinity of the void a stronger stress gradients take place, therefore the mesh around the void should be properly refined. 
It can be easily achieved by surrounding the void with a finer patch mesh, as done here, and by embedding this refined geometry in a coarse host mesh.  
The geometric dimensions used in the problem are the following: the plate's side is $L^H = 12$ mm, the hole's radius is
$R=0.75$ mm and the patch's side is $L^P = 4.5$ mm [Fig.~\ref{fig:plate_with_hole_setup}(a)]. The patch and the host are made of the same material with
Young's modulus $E=1000$ {MPa} and Poisson's ratio $\nu=0.3$. 
The left edge of the host domain is fixed in $x$ ($u_x=0$), the lower left corner is fixed, and a uniform traction
$\sigma_0=1$ MPa is applied on the right edge, the upper and lower boundaries remain
free $\sigma_{xy} = \sigma_{yy} = 0$. For comparison
purposes, a reference solution is obtained with a classical monolithic mesh  [Fig.~\ref{fig:plate_with_hole_setup}(c)].

\begin{figure}[htb!]
   \centering
   \includegraphics[width=\textwidth]{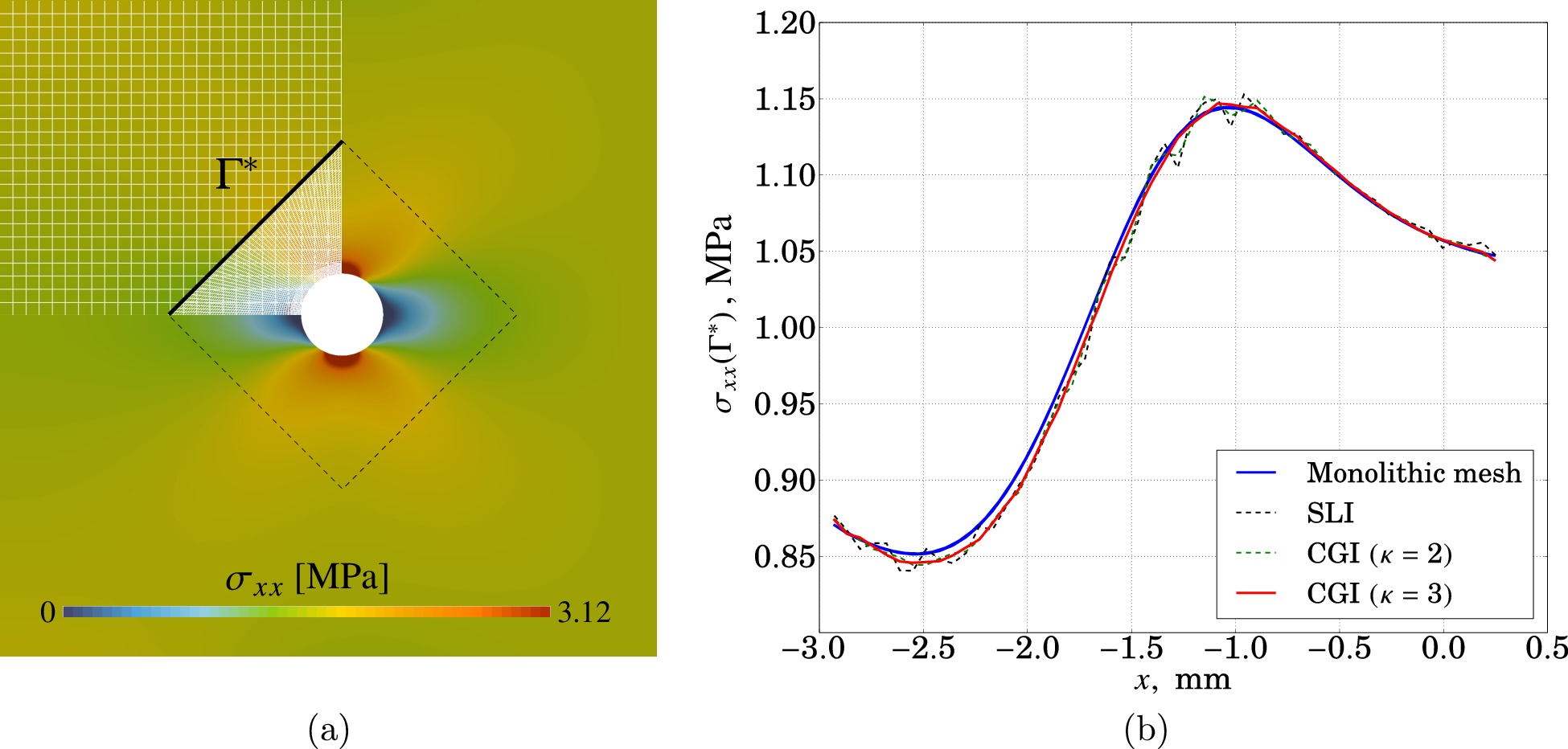}
   \caption{Regular host mesh with an embedded patch containing a circular hole
   ($m_c\approx 3$):
   (a)  contour plot of $\sigma_{xx}$ stress component obtained using  MorteX
   (SLI scheme), (mesh is shown only in the second quadrant, the interface is
   marked with a dashed line; (b) comparison of $\sigma_{xx}$ distributions
   along $\Gamma^*$ between the solution obtained with a monolithic mesh
   [Fig.~\ref{fig:plate_with_hole_setup}(b)] and MorteX solution obtained with
   overlapping meshes [Fig.~\ref{fig:plate_with_hole_setup}(a)] using SLI and
   CGI schemes.}
   \label{fig:plate_with_hole_result}
\end{figure}

Fig.~\ref{fig:plate_with_hole_result}(a) shows a rather smooth contour of stress
component $\sigma_{xx}$,  which was obtained using MorteX tying with SLI scheme.
However, the seemingly smooth stress field near the interface, 
exhibits oscillations near the interface as can be seen in
Fig.~\ref{fig:plate_with_hole_result}(b),
where $\sigma_{xx}$ was plotted over a part of the interface $\Gamma^*$. These
oscillations have a smaller amplitude than in cases with high material contrast,
and as previously, they can be efficiently removed when coarse-grained
interpolation CGI is used, what is shown in the same figure. Coarse-graining
parameter $\kappa=3$ appears to be sufficient to remove them. Note that the slight
difference between the MorteX tying and the monolithic mesh comes from
inherently different extrapolation/interpolation of stresses to the
interface nodes.

\subsection{Crack inclusion in a complex mesh}
\begin{figure}[htb!]
    \centering
    \includegraphics[width=\textwidth]{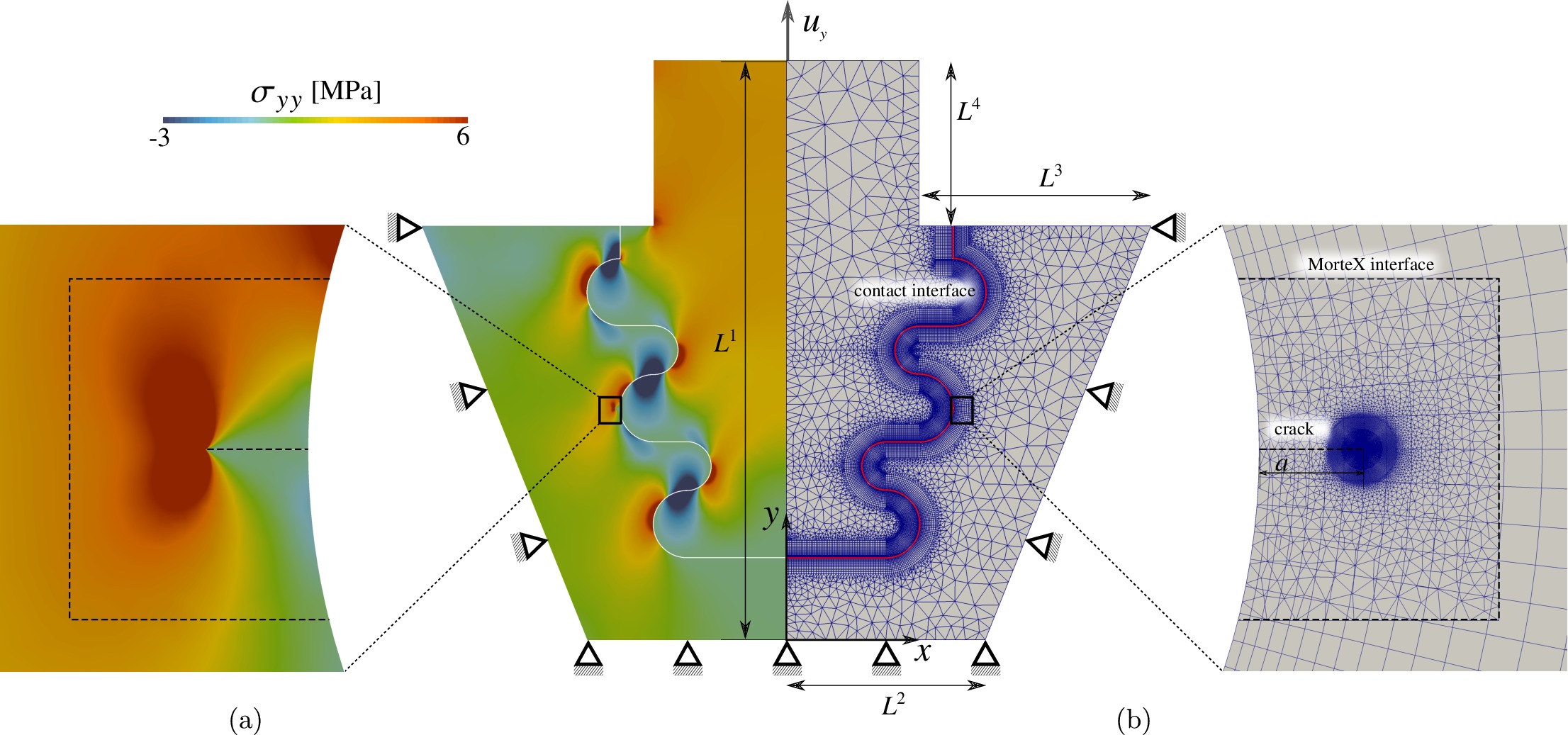}
    \caption{Problem setting for the model blade-disk assembly with an embedded
    patch mesh with a crack: (a) shows stress component $\sigma_{yy}$ contour plot and a zoom near the tying interface; (b) shows used host and patch mesh.}
    \label{fig:blade_disc_sigYY}
\end{figure}

In many engineering applications, the solids are subjected to cyclic loads
and therefore modeling of structures with fatigue cracks appears essential for
computational lifespan prediction. The structural finite element analysis can
indicate potential locations of the onset of fatigue cracks, however, insertion
of cracks is not always
trivial~\cite{proudhon_3d_2016,feld-payet_new_2015}, especially,
in the common case where the original CAD model is not
available 
. Moreover, the position of the onset of the crack is subjected to
statistical perturbations, therefore it is often of interest to probe
various scenarios in which the crack starts at different locations. 
Within the proposed framework, studying various fracture scenarios (crack in
this case) merely
implies placing the patch at a different location on a host mesh, avoiding
potential creation of conformal geometries. 
Here we demonstrate an example of incorporating a crack in a model blade-disk
fir-tree connection subject to a vertical tensile load.  The frictionless contact is handled
using the augmented Lagrangian method in the framework of the standard mortar method.
The following dimensions [see Fig.~\ref{fig:blade_disc_sigYY}(a)] are used for the
blade disk assembly: $L^1=35$ mm, $L^2=12$ mm, $L^3=14$ mm and $L^4=10$ mm.
The Young's modulus is $E=1000$ MPa and $\nu=0.3$ for the blade, disk, and the
patch containing the crack of length $a=0.3$ mm. A vertical displacement
$u_y=0.2$ mm is applied on the top
surface of the blade.  In Fig.~\ref{fig:blade_disc_sigYY}(a) we
present
the resulting stress field for the case of intact structure and for the case of
a structure with embedded crack, respectively. As seen in the later case, the stress fields
are very smooth across the tying interface  (shown as white dashed boundary)
ensured by MorteX with SLI only.
The coarse graining is not needed here as the mesh densities are comparable and the same material is used for the patch and for the host.
Similarly to the presented case of crack insertion, the method
can be used in general for introducing various other geometric features into the existing mesh. 
Using the MorteX method, the location/orientation of these features can be adjusted with ease and without remeshing to perform a sensitivity analysis.

\subsection{Multi-level submodeling: patch in a patch}

\begin{figure}[htb!]
    \centering
    \includegraphics[width=\textwidth]{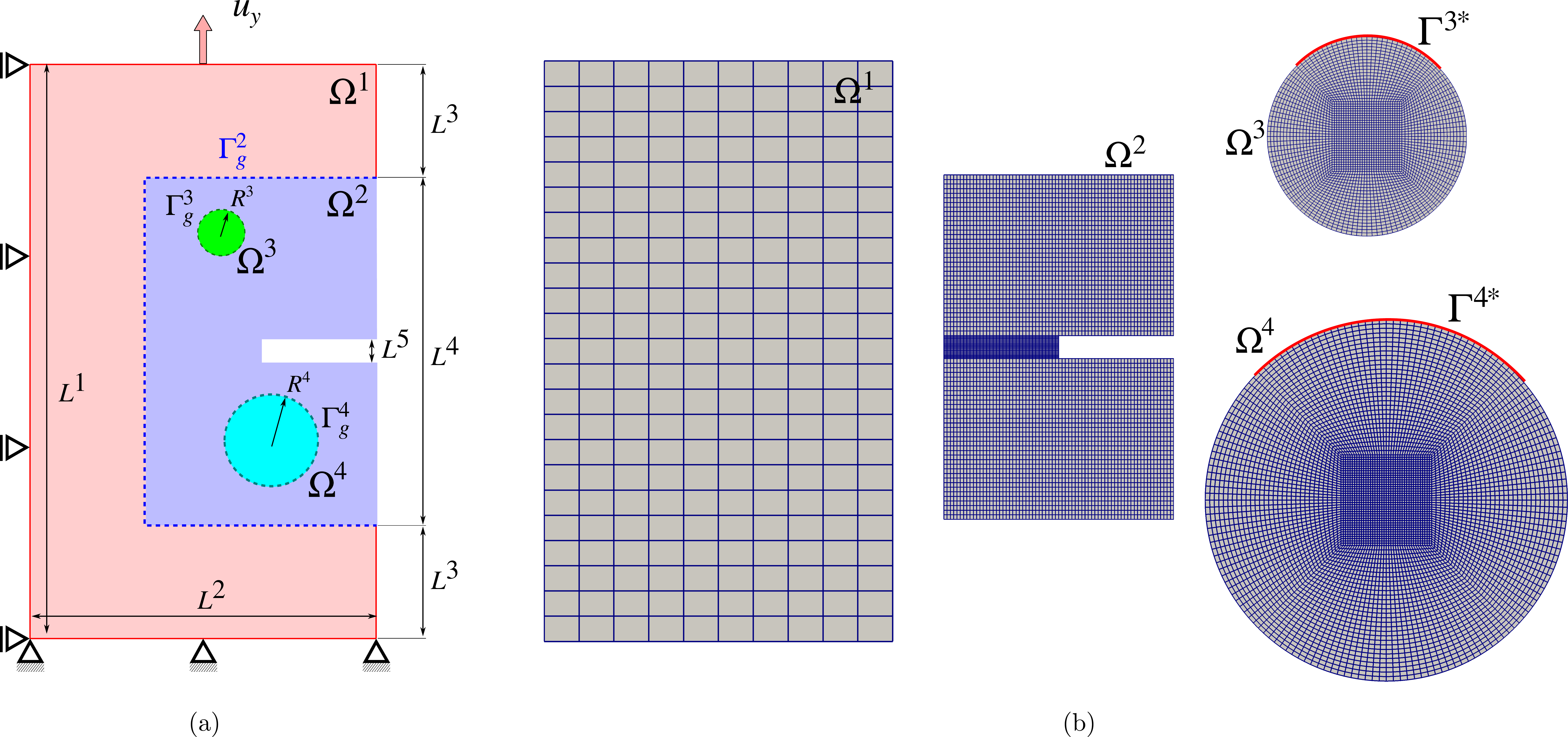}
    \caption{(a) Multi-level overlapping domain set-up; (b) finite element meshes to be coupled (shown not in proportion), $\Gamma^{3*}$, $\Gamma^{4*}$ denote interfaces over which CGI and SLI are compared in Fig.~\ref{fig:multi_patch_incl1_plot}.}
    \label{fig:multi_patch_SETTING_1}
\end{figure}

\begin{figure}[htb!]
    \centering
    \includegraphics[width=\textwidth]{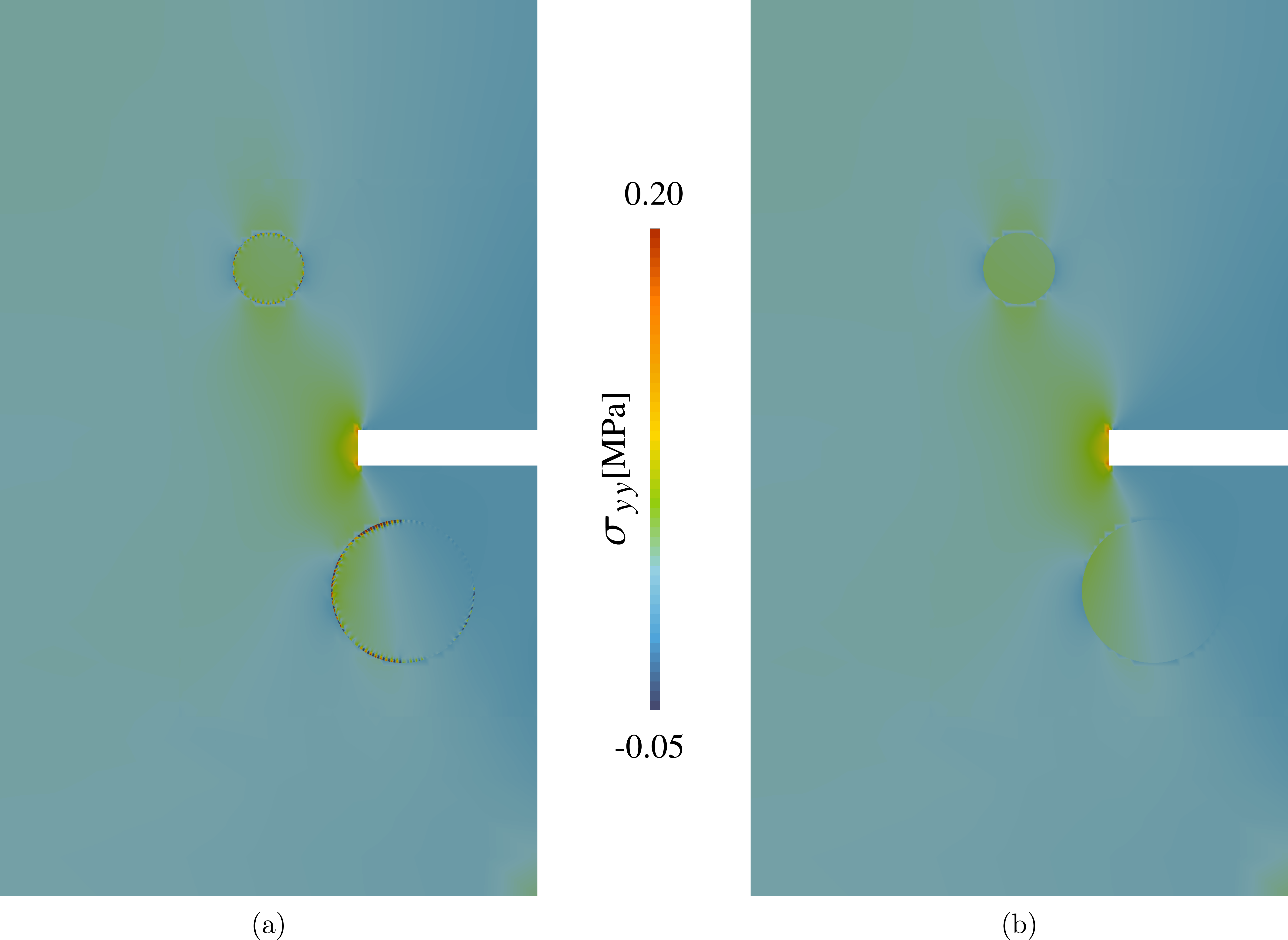}
    \caption{Contour plots of $\sigma_{yy}$ for multi-level overlapping domains: (a) SLI, (b) CGI used with $\kappa = 4$.}
    \label{fig:multi_patch_stress}
\end{figure}

\begin{figure}[htb!]
    \centering
    \includegraphics[width=\textwidth]{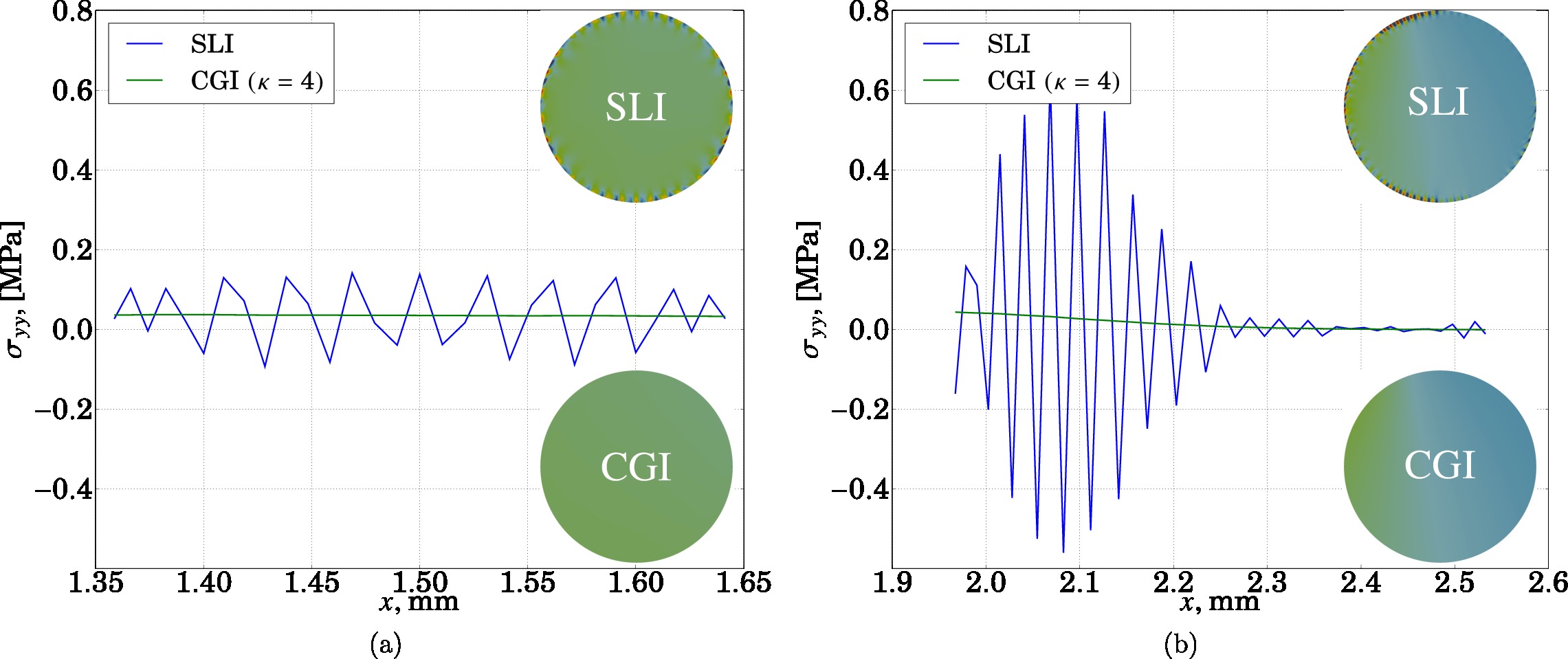}
    \caption{Comparison of $\sigma_{yy}$ stress for standard (SLI) and coarse grained (CGI) 
    interpolations along : (a) $\Gamma^{3*}$, (b) $\Gamma^{4*}$.}
    \label{fig:multi_patch_incl1_plot}
\end{figure}

In this example we demonstrate the ability of the MorteX method to handle
multi-level overlapping domains, i.e. when an embedded patch mesh hosts other
domains.  In Fig.~\ref{fig:multi_patch_SETTING_1}(a) we present such a scenario
where a patch with a notch ($\Omega^2$) is embedded into a host domain
$\Omega^1$, both made of the same material.  At the same time, the patch
$\Omega^2$ itself hosts 2 circular inclusions ($\Omega^{3,4}$), which are
stiffer than the surrounding material.  The following dimensions are used:
$L^1=5$ mm, $L^2=3$ mm, $L^3=1$ mm, $L^4=3$ mm, $R^3=0.2$ mm, and $R^4=0.4$
mm.
The material properties used are: $E^1=1.0$ MPa, $E^2=1.0$ MPa, $E^3=100.0$ MPa,
$E^4=1000.0$ MPa (the upper indices correspond to the domains
$\Omega^1,\,\Omega^3,\,\Omega^3,\,\text{and }\Omega^4$ respectively).
A Poisson's ratio of $\nu=0.3$ is used
for all the domains. A vertical displacement $u_y=0.1$ mm is applied on
    top surface of the $\Omega^1$, while the left surface is fixed in
    the
    $x$ direction and bottom is fixed in all directions.
    The contour plots of
    $\sigma_{yy}$ for the cases of SLI and CGI are shown in
    Fig.~\ref{fig:multi_patch_stress}(a) and (b), respectively.  The
    oscillations in the stress are distinctly seen in case of SLI, but they are removed by applying CGI with $\kappa = 4$. Fig.~\ref{fig:multi_patch_incl1_plot} compares 
    $\sigma_{yy}$ along $\Gamma^{3*}$, $\Gamma^{4*}$, which form $\pi/2$ portions of matrix/inclusion interfaces.
This example illustrates the case where a host mesh with embedded domains of
different material properties can be dealt within CGI MorteX scheme.
Note that in contrast to the Nitsche method where the stabilization parameter needed to avoid mesh locking is
dependent on the local material contrasts~\cite{sanders_nitsche_2012}, the CGI stabilization does not
require the knowledge of this contrast.
In the CGI scheme, knowing a local or a global contrast of mesh densities across the tying interface $m_c$ is enough to automatically select the coarse graining parameter $\kappa$, which 
efficiently stabilizes the mixed formulation and removes spurious oscillations present in the standard mortar scheme.

\section{Conclusion}\label{sec:conclusion}

We presented a unified framework for mesh tying between overlapping domains.
This framework was entitled MorteX as it combines features of the mortar and
X-FEM methods.  As known, the resulting mixed finite element problem may be
prone to mesh locking phenomena especially for high material or mesh-density
contrasts between the host and the patch meshes.  Manifestation of
the emerging spurious oscillations for different element types and various
material as well as mesh contrasts was illustrated on two patch tests (bending and
compression) and on selected examples.  These oscillations strongly deteriorate
solution in the vicinity of interfaces resulting in poor mesh convergence.
Even though triangular elements help to avoid oscillations in compression patch
tests, they do not perform well in bending patch test, nor in more complicated
examples. These oscillations comes from the
over-constraining of the interface in case
of mesh-density contrast, when few mortar-side nodes located on the patch mesh
are tied to displacement field of a single host element. To get rid of the
resulting mesh locking, we suggested to coarse-grain interpolation (CGI) of
Lagrange multipliers by interpolating the associated field along few mortar
edges. It implies that only every $(\kappa+1)$ node along the mortar side stores a
Lagrange multipliers and a linear interpolation is used in between. The value of
coarse-graining spacing parameter $\kappa$ controls the
performance of the scheme.
If $\kappa$ is too small compared to mesh-density contrast, the spurious
oscillations persist as in standard interpolation of Lagrange multipliers (SLI).
If the value of $\kappa$ is too high, the spatial variation of resulting
interface tractions cannot be captured properly. Therefore, in general problem,
there exists an optimal choice for the spacing parameter $\kappa$ which can be
automatically determined either by local mesh-density contrast between the patch
and the host mesh or by the global mesh-density contrast.  The performance of
the MorteX method with coarse-grained interpolation was demonstrated on Eshelby
problem for a stiff inclusion in a softer matrix (elastic contrast of 1000). Few
other examples, demonstrating the ease with which the method can be used for:
submodeling, local mesh refinement and inclusion of arbitrary geometrical
features in the existing mesh, without remeshing. Among these examples, a
multi-level/hierarchical overlapping is shown,
where a patch is inserted
into a host mesh, which in turn is inserted into another host mesh. The MorteX
method equipped with CGI demonstrates a very good performance, removes the mesh
locking oscillations, and ensures optimal convergence.  The important feature of
the method is that its stabilization requires knowledge of local mesh densities
only, thus it presents a good alternative to the Nitsche method, which requires
stabilization constructed on a priori knowledge of material stiffness in the
interface. In analogy with the classical mortar method, the MorteX method can be
extended to handle contact problems along virtual interfaces embedded in a mesh;
this extension is presented in a separate
paper~\cite{akula_mortex_2018}.

\section*{Acknowledgment}
The authors acknowledge financial support of the ANRT (grant CIFRE no
2015/0799). We are also grateful to Nikolay Osipov and St\'ephane Quilici from the
Z-set team for their help with the implementation of the method.


\begin{thebibliography}{10}
\expandafter\ifx\csname url\endcsname\relax
  \def\url#1{\texttt{#1}}\fi
\expandafter\ifx\csname urlprefix\endcsname\relax\def\urlprefix{URL }\fi
\expandafter\ifx\csname href\endcsname\relax
  \def\href#1#2{#2} \def\path#1{#1}\fi

\bibitem{fries_higher-order_2018}
T.~P. Fries, Higher-order conformal decomposition {FEM} ({CDFEM}), Computer
  Methods in Applied Mechanics and Engineering 328 (2018) 75 -- 98.

\bibitem{burman_cutfem:_2015}
E.~Burman, S.~Claus, P.~Hansbo, M.~G. Larson, A.~Massing, {CutFEM}:
  {Discretizing} geometry and partial differential equations, International
  Journal for Numerical Methods in Engineering 104~(7) (2015) 472--501.

\bibitem{claus_stable_2018}
S.~Claus, P.~Kerfriden, A stable and optimally convergent {LaTIn}-{CutFEM}
  algorithm for multiple unilateral contact problems, International Journal for
  Numerical Methods in Engineering 113~(6) (2018) 938--966.

\bibitem{belytschko_structured_2003}
T.~Belytschko, C.~Parimi, N.~Mo{\"e}s, N.~Sukumar, S.~Usui, Structured extended
  finite element methods for solids defined by implicit surfaces, International
  journal for numerical methods in engineering 56~(4) (2003) 609--635.

\bibitem{duboeuf_embedded_2017}
F.~Duboeuf, E.~B{\'e}chet, Embedded solids of any dimension in the {X}-{FEM}.
  {Part} {I} {\textendash} {Building} a dedicated {P}1 function space, Finite
  Elements in Analysis and Design 130 (2017) 80 -- 101.

\bibitem{duboeuf_embedded_2017-1}
F.~Duboeuf, E.~B{\'e}chet, Embedded solids of any dimension in the {X}-{FEM}.
  {Part} {II} {\textendash}{Imposing} {Dirichlet} boundary conditions, Finite
  Elements in Analysis and Design 128 (2017) 32 -- 50.

\bibitem{baaijens_fictitious_2001}
F.~P.~T. Baaijens, A fictitious domain/mortar element method for
  fluid{\textendash}structure interaction, International Journal for Numerical
  Methods in Fluids 35~(7) (2001) 743--761.

\bibitem{fournie_fictitious_2014}
M.~Fourni{\'e}, A.~Lozinski, {others}, A fictitious domain approach for
  {Fluid}-{Structure} {Interactions} based on the {eXtended} {Finite} {Element}
  {Method}., ESAIM: Proceedings and Surveys 45 (2014) 308--317.

\bibitem{puso_embedded_2015}
M.~Puso, E.~Kokko, R.~Settgast, J.~Sanders, B.~Simpkins, B.~Liu, An embedded
  mesh method using piecewise constant multipliers with stabilization:
  mathematical and numerical aspects, International Journal for Numerical
  Methods in Engineering 104~(7) (2015) 697--720.

\bibitem{moes_imposing_2006}
N.~Mo{\"e}s, E.~B{\'e}chet, M.~Tourbier, Imposing {Dirichlet} boundary
  conditions in the extended finite element method, International Journal for
  Numerical Methods in Engineering 67~(12) (2006) 1641--1669.

\bibitem{sanders_methods_2009}
J.~D. Sanders, J.~E. Dolbow, T.~A. Laursen, On methods for stabilizing
  constraints over enriched interfaces in elasticity, International Journal for
  Numerical Methods in Engineering 78~(9) (2009) 1009--1036.

\bibitem{haslinger_new_2009}
J.~Haslinger, Y.~Renard, A new fictitious domain approach inspired by the
  extended finite element method, SIAM Journal on Numerical Analysis 47~(2)
  (2009) 1474--1499.

\bibitem{ramos_new_2015}
A.~C. Ramos, A.~M. Arag{\'o}n, S.~Soghrati, P.~H. Geubelle, J.-F. Molinari, A
  new formulation for imposing {Dirichlet} boundary conditions on non-matching
  meshes, International Journal for Numerical Methods in Engineering 103~(6)
  (2015) 430--444.

\bibitem{babuska_partition_1997}
I.~Babu{\v s}ka, J.~M. Melenk, The partition of unity method, International
  journal for numerical methods in engineering 40~(4) (1997) 727--758.

\bibitem{daux_arbitrary_2000}
C.~Daux, N.~Mo{\"e}s, J.~Dolbow, N.~Sukumar, T.~Belytschko, Arbitrary branched
  and intersecting cracks with the extended finite element method,
  International Journal for Numerical Methods in Engineering 48~(12) (2000)
  1741--1760.

\bibitem{sukumar_modeling_2001}
N.~Sukumar, D.~L. Chopp, N.~Mo{\"e}s, T.~Belytschko, Modeling holes and
  inclusions by level sets in the extended finite-element method, Computer
  methods in applied mechanics and engineering 190~(46) (2001) 6183--6200.

\bibitem{diez_stable_2013}
P.~Diez, R.~Cottereau, S.~Zlotnik, A stable extended {FEM} formulation for
  multi-phase problems enforcing the accuracy of the fluxes through {Lagrange}
  multipliers, International Journal for Numerical Methods in Engineering
  96~(5) (2013) 303--322.

\bibitem{gross_extended_2007}
S.~Gross, A.~Reusken, An extended pressure finite element space for two-phase
  incompressible flows with surface tension, Journal of Computational Physics
  224~(1) (2007) 40 -- 58.

\bibitem{ji_hybrid_2002}
H.~Ji, D.~Chopp, J.~Dolbow, A hybrid extended finite element/level set method
  for modeling phase transformations, International Journal for Numerical
  Methods in Engineering 54~(8) (2002) 1209--1233.

\bibitem{savvas_homogenization_2014}
D.~Savvas, G.~Stefanou, M.~Papadrakakis, G.~Deodatis, Homogenization of random
  heterogeneous media with inclusions of arbitrary shape modeled by {XFEM},
  Computational mechanics 54~(5) (2014) 1221--1235.

\bibitem{faivre_2d_2016}
M.~Faivre, B.~Paul, F.~Golfier, R.~Giot, P.~Massin, D.~Colombo, 2d coupled
  {HM}-{XFEM} modeling with cohesive zone model and applications to
  fluid-driven fracture network, Engineering Fracture Mechanics 159 (2016)
  115--143.

\bibitem{ferte_3d_2016}
G.~Fert{\'e}, P.~Massin, N.~Mo{\"e}s, 3d crack propagation with cohesive
  elements in the extended finite element method, Computer Methods in Applied
  Mechanics and Engineering 300 (2016) 347 -- 374.

\bibitem{sanchez-rivadeneira_stable_2019}
A.~G. Sanchez-Rivadeneira, C.~A. Duarte, A stable generalized/{eXtended} {FEM}
  with discontinuous interpolants for fracture mechanics, Computer Methods in
  Applied Mechanics and Engineering 345 (2019) 876 -- 918.

\bibitem{wohlmuth_discretization_2001}
B.~I. Wohlmuth, Discretization {Methods} and {Iterative} {Solvers} {Based} on
  {Domain} {Decomposition}, Vol.~17 of Lecture {Notes} in {Computational}
  {Science} and {Engineering}, Springer Berlin Heidelberg, Berlin, Heidelberg,
  2001.

\bibitem{gosselet_non-overlapping_2006}
P.~Gosselet, C.~Rey, Non-overlapping domain decomposition methods in structural
  mechanics, Archives of computational methods in engineering 13~(4) (2006)
  515.

\bibitem{keyes_domain_2007}
D.~E. Keyes, O.~B. Widlund, Domain decomposition methods in science and
  engineering {XVI}, Springer, 2007.

\bibitem{mathew_domain_2008}
T.~Mathew, Domain decomposition methods for the numerical solution of partial
  differential equations, Vol.~61, Springer Science \& Business Media, 2008.

\bibitem{bernardi_new_1994}
C.~Bernardi, A new nonconforming approach to domain decomposition: the mortar
  element method, Nonliner Partial Differential Equations and Their
  Applications.

\bibitem{belgacem_spectral_1994}
F.~B. Belgacem, Y.~Maday, A spectral element methodology tuned to parallel
  implementations, Computer Methods in Applied Mechanics and Engineering
  116~(1-4) (1994) 59--67.

\bibitem{bernardi_coupling_1990}
C.~Bernardi, N.~Debit, Y.~Maday, Coupling finite element and spectral methods:
  {First} results, Mathematics of Computation 54~(189) (1990) 21--39.

\bibitem{belgacem_mortar_1998}
F.~B. Belgacem, P.~Hild, P.~Laborde, The mortar finite element method for
  contact problems, Mathematical and Computer Modelling 28~(4) (1998) 263--272.

\bibitem{mcdevitt_mortar-finite_2000}
T.~McDevitt, T.~Laursen, A mortar-finite element formulation for frictional
  contact problems, International Journal for Numerical Methods in Engineering
  48~(10) (2000) 1525--1547.

\bibitem{puso_mortar_2004}
M.~A. Puso, T.~A. Laursen, A mortar segment-to-segment contact method for large
  deformation solid mechanics, Computer methods in applied mechanics and
  engineering 193~(6) (2004) 601--629.

\bibitem{gitterle_finite_2010}
M.~Gitterle, A.~Popp, M.~W. Gee, W.~A. Wall, Finite deformation frictional
  mortar contact using a semi-smooth {Newton} method with consistent
  linearization, International Journal for Numerical Methods in Engineering
  84~(5) (2010) 543--571.

\bibitem{farah_mortar_2018}
P.~Farah, W.~Wall, A.~Popp, A mortar finite element approach for point, line,
  and surface contact, International Journal for Numerical Methods in
  Engineering 114~(3) (2018) 255--291.

\bibitem{babuska_finite_1973}
I.~Babu{\v s}ka, The finite element method with {Lagrangian} multipliers,
  Numerische Mathematik 20~(3) (1973) 179--192.

\bibitem{brezzi_mixed_2012}
F.~Brezzi, M.~Fortin, Mixed and hybrid finite element methods, Vol.~15,
  Springer Science \& Business Media, 2012.

\bibitem{barbosa_finite_1991}
H.~J. Barbosa, T.~J. Hughes, The finite element method with {Lagrange}
  multipliers on the boundary: circumventing the {Babu{\v s}ka}-{Brezzi}
  condition, Computer Methods in Applied Mechanics and Engineering 85~(1)
  (1991) 109--128.

\bibitem{burman_fictitious_2010}
E.~Burman, P.~Hansbo, Fictitious domain finite element methods using cut
  elements: {I}. {A} stabilized {Lagrange} multiplier method, Computer Methods
  in Applied Mechanics and Engineering 199~(41-44) (2010) 2680--2686.

\bibitem{fernandez-mendez_imposing_2004}
S.~Fern{\'a}ndez-M{\'e}ndez, A.~Huerta, Imposing essential boundary conditions
  in mesh-free methods, Computer methods in applied mechanics and engineering
  193~(12-14) (2004) 1257--1275.

\bibitem{bechet_stable_2009}
{\'E}.~B{\'e}chet, N.~Mo{\"e}s, B.~Wohlmuth, A stable {Lagrange} multiplier
  space for stiff interface conditions within the extended finite element
  method, International Journal for Numerical Methods in Engineering 78~(8)
  (2009) 931--954.

\bibitem{hautefeuille_robust_2012}
M.~Hautefeuille, C.~Annavarapu, J.~E. Dolbow, Robust imposition of {Dirichlet}
  boundary conditions on embedded surfaces, International Journal for Numerical
  Methods in Engineering 90~(1) (2012) 40--64.

\bibitem{chahine_etude_2008}
E.~Chahine, Etude math{\'e}matique et num{\'e}rique de m{\'e}thodes
  d'{\'e}l{\'e}ments finis {\'e}tendues pour le calcul en domaines
  fissur{\'e}s, {PhD} {Thesis}, Institut National des Sciences Appliqu{\'e}es
  de Toulouse (2008).

\bibitem{chahine_non-conformal_2011}
E.~Chahine, P.~Laborde, Y.~Renard, A non-conformal {eXtended} {Finite}
  {Element} approach: {Integral} matching {Xfem}, Applied Numerical Mathematics
  61~(3) (2011) 322--343.

\bibitem{mayer_3d_2010}
U.~M. Mayer, A.~Popp, A.~Gerstenberger, W.~A. Wall, 3d
  fluid{\textendash}structure-contact interaction based on a combined {XFEM}
  {FSI} and dual mortar contact approach, Computational Mechanics 46~(1) (2010)
  53--67.

\bibitem{dhia_arlequin_2005}
H.~B. Dhia, G.~Rateau, The {Arlequin} method as a flexible engineering design
  tool, International journal for numerical methods in engineering 62~(11)
  (2005) 1442--1462.

\bibitem{zamani_embedded_2011}
A.~Zamani, M.~R. Eslami, Embedded interfaces by polytope {FEM}, International
  Journal for Numerical Methods in Engineering 88~(8) (2011) 715--748.

\bibitem{sanders_nitsche_2012}
J.~D. Sanders, T.~A. Laursen, M.~A. Puso, A {Nitsche} embedded mesh method,
  Computational Mechanics 49~(2) (2012) 243--257.

\bibitem{fischer_frictionless_2005}
K.~Fischer, P.~Wriggers, Frictionless 2d contact formulations for finite
  deformations based on the mortar method, Computational Mechanics 36~(3)
  (2005) 226--244.

\bibitem{fischer_mortar_2006}
K.~A. Fischer, P.~Wriggers, Mortar based frictional contact formulation for
  higher order interpolations using the moving friction cone, Computer methods
  in applied mechanics and engineering 195~(37) (2006) 5020--5036.

\bibitem{popp_mortar_2012}
A.~Popp, Mortar methods for computational contact mechanics and general
  interface problems, {PhD} {Thesis}, Technische Universit{\"a}t M{\"u}nchen
  (2012).

\bibitem{yastrebov_numerical_2013}
V.~Yastrebov, Numerical methods in contact mechanics, ISTE/Wiley, 2013.

\bibitem{akula_mortex_2018}
B.~R. Akula, J.~Vignollet, V.~A. Yastrebov, {MorteX} method for contact along
  real and embedded surfaces: coupling {X}-{FEM} with the {Mortar} method
  (2018).

\bibitem{sethian_level_1999}
J.~A. Sethian, Level set methods and fast marching methods: evolving interfaces
  in computational geometry, fluid mechanics, computer vision, and materials
  science, Vol.~3, Cambridge university press, 1999.

\bibitem{mei_ear-clipping_2013}
G.~Mei, J.~C. Tipper, N.~Xu, Ear-clipping based algorithms of generating
  high-quality polygon triangulation, in: Proceedings of the 2012
  {International} {Conference} on {Information} {Technology} and {Software}
  {Engineering}, Springer, 2013, pp. 979--988.

\bibitem{besson_large_1997}
J.~Besson, R.~Foerch, Large scale object-oriented finite element code design,
  Computer methods in applied mechanics and engineering 142~(1-2) (1997)
  165--187.

\bibitem{sharma_circular_1979}
C.~Sharma, Circular inclusion in an infinite elastic strip, Zeitschrift f{\"u}r
  angewandte Mathematik und Physik ZAMP 30~(6) (1979) 983--990.

\bibitem{kachanov_handbook_2013}
M.~L. Kachanov, B.~Shafiro, I.~Tsukrov, Handbook of elasticity solutions,
  Springer Science \& Business Media, 2013.

\bibitem{herve_elastic_1995}
E.~Herv{\'e}, A.~Zaoui, Elastic behaviour of multiply coated fibre-reinforced
  composites, International Journal of Engineering Science 33~(10) (1995)
  1419--1433.

\bibitem{eshelby_elastic_1959}
J.~D. Eshelby, The elastic field outside an ellipsoidal inclusion, Proceedings
  of the Royal Society of London. Series A. Mathematical and Physical Sciences
  252~(1271) (1959) 561--569.

\bibitem{muskhelishvili_basic_nodate}
N.~Muskhelishvili, Some basic problems of the mathematical theory of
  elasticity.

\bibitem{proudhon_3d_2016}
H.~Proudhon, J.~Li, F.~Wang, A.~Roos, V.~Chiaruttini, S.~Forest, 3d simulation
  of short fatigue crack propagation by finite element crystal plasticity and
  remeshing, International Journal of Fatigue 82 (2016) 238--246.

\bibitem{feld-payet_new_2015}
S.~Feld-Payet, V.~Chiaruttini, J.~Besson, F.~Feyel, A new marching ridges
  algorithm for crack path tracking in regularized media, International Journal
  of Solids and Structures 71 (2015) 57--69.

\end{thebibliography}

\end{document}